\documentclass[rmp,aps,amsmath,amssym,amsfonts,twocolumn]{revtex4}
\usepackage[dvips]{epsfig}
\usepackage{color}
\usepackage{amssymb}

\newcommand{\phiStar}{\phi^{*}}

\newcommand{\fc}{f_\text{T}}

\newcommand{\eCap}{{\epsilon_\text{cap}}}

\newcommand{\DeltaMu}{{\Delta \mu}}

\newcommand{\epsilonBulk}{\epsilon_\text{bulk}}

\newcommand{\depsMax}{{\Delta \epsilon_{\rm max}}}

\newcommand{\rSphOpt}{\bar{r}_\text{sph}}
\newcommand{\nSphOpt}{\bar{n}_\text{sph}}
\newcommand{\rCylOpt}{\bar{r}_\text{cyl}}
\newcommand{\eCylOpt}{\bar{\epsilon}_\text{cyl}}
\newcommand{\eSphOpt}{\bar{\epsilon}_\text{sph}}
\newcommand{\eLamOpt}{\bar{\epsilon}_\text{layer}}
\newcommand{\rSCOpt}{\bar{r}_\text{SC}}
\newcommand{\eSC}{\epsilon_\text{SC}}
\newcommand{\eSPH}{\epsilon_\text{sph}}
\newcommand{\nSPH}{n_\text{sph}}
\newcommand{\eCYL}{\epsilon_\text{cyl}}
\newcommand{\nCYL}{n_\text{cyl}}
\newcommand{\argmax}[1]{\underset{#1}{\operatorname{arg}\,\operatorname{max}}\;}
\newcommand{\argmin}[1]{\underset{#1}{\operatorname{arg}\,\operatorname{max}}\;}
\newcommand{\nBar}{{n_\text{B}}}
\newcommand{\rBar}{r_\text{B}}
\newcommand{\lBar}{{L_\text{B}}}
\newcommand{\lSCOpt}{L_\text{SC}}
\newcommand{\nMax}{{n_\text{\rm max}}}

\newcommand*{\nT}{n_{\rm T}}
\newcommand*{\RT}{{R_{\rm T}}}
\newcommand*{\eT}{{\epsilon_{\rm T}}}

\newcommand{\kt}{k_\text{B}T}
\newcommand*{\fAssem}{f_\text{assem}}
\newcommand*{\lamTilde}{\tilde{\lambda}}
\newcommand*{\tElong}{\tau_\text{grow}}
\newcommand*{\tNuc}{\tau_\text{nuc}}
\newcommand*{\tHalf}{\tau_{1/2}}
\newcommand*{\nCrit}{n_\text{nuc}}
\newcommand*{\nCritZero}{n_{\text{nuc}}(\Phi)}
\newcommand*{\phiSat}{\Phi_\text{s}}
\newcommand*{\eMin}{\epsilon_\text{min}}

\newcommand{\KG}{K_\text{G}}
\newcommand{\Wc}{W_\text{c}}

\begin{document}
\title{Equilibrium mechanisms of self-limiting assembly}

\author{Michael F. Hagan}
\affiliation{Martin Fisher School of Physics, Brandeis University, Waltham, MA 02454, USA}
\author{Gregory M. Grason}
\affiliation{Department of Polymer Science and Engineering, University of Massachusetts, Amherst, MA 01003, USA}
\begin{abstract}
 Self-assembly is a ubiquitous process in synthetic and biological systems, broadly defined as the spontaneous organization of multiple subunits (e.g. macromolecules, particles) into ordered multi-unit structures.  The vast majority of equilibrium assembly processes give rise to two states:  one consisting of dispersed disassociated subunits, and the other, a bulk-condensed state of unlimited size.  This review focuses on the more specialized class of {\it self-limiting assembly}, which describes equilibrium assembly processes resulting in finite-size structures.  These systems pose a generic and basic question, how do thermodynamic processes involving non-covalent interactions between identical subunits ``measure'' and select the size of assembled structures?  In this review, we begin with an introduction to the basic statistical mechanical framework for assembly thermodynamics, and use this to highlight the key physical ingredients that ensure equilibrium assembly will terminate at finite dimensions.  Then, we introduce examples of self-limiting assembly systems, and classify them within this framework based on two broad categories:  {\it self-closing assemblies} and {\it open-boundary assemblies}.  These include well-known cases in biology and synthetic soft matter --- micellization of amphiphiles and shell/tubule formation of tapered subunits ---  as well as less widely known  classes of assemblies, such as short-range attractive/long-range repulsive systems and geometrically-frustrated assemblies.  For each of these self-limiting mechanisms, we describe the physical mechanisms that select equilibrium assembly size, as well as potential limitations of finite-size selection.  Finally, we discuss alternative mechanisms for finite-size assemblies, and draw contrasts with the size-control that these can achieve relative to self-limitation in equilibrium, single-species assemblies.
\end{abstract}
\pacs{}
\date{\today}

\maketitle
\tableofcontents

\section{Introduction}
\subsection{Overview}

{\it Self-assembly} is a process in which multiple subunits, or ``building blocks'', spontaneously organize into collective and coherent structures.  This process is ubiquitous in living systems, where it underpins a wide range of structures at the cellular and sub-cellular scale, from lipid membranes to multi-protein filaments and capsules~\cite{Alberts2002}.   Inspired by biology's successful strategies to build functional nanostructures, self-assembly is forming the basis of modern approaches to generate materials from the ``bottom up''~\cite{Hamley2003}.  Chemical techniques enable synthesizing a bewildering array of small-molecule, macromolecular, or particulate subunits that are engineered to self-assemble into high-order architectures~\cite{Klok2001, Stupp2014, Boles2016}.  As in the biological context, the assemblies bridge between the scales of molecules and chemical function (nanometer and sub-nanometer) to size scales that are useful for controlling material properties (microns and beyond).  

In different domains of science and engineering, the term ``self-assembly'' often connotes a range of distinct, if overlapping, physical processes.  In its broadest usage, self-assembly implies the collective association of multiple elements into organized configurations, by dynamics that start from a relatively ``disorganized" state and evolve with at least some degree of randomness. The great conceptual appeal of self-assembly in materials science is that the instructions for a desirable or useful structure may somehow be imprinted into the assembling subunits themselves, such that in a simple mixture, the desired target structures emerge from the random processes of Brownian motion and subunit association.  

\begin{figure}
\centering
\includegraphics[width=1\linewidth]{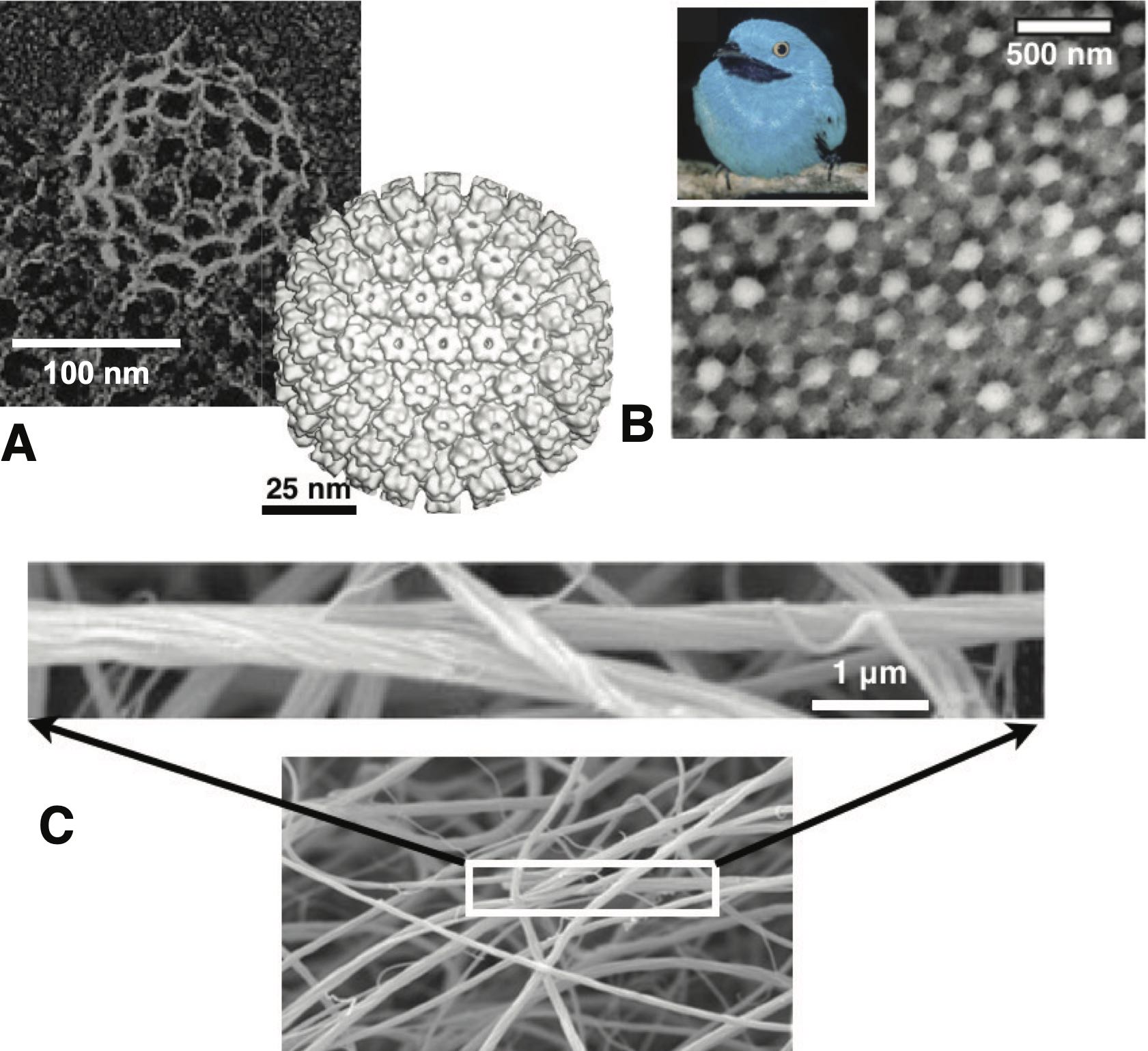}
\caption{\footnotesize  Functional, finite-sized assemblies of proteins in biology:  (A) protein shells of clathrin (left) and viral capsids of Herpes simplex (right); (B) photonic nanostructures form by keratin aggregates in feather barbs of Plum-throated Continga (inset); and (C) finite-diameter fibers in reconstituted fibrin clot.  Figures are adapted from \cite{Royle2012} (A, left) and \cite{Baker2000} (A, right); \cite{Dufresne2009} (B); and \cite{Weisel2004} (C).  }
\label{fig: finite_bio}
\end{figure}

In this article, we focus on {\it self-limiting assembly} (SLA), defined as self-assembly processes that terminate at an equilibrium state in which superstructures have a well-defined and finite spatial extent in one or more dimensions.  Many examples of SLA can be found in biological systems, where the assembly of identical subunits into larger, yet finite-sized, superstructures is common and functionally vital. As shown in Fig.~\ref{fig: finite_bio}, examples include i) the protein shells that enclose viruses \cite{Caspar1962,Mateu2013,Perlmutter2015} and microcompartments \cite{Kerfeld2010,Rae2013,Tanaka2008}, ii) finite-size protein superstructures in photonic tissues \cite{Prum2009, Saranathan2012, McPhedran2015}, and iii) finite-diameter bundles and fibers of cytoskeletal or extracellular protein filaments~\cite{Neville1993, Fratzl2003, Popp2012}.  Each of these examples shares the notable feature that the finite size of the assembled structure far exceeds the nanometer size scale of the protein subunits.  Crucial to their biological roles, the functional properties of these protein superstructures are regulated through the control of their finite size: respectively, (i) selective encapsulation and transport; (ii) optical response; and (iii) stiffness and strength.  In this way, Nature exploits self-assembly to deploy structures, built from the same or similar subunits, in diverse intra-cellular and extra-cellular environments, and adapts their performance and functions by controlling the size of the assembled structure.

In contrast to these examples of SLA, most typical mechanisms of self-assembly in synthetic systems result in \emph{unlimited} organized states, such as crystalline or liquid crystalline mesophases.   In these states, structure may be well-defined on some microscopic scale, such as the unit cell dimension, but its overall size is uncontrolled by assembly thermodynamics.  This result, which may be described as bulk phase separation, is a generic consequence of the thermodynamic trade-off between entropic and energetic drives.  In the most general case, once the net cohesive drive for a subunit to join an assembled structure exceeds the entropic penalty for giving up its higher configurational freedom as a disassociated unit, there is no thermodynamic reason to stop this process. Thus, subunits continually add to the aggregate until it reaches macroscopic proportions and the subunits are nearly depleted. 

This review aims to describe the basic physical ingredients and common outcomes of assembly mechanisms that terminate at well-defined, finite sizes. We draw upon examples of SLA from biological systems, and consider the requirements to achieve such assemblies in synthetic systems.  For clarity of presentation, we specifically focus on assemblies comprising a {\it single species of identical subunits}.  Moreover, we restrict our definition of SLA to \textit{equilibrium} assembly mechanisms, meaning that assembly terminates at a finite-sized \textit{free energy} minimum structure.

Equilibrium assembly processes deserve special focus for both conceptual and practical reasons.  A key advantage is that they are described by well-defined and generic statistical mechanical principles.  This allows one, as we attempt to do in this article, to draw sharp distinctions between assemblies that either {\it are} or {\it are not} self-limiting. Of course, reaching thermodynamic equilibrium requires subunits to associate and disassociate from aggregates sufficiently freely that a thermodynamically large collection of subunits behaves ergodically, sampling a sufficiently large ensemble of aggregation states in an experimentally relevant time.  For systems at or near room temperature, such conditions are accessible when assembly is driven by non-covalent and reversible interactions, of the type that characterize physical association between macromolecules and colloidal particles in solutions~\cite{Israelachvili2011, Russel1989}, including van der Waals, electrostatic, hydrophobic, hydrogen-bonding, and depletion forces.  

The ability of reversibly associating assemblies, if given sufficient time, to proceed toward one specific, thermodynamically-defined state, points to practical advantages of equilibrium assembly.  As evidenced by the synthetic approaches to size-controlled structures referenced above, non-equilibrium control over finite-dimensions of assemblies requires extensive protocols to control the assembly environment, for example, precisely regulating the temporal sequence of temperatures and subunit concentrations.  This makes it exceedingly difficult, if not impossible, to deploy these non-equilibrium size-control strategies in uncontrolled environments, such as the complex and dynamic milieu of living organisms.  In such scenarios where assembly cannot be carefully ``supervised'', equilibrium mechanisms of assembly offer the distinct advantage that the final states may still be well defined.  For example, viruses can exert only limited control over the inter-cellular media of their host organisms.  Nevertheless, to be infectious, size-controlled capsid shells must assemble with high-fidelity from the capsomer subunits.  While this assembly process is in general not purely equilibrium, biology often achieves such high fidelity by building upon equilibrium processes. For example, many viral capsids can spontaneously assemble from their purified components under (near) equilibrium conditions, with structures that are indistinguishable from capsids formed within a host cell~\cite{Fox1998,Wang2015,Wingfield1995,Johnson1997}, and in some cases are even infectious (e.g. \cite{Fraenkel-Conrat1955}).

At the center of this specialized focus on equilibrium mechanisms for self-limiting single-species assembly is the puzzle:  \emph{how can equilibrium association processes ``measure'' the assembly to select a thermodynamic preferred state that is larger than a single subunit, yet less than infinite (i.e. bulk)?}  Because thermodynamic equilibrium is independent of the history of system, this state cannot be defined by the temporal process in which subunits arrive to the aggregate.  Nor do these identical subunits have specific ``addresses'' that prescribe where they are supposed to sit in a particular  aggregation state.  The answers, not surprisingly, lie in how the shape and interactions of subunits conspire to determine the dependence of assembly energetics on size.  For example, in the canonical example of SLA, formation of spherical micelles from amphiphillic molecules~\cite{Israelachvili1976}, the assembly motif favors individual subunits to span the assembly from the solvophobic core to the solvophillic surface.  Hence, in this case it is intuitive that energetics favors aggregates that are limited to sizes that are comparable to the length of amphiphilies themselves.  Far less intuitive is how single-species assemblies select finite equilibrium sizes that are much bigger than the subunit dimensions, or the range of their interactions.    That is to say, what are ``limits of self-limitation'' --- i.e., how large can a self-limited structure be, and how does this size limit depend on the physical characteristics (e.g. shape, interactions) of the subunits?

\subsection{Outline}
With these basic questions in mind, this review has two broad aims.  We first overview the generic statistical and thermodynamic elements of SLA, and then present a broad classification for known mechanisms of SLA of identical subunits.  The article is organized into two main sections based on these aims.  In Sec. \ref{sec: thermo}, we present the thermodynamic principles of SLA based on the statistical mechanics of ideal aggregation of identical subunits.  This begins with an introduction to ideal aggregation theory and illustration of the more generic case of {\it unlimited assembly}.  Following this, we introduce a generic description of the ingredients for SLA. We review how the onset of aggregation -- known as the {\it critical aggregation concentration} (CAC) -- the self-limiting size of aggregates, and the statistics of aggregate size fluctuations depend on the functional form for the size-dependence of the intra-aggregate interaction energy.  We then review the conditions for  ``polymorphic'' SLA, in which assemblies exhibit multiple states of aggregration (some finite, some not). These systems are characterized by so-called secondary CACs, in which increasing concentration sufficiently far above the CAC leads to additional transitions between aggregation states.

\begin{figure}
\centering
\includegraphics[width=1\linewidth]{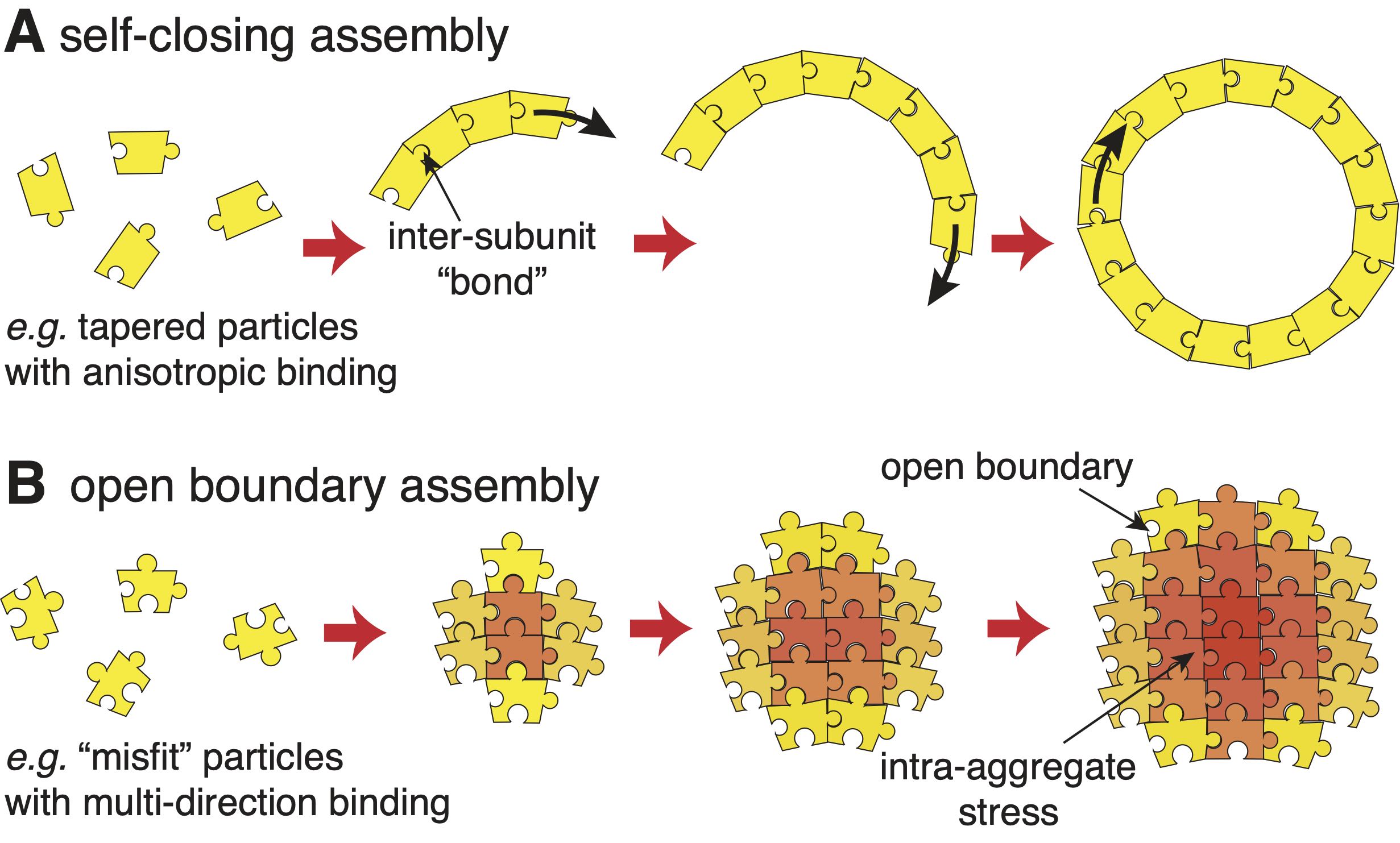}
\caption{\footnotesize Schematic illustrations of two classes of SLA described in Sec. ~\ref{sec: models}: (A) Self-closing assembly, in which inter-subunit rotations lead to cohesive assembly into closed, boundary-free aggregations; and (B) Open boundary (self-limiting) assembly, in which intra-aggregate stress accumulates with assembly and restrains the cohesive drive toward unlimited size.   }
\label{fig: classes}
\end{figure}

Sec. \ref{sec: models} describes physical systems that exhibit SLA, and classifies the models that capture their behavior into two categories illustrated schematically in Fig.~\ref{fig: classes}: {\it self-closing} and {\it open boundary} assembly.  The former category describes assembly processes that terminate because they close upon themselves (Fig.~\ref{fig: classes}A), and applies to shell and tubule formation, as well as the micellar assembly of surfactants, co-polymers, and other amphiphiles.  The latter category applies to arguably lesser known classes of systems that have short-range attractions and long-range repulsions, or form geometrically-frustrated assemblies. These conditions enable assembly to terminate when the aggregate still has ``open boundaries'' characterized by a finite surface energy (Fig.~\ref{fig: classes}B).  For this case, we introduce a generic framework for understanding how the interplay between intra-aggregate stress accumulation and aggregate surface energy controls the finite-size of aggregates and the phase boundary between self-limiting and bulk aggregation. 

While the core focus of this review is on the ingredients and outcomes of SLA from the point of view of thermodynamic equilibrium, the kinetic processes by which such systems reach  equilibrium states (or in some cases, not) are essential to their study, particularly from the experimental point of view.  A comprehensive review of kinetic limitations on assembly, which is relevant to both self-limited and unlimited assembly, is beyond the scope of this article. Nevertheless, we provide a basic introduction to some key considerations of assembly kinetics in Sec. \ref{sec: kinetics}.  In that section, our purpose is to illustrate how those features of the assembly energetics that give rise to size-selection in equilibrium influence the principle kinetic pathways of their formation.

Before concluding, we provide a discussion in Sec. \ref{sec: othermeans} of physical mechanisms leading to finite-size aggregates that fall outside of the major scope of the review, namely non-equilibrium and multi-species SLA, and questions these pose to the forgoing discussion for the more limited focus on single-species equilibrium SLA.  We conclude with some remarks about open challenges in the application of the mechanisms and principles of SLA.

\subsection{Scope of review}
This review considers equilibrium assembly mechanisms that terminate at well-defined, finite sizes.  As this focus suggests, we will leave out discussion of non-equilibrium processes in general, and more specifically, what might be called active-assembly processes, such as the steady-state length of treadmilling and severing cytoskeleltal filaments~\cite{Desai1997,Pollard2016,Mohapatra2016}.  Beyond that, we specifically consider assembly mechanisms of a {\it single species of identical subunits}.  To be sure, this leaves out an emerging and fascinating area of research on so-called ``addressable assemblies~\cite{Jacobs2016, Zeravcic2017}, where mixtures of multiple distinct subunit species may be ``programmed'' to assemble into a specifically defined 3D structure in equilibrium.  In this article, we provide only a limited discussion about size-controlled multi-species assembly and possible trade-offs with single-species mechanisms, particularly how the number of required species increases with target size.  

Although the fabrication and synthesis of finite, size-controlled structures is well-known in synthetic materials, for example, size-controlled nanoparticles of atoms~\cite{Yin2005, Cozzoli2006} and macromolecules~\cite{Hiemenz2007}, these examples raise a key distinction between equilibrium and non-equilibrium assembly.  The control over finite size in all of these foregoing examples relies on the non-equilibrium process by which they form. For example, the size distribution of metal nanoparticles~\cite{Yin2005, OBrien2016} is selected through spatio-temporal control of the physical-chemical factors that control nanocrystal growth (e.g. concentrations, temperature, ionic conditions).  Indeed, as we discuss below, in generic conditions under which such assemblies form, allowing these assemblies to proceed to thermodynamic equilibrium would destroy the size control. Finite-sizes are only possible when these processes are driven, maintained, and arrested out of equilibrium.  In this sense, we reserve the term ``self-limiting'' for those rarefied assembly processes that result in finite-size  structures in thermodynamic equilibrium.   The physical mechanisms of equilibrium assembly that achieve such size control are the central focus of this article.

\section{Thermodynamic elements}

\label{sec: thermo}

We begin with a review of the elementary statistical mechanical framework to describe equilibrium aggregation.  We then illustrate the statistical thermodynamics of aggregation in models of what we will call {\it canonical aggregation}, where assembly proceeds via cohesive (short-range and stress free) assembly of elemental units into 1D, 2D, and 3D aggregates.  We illustrate how so-defined canonical assemblies do not exhibit self-limitation.  We then describe the generic conditions for self-limiting (\emph{finite}) equilibrium assembly, and give an overview of the concentration-dependent thermodynamics of self-limiting assembly. Finally, we discuss models of {\it competing} finite aggregates and {\it polymorphic} transitions between finite to unlimited assembly, both of which may by characterized multiple aggregation thresholds in the ideal theory.

\subsection{Equilibrium principles}

In this review we concern ourselves with equilibrium association of single subunits, or {\it monomers}, into states with aggregation number $n$ subunits, or {\it $n$-mers}.  Our purpose is to describe the minimal ingredients of assembly dominated by structures with finite aggregation number $n$.  To this end, we restrict our presentation to {\it ideal} aggregation theory, where interactions between distinct aggregates are neglected.   This is not to say that interactions among subunits within the same aggregate are neglected.  Quite the contrary, as we describe below, the intra-aggregate energetics, and its $n$-dependence, are critical for determining whether or not association leads to self-limited states, or instead, more canonical states of bulk aggregation.  

\subsubsection{Classical aggregation theory: fixed total concentration, non-interacting aggregates}

Ideal aggregation theory is well established for certain classes of self-assembly systems, particularly in the context of amphiphiles and surfactants.  As such, this theory is better described, and in greater depth, in references such as \cite{Tanford1974, Israelachvili1976, Safran1994, Gelbart1994}.  Here, our purpose is to consider the application and implications of ideal aggregation theory to a broader class of self-limiting assembly systems.  Hence, we only present a minimal introduction to the elements necessary to describe aggregation to self-limiting states.

We consider a solution of $N$ total subunits in a fixed total volume $V$.  In what follows, we refer to unnassembled single subunits as {\it monomers}~\footnote{In literature on amphiphile aggregation the term ``unimer" is often used to describe the single subunit, to avoid overlap with the connotation of ``monomer'' as the chemical repeat of macromolecular chain, which is often a component of self-assembling molecular subunits.}.  To describe the concentration of subunits, it is convenient  to scale concentration by the reference state concentration $v_0^{-1}$; i.e., $v_0$ is the volume per subunit in the reference state~\footnote{For concreteness, we may take $v_0$ to be the subunit volume in the disassociated state, which provides a convenient and non-dimensional measure of concentration which will be much less than unity for dilute conditions. However, the formalism is independent of choice of reference state; for example, $v_0=1/N_\mathrm{AV}$ liter with the reference state concentration of 1 mol/liter commonly used in the life sciences. Changing the definition of the reference state only uniformly rescales the free energy of intra-aggregate interactions $\epsilon(n)$.}. In this way, we can define concentration in non-dimensional terms as the {\it total volume fraction} of subunits, $\Phi = N v_0/V$~\footnote{Throughout, we also refer to a $\Phi$ more colloquially as the ``concentration'' of subunits}.   The $N$ subunits are distributed  among distinct $n$-mer aggregates, with the volume fraction of subunits in $n$-mers defined as $\phi_n$.  In the following, we refer to $\phi_n$ as the {\it subunit fraction distribution}.  Throughout this review, it is also useful to introduce a separate variable for the {\it aggregate distribution} $\rho_n \equiv \phi_n/n$, which describes the relative count of $n$-mers in the mixture.  Defined in this way,  $\phi_n$  and $\rho_n$ are all less than unity for all $n$.  Moreover, for the particular assumptions of ideal aggregation to hold (i.e. two-body contacts between aggregates are vanishingly rare), these quantities must all remain much less than unity. To simplify the nomenclature, throughout the review we refer to the non-dimensional concentration (volume-fraction) simply as the concentration.

We define $n \epsilon(n)$ as the free energy of intra-aggregate interactions; i.e., $\epsilon(n)$ is the {\it per subunit} aggregation free energy in an $n$-mer.  The total free energy $F$ for the ideal distribution of aggregates is given by
\begin{equation}
\label{eq: F}
\frac{F}{ (V/v_0)} = \sum_{n=1}^\infty \phi_n \Big( \epsilon(n) + \frac{\kt}{n}\big[ \ln ( \phi_n/n)-1 \big] \Big) ,  
\end{equation}
with the two terms in the parentheses respectively representing  the intra-aggregate interaction free energy and translational entropy (in the ideal solution approximation) of $n$-mers, with the $1/n$ in the latter term reflecting the critical fact that all subunits of an $n$-mer share a common, single center-of-mass degree of freedom. 

To obtain the equilibrium aggregate size distribution, we minimize $F$ with respect to $\phi_n$, subject to the constraint that the total subunit concentration is fixed: 
\begin{align}
\label{eq: fixN}
\sum_{n=1}^\infty \phi_n = \Phi,
\end{align}
i.e.,
\begin{align}
\frac{\partial }{\partial \phi_n} \Big[F  + \mu \Big(\Phi - \sum_{n=1}^\infty \phi_n \Big) \Big]= 0
\end{align}
with $\mu$ playing the role of a Lagrange multiplier. This yields
\begin{multline}
\label{eq: chemeq}
\mu =  \epsilon(1)+ \kt \ln \phi_1 = \epsilon(2)+ \frac{\kt}{2} \ln (\phi_2/2) = \\ \ldots  = \epsilon(n)+ \frac{\kt}{n} \ln (\phi_n/n) = \frac{\partial{F}}{\partial \phi_n}
\end{multline}
showing that $\mu$ is the subunit chemical potential.
This condition requires that subunits have the same chemical potential in all aggregates, and it derives from both the energetics of the assembly and the (ideal) translational entropy of the $n$-mer.  Note that the prefactor of $1/n$ of the translational entropy,  deriving from the sharing of a single center of mass in an $n$-mer, reflects a generically higher translational entropy of disaggregated states. The limit $n\to\infty$ gives $\mu=\epsilon(\infty)$, describing the equilibrium between monomers and a bulk phase-separated condensate which has no translational entropy.  For simplicity of notation, throughout this article, we choose to define energies such that $\epsilon(1) =0$, in which case $\epsilon(n)$ is defined as the {\it difference} of the per subunit energy between an $n$-mer and disassembled monomers.  

It is convenient to use the first equality in eq. (\ref{eq: chemeq}) to recast the chemical potential in terms of the (unknown) monomer concentration, from which we can reformulate  the generic chemical equilibrium conditions in terms of the {\it law of mass action}
\begin{equation}
\label{eq: lma}
\phi_n = n\Big( \phi_1 e^{- \beta \epsilon(n) } \Big)^n.
\end{equation}
where $\beta^{-1} \equiv \kt$.  Inserting the expression for $\phi_n(\phi_1)$ into the fixed number concentration eq.~(\ref{eq: fixN}) and summing over $n$ then results in an equation of state relating the total concentration $\Phi$ to the monomer concentration $\phi_1$.  This equation of state and the underlying distribution of assemblies, $\phi_n$, derive from the specific $n$-dependence of aggregate interactions, with equilibrium states that are dominated by self-limited aggregates occurring only for certain forms of $\epsilon(n)$.

\begin{figure*}
\centering
\includegraphics[width=1\linewidth]{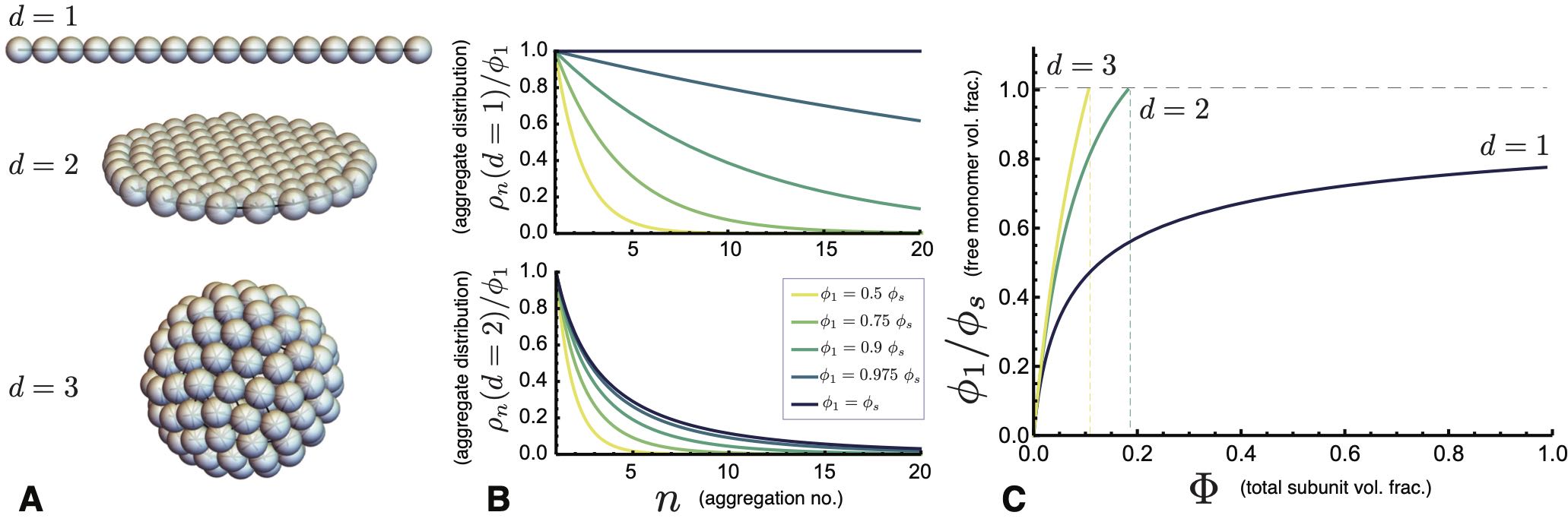}
    \caption{\footnotesize  (A) Examples of $d$-dimensional (linear, planar and spherical) short-range cohesive aggregation.  (B) Plots of the aggregation distributions (relative counts of $n$-mers) for $d=1$ (top) and $d=2$ (bottom) for monomer concentrations increasing to saturation (i.e. $\phi_1 = \phi_{\rm s}$) for $\Delta_0 = 1 ~\kt$.  While the dispersity (and mean size) of linear aggregates diverges as $\phi_1 \to \phi_{\rm S}$, it remains finite for $d\ge2$  at saturation. (C) The equation of state (ideal aggregation theory) for the free monomer population $\phi_1$ as a function of the total concentration $\Phi$ for linear, planar and spherical aggregates for $\Delta_0 = 2.75 ~ \kt$.  For $d=2$ and $d=3$ the free monomer concentration saturates at a finite $\Phi$ where $\phi_1=\phi_{\rm S}$.  For linear aggregation, saturation is not reached in the ideal theory. }
\label{fig: unlimited}
\end{figure*}

\subsubsection{Unlimited assembly: short-range cohesive aggregation}
\label{sec: unlimited}

Before describing models that give rise to SLA, we first consider the thermodynamics of the simplest models of physical association, described by  short-range attraction between  subunits. Crucially,  while these models are relevant to a broad range of physical scenarios like colloidal crystallization~\cite{Manoharan2015, Morphew2018}, they do not exhibit SLA. Yet, they will  serve as a useful reference point for illuminating the necessary conditions for SLA.  Specifically, these models result in either a single dispersed state whose most populous aggregate state is $n=1$ (the free monomer), or coexistence between the dispersed monomer-dominated state and an unlimited aggregate (macrophase separation). The absence of equilibrium finite-sized states will be traced to the generic size-dependence of the cost of open boundaries at the edges of cohesive aggregates. In subsequent sections we will show mechanisms that compete against the open boundary cost to enable SLA.

Here, we consider models where inter-subunit association promotes uniform $d$-dimensional aggregates, e.g. 1-dimensional chain-like or 2-dimension sheet-like aggregates.  Every internal subunit forms, on average, $z$ attractive bonds of strength $-u_0$, and subunits at the free boundary  have $\delta z$ fewer contacts (e.g. Fig.~\ref{fig: unlimited}A). For example, in $d=1$ chain-like assembly $z=2$ and $\delta z =1$.  For the generic dimensionality, the interaction free energy takes the form
\begin{equation}
\epsilon(n) = - \epsilon_0 + \frac{\Delta_0} {n^{1/d}} ,
\label{eq:epsilonUnlimited}
\end{equation}
where $\epsilon_0 = u_0 z/2$ and represents the per subunit cohesive free energy in the bulk (i.e. $n\to \infty$) structure.  The second term derives from the growth of number of particles at the boundary ($\sim n^{(d-1)/d}$) and their deficit of cohesive bonds ($\delta z$), so that $\Delta_0$ is equal to $u_0(\delta z)$ times a geometric factor accounting for the mean bond geometry at the  boundary.   Notably, the bonding-geometry in these assemblies permits the structure to grow uniformly without disrupting this local contact structure at any size scale, a condition that we revisit when describing examples of self-limiting assembly in Sec. \ref{sec: models}.

We define the concentration $\phi_{\rm s} \equiv e^{-\beta \epsilon_0}$, so that the law of mass action, eq. (\ref{eq: lma}), takes the form
\begin{equation}
\label{eq: phisr}
\phi_n = n\big( \phi_1  / \phi_{\rm s} \big)^n e^{- \beta \Delta_0 n^\alpha}  ,
\end{equation}
where $\alpha = 1-1/d$ is an exponent that characterizes the geometric growth of the exposed boundary with $n$. As shown in Fig.~\ref{fig: unlimited}B, $\phi_1 \leq \phi_{\rm s}$ and the distribution $\rho_n=\phi_n/n$ decreases exponentially with aggregate size for large $n$, and for any $d$.  Hence, under these conditions  $\Phi$, the sum over the subunit fraction distribution in eq. (\ref{eq: fixN}), is finite, implying the existence of conditions where the concentration of subunits achieves equilibrium in the suspension.  However, no such equilibrium exists for  $\phi_1 > \phi_{\rm s}$, implying that $\phi_1 \to \phi_{\rm s}$ is an upper limit to concentrations that may be in equilibrium in a dispersed state.  In other words, when $\Phi$ is sufficiently large that $\phi_1 = \phi_{\rm s}$ the solution is {\it saturated}, and additional subunits (further increasing $\Phi$) must phase separate to the macroscopic state (i.e. $n \to \infty$).

First consider the linear case ($d=1$), where the equation of state can be ready computed from eq. (\ref{eq: phisr}) with $\alpha =0$ and the geometric series,
\begin{equation}
\Phi = e^{-\beta \Delta_0} \frac{ \phi_1 \phi_{\rm s}}{(\phi_{\rm s} - \phi_1)^2 } \ \ \ \ {\rm for } \ d=1
\end{equation}
which notably diverges as $\phi_1 \to \phi_{\rm s}$.  As plotted in Fig. Fig.~\ref{fig: unlimited}C, this divergence indicates that the monomer concentration increases with total concentration, but never reaches the point of saturation (i.e. $\phi_1 < \phi_{\rm s}$ for any finite $\Phi$).  Hence, for all subunit concentrations the system maintains $ \phi_1  / \phi_{\rm s} <1$, implying that the distribution of linear aggregates is always exponential, $\phi_n / n \propto e^{-n/ \langle n \rangle}$, where the number-average length is $\langle n \rangle = 1/\ln (\phi_{\rm s}/\phi_1)$.~\footnote{The {\it number-average} aggregate size is $\langle n \rangle= (\sum_n n \rho_n)/(\sum_n \rho_n) = (\sum_n \phi_n)/(\sum_n \phi_n/n)$, where $\phi_n/n$ is the distribution of $n$-mers, while the {\it mass-average} aggregate size is $\langle n \rangle_\text{M} = (\sum_n n \phi_n)/(\sum_n \phi_n)$.} Noting that $\epsilon_0=\Delta_0$ for a 1D chain assembly, the growth of mean length with end energy in the limit of high concentration $\langle n \rangle  \simeq e^{\beta \Delta_0/2} \sqrt{\Phi}$, which is well-known for equilibrium polymers ~\cite{Hiemenz2007} and cylindrical micelles \cite{Safran1994,Gelbart1994}, highlights the mechanism that prevents ``bulk'' assembly for 1D aggregation.  In this dimension, the probability to introduce a free end remains finite $\sim e^{\beta \Delta_0}$ in the $n\to \infty$ limit, analogous to the statistics of domain walls in the 1D Ising model at finite temperature~\cite{Fisher1984}.  However, while the mean-size is finite, this case is distinct from what we will describe as {\it self-limiting assembly}, in both the strong dependence of $\langle n \rangle$ on total concentration, and perhaps more significantly, the fact that fluctuations in aggregate size are always comparable to the mean; that is, $\big\langle (n - \langle n \rangle)^2 \big\rangle^{1/2} \propto \langle n \rangle$.

For higher assembly dimensionality, ($d>1$), the geometric growth of the free boundary cost restrains the $n\to \infty$ divergence in the distribution eq. (\ref{eq: phisr}) as the solution approaches saturation.  For $\phi_1=\phi_{\rm s}$ the distribution takes the form $\phi_n (\phi_1 \to \phi_{\rm s}) = n  e^{- \beta \Delta_0 n^\alpha}$, the sum over which converges for $\alpha>0$ when $d>1$.  For example, we can approximate the sum for planar aggregates ($d=2$) by replacing the sum over aggregation number $n$ with dimensionless aggregate radius $n = \pi r^2$, i.e. $\sum_n \phi_n \to 2 \pi \int dr r ~\phi(r)$.  At saturation, aggregates sizes are  exponentially distributed, $\phi(r) \simeq \pi r^2 e^{- \sqrt{\pi} \beta \Delta_0 r}$, yielding a saturation concentration
\begin{equation}
\Phi_{\rm s} (d=2) \simeq 12 \Big(\frac{\kt}{\Delta_0} \Big) ^4
\end{equation}
at which point the ideal solution of aggregates reaches equilibrium with the bulk condensate.  Thus, with the exception of the special case of $d=1$ assembly, short-range cohesive aggregation is characterized by a finite saturation concentration, $\Phi_{\rm s}=\Phi(\phi_1=\phi_{\rm s})$, above which subunits phase separate into an unlimited bulk structure.

The thermodynamics of these examples are plotted in Fig.~\ref{fig: unlimited}C in terms of the equation of state $\phi_1(\Phi)$ for each dimensionality.  Note that while the distributions of short-range interacting systems have finite mean sizes in the absence of macrophase separation, their distributions are dominated by monomers; i.e., the concentration of $n$-mers $\rho_n$ is always maximal for $n=1$, a property that sharply contrasts with the SLA behavior described next in Sec.~\ref{sec:self-limiting}.

\begin{figure}
\centering
\includegraphics[width=0.85\linewidth]{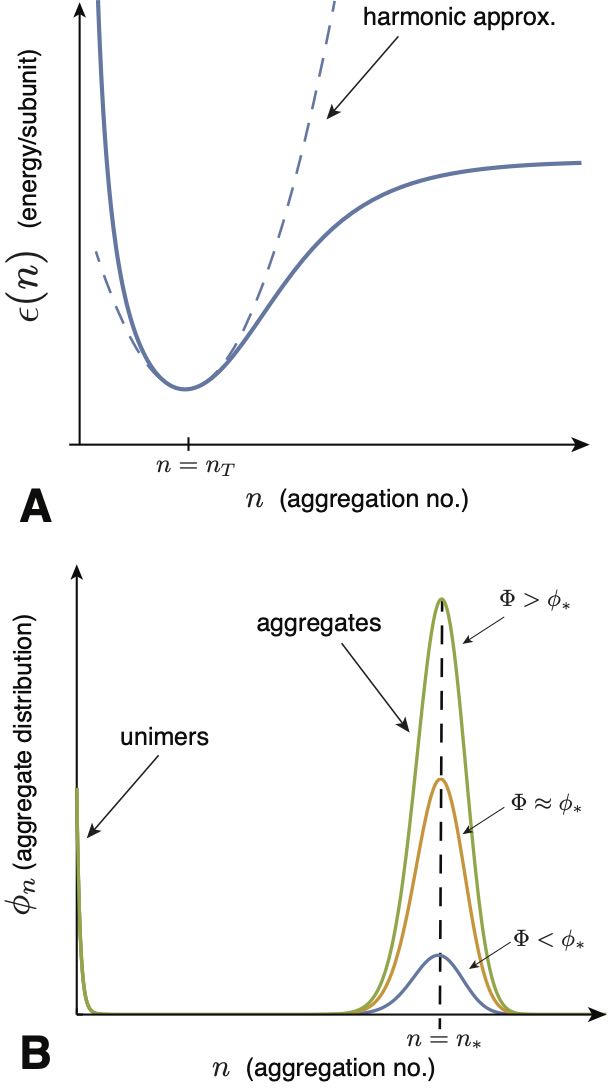}
\caption{\footnotesize (A) A schematic plot of the aggregation (free) energy per subunit is shown as a continuous function of aggregation number $n$.  The dashed line shows a harmonic expansion around a local minimum at the {\it target size} $n=n_{\rm T}$.  (B) Schematic plots of the aggregate distribution for a model of the type sketched in (A).  }
\label{fig: SLA_generic}
\end{figure}

\subsection{Self-limiting assembly: Elements and outcomes}
\label{sec:self-limiting}

In this section, we describe the generic ingredients and thermodynamic outcomes of assembly models that exhibit {\it self-limitation}.  That is, unlike the short-range cohesive models described above, these systems undergo  ideal assembly into self-limiting states  dominated by aggregates with a finite size $n_*$ that is larger than 1, yet smaller than bulk (unlimited) states.  

The physical mechanisms that give rise to this behavior will be discussed in detail in Sec. ~\ref{sec: models}.  Here, we give an overview of the essential thermodynamic ingredients and behavior based on a generic description of the energetics of a self-limiting system.  We consider the assembly behavior in terms of a generic function for the interaction free energy per subunit, $\epsilon(n)$, which, as sketched in Fig.~\ref{fig: SLA_generic}A, favors aggregation at a particular finite size, or possibly several distinct finite sizes.  To highlight its distinct role from the translational entropy of aggregation, we use the term {\it aggregation energetics} to refer to $\epsilon(n)$, but we note that  this describes a free energy per subunit, as interactions in general have both energetic and entropic contributions.  

Given a known form of the $\epsilon(n)$, the discussion of this section addresses several key questions.  First, what determines the onset of aggregation from the dispersed state?  Second, what selects the (dominant) size of finite aggregates?  And third, what are the conditions for driving transitions between different states of self-limiting aggregation, or between self-limited and unlimited aggregation states?

\subsubsection{Aggregation threshold}
\label{sec:CAC}

We first describe the simplest picture of the concentration dependence of ideal aggregation.  We begin with the assumption of an energy per subunit $\epsilon(n)$ of the form shown in Fig.~\ref{fig: SLA_generic}A, which has a single energy minimum at a finite aggregation number $n_{\rm T}$, which we call the {\it target size}.  The basic dependence of aggregation on concentration for such a model is sketched in Fig.~\ref{fig: SLA_generic}B.  There are two dominant populations of aggregates, {\it monomers} and $n$-{\it mers}, with the $n$-mers narrowly distributed around the most populous state $n_{*}\approx n_{\rm T}$.   

\begin{figure}
\centering
\includegraphics[width=0.85\linewidth]{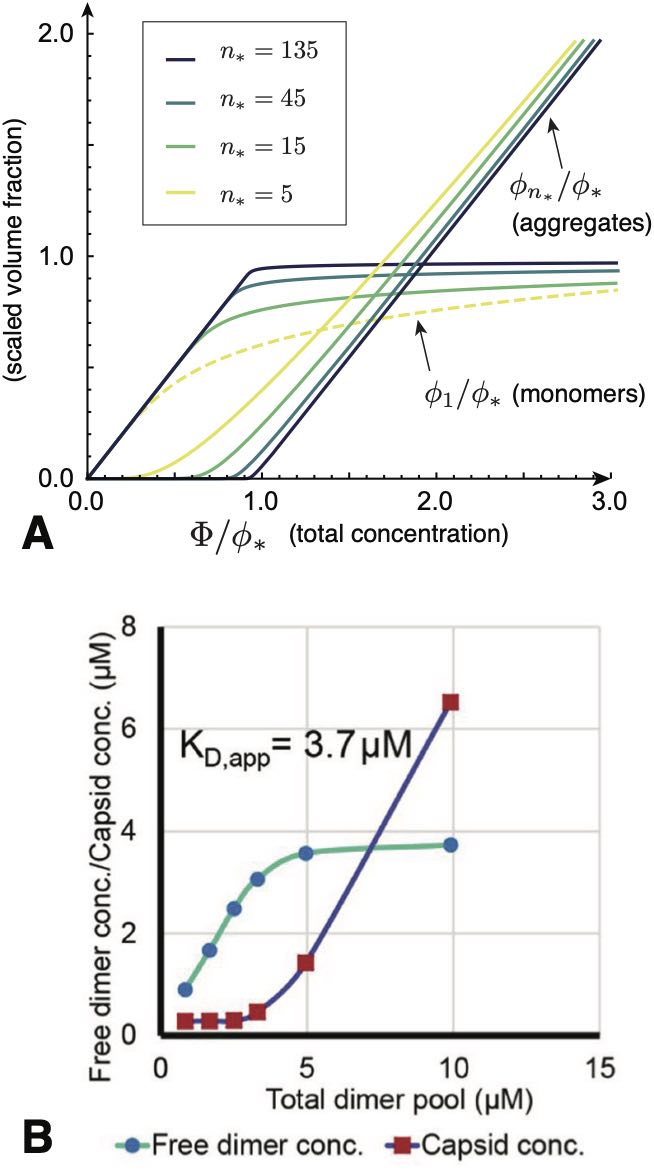}
\caption{\footnotesize (A) Plots of a ``two-state'' model composed of only monomers ($n=1$) and aggregates of a single peak size ($n=n_* \approx n_{\rm T}$) as functions of the total concentration $\Phi$ and for different finite aggregate sizes.  The {\it critical aggregation concentration} (CAC), here $\phi_*$, characterizes the concentration range beyond which aggregates dominate the subunit population.   (B) Assembly behavior of hepatitis B virus (HBV) capsid protein in vitro as a function of concentration. The plot shows the concentration-dependence of the fraction of subunits (protein dimer) in two states: free subunits and assembled capsids composed of 120 subunits. Notice that as the total concentration crosses the CAC (labeled $K_\text{D,app}$), the concentration of free subunits is nearly constant, with almost all additional subunits assembling into capsids. Figure in (B) reprinted from \cite{Ruan2018}.}
\label{fig: CAC}
\end{figure}

For large enough $n_*$, the thermodynamics of aggregation can be captured, to a first approximation, by a {\it two-state}, or bimodal, distribution, in which fluctuations around free monomers and the $n$-mer aggregate peak are neglected. We will self-consistently test the validity of this approximation below (see eq.~\ref{eq: cac_guass}). When subunits are distributed strictly between the $n=1$ and $n=n_*$ states, the conservation of subunit mass is simply $\Phi = \phi_1 + \phi_{n_*}$.  Chemical equilibrium then gives the concentration in preferred aggregates $\phi_{n_*}=n_* \big(\phi_1 e^{-\beta \epsilon_*} \big)^{n_*}$, where $ \epsilon_*\equiv \epsilon(n_{*}) <0$ is the per subunit energy gain upon aggregation into the optimal size.  Defining the concentration scale
\begin{equation}
\phi_* \equiv \big[n_* e^{-n_* \beta \epsilon_*} \big]^{-1/(n_* -1)}\approx e^{  \beta \epsilon_*}/n_*^{1/n_*},
\label{eq: cac}
\end{equation}
yields the following equation of state, relating total concentration to monomer concentration
\begin{equation}
    \frac{ \Phi}{\phi_*} = \frac{\phi_1}{\phi_*} + \Big(\frac{\phi_1}{\phi_*} \Big)^{n_*} ,
    \label{eq: massaction}
\end{equation}
a result derived originally by Debye to explain scattering in soap solutions~\cite{Debye1949}.  The dependence of the populations of monomers and $n_*$-mers on total concentration is plotted in Fig.~\ref{fig: CAC}A and can be summarized as follows.  For low concentration, $\Phi \ll \phi_*$, and additional subunits added to the system go predominantly to monomers, since $\phi_1 \approx \Phi$ while $\phi_{n_*} \approx \phi_* (\Phi/\phi_*)^{n_*}  \ll \Phi$.   Notably, the population of aggregates in this regime $\phi_{n_*}$ is simply proportional to the random probability of $n_*$ free subunits to spatially coincide, $\Phi^{n_*}$, times the enhanced Boltzmann factor for aggregation, $e^{-n_* \beta \epsilon_*} \approx \phi_*^{-n_*}$, and hence is diminishingly small. In the large concentration regime $\Phi \gg \phi_*$, the dominant populations are reversed: the $n_*$-mer population increases in proportion to total concentration, $\phi_{n_*} \approx \Phi$, while monomers increase much more slowly, $\phi_1 \approx \phi_* (\Phi/\phi_*)^{1/n_*} \ll \Phi$.  These two regimes are characterized by a crossover near $\Phi \approx \phi_*$, which is known as the {\it critical aggregation concentration} (CAC), although it is not strictly a phase transition for finite $n_*$.\footnote{ The CAC is commonly referred to as the \emph{critical micelle concentration} (CMC) in the amphiphile literature. In the virus assembly literature it is often called the \emph{pseudo-critical concentration} to emphasize that it does not correspond to a true phase transition for finite $n_*$, and because the CAC observed in finite-time experiments typically exceeds the equilibrium CAC due to nucleation barriers.}  As illustrated in Fig.~\ref{fig: CAC}A, the aggregation crossover becomes increasingly sharp as the aggregation number $n_*$ increases.  Fig. Fig.~\ref{fig: CAC}B shows an example of CAC behavior observed in experiments  of hepatitis B virus capsid protein assembly~\cite{Ruan2018}.

Underlying the transition is a thermodynamic trade off between translational entropy and the interaction free energy that drives aggregation.  Maximizing the ideal translational entropy of aggregates favors maximizing the number of independent translational degrees of freedom, i.e. the number of independent centers of mass in the mixture.  To form an aggregate, $n_*$ monomers must give up their $n_*$ centers of mass, for the single center of mass of the aggregate.  Only when the aggregation free energy is sufficient to ``pay'' this entropic price (i.e. when the concentration of ``excess'' monomers is sufficiently large), does aggregation become thermodynamically favorable.  Hence, the CAC depends not only on aggregation energetics, but also the aggregation number $n_*$.  According to eq. (\ref{eq: massaction}), $\phi_*$ exhibits a modest increase with $n_*$ (as $\sim n_*^{-1/n_*}$) due to the increased translational entropy loss for when joining larger aggregates.  We return to the implications of the $n_*$-dependence of aggregation thresholds in the discussion of competing aggregate states below.   Notice that although the model ignores physical interactions between distinct aggregates, the change in translational entropy couples $n_*$ units, making aggregation a cooperative process.  For this reason, the CAC becomes progressively sharper and tends toward a thermodynamic transition as $n_* \to \infty$ (Fig.~\ref{fig: CAC}A).

\subsubsection{Finite aggregates: Mean size and size dispersity}
\label{sec:maxClusterSize}

Here we review the conditions for the most probable, or {\it optimal aggregate size} $n_*$  given a known form of $\epsilon(n)$, which we assume for the moment to have a single minimum at target size $n_{\rm T}$.  The optimal size $n_*$ corresponds to the maximum in the aggregate distribution $\rho_n = \big(e^{ \beta(\mu - \epsilon(n))}\big)^n$, or equivalently, the minimum in the free energy $n\big[ \epsilon(n) - \mu\big]$, which includes the total interaction free energy and entropy cost of forming an $n$-mer from free monomers.  However, except under conditions where monomers are buffered to a fixed concentration, the chemical potential $\mu= \ln \phi_1$  varies as the equilibrium monomer population changes with total concentration.  Naively, this might suggest that the optimal aggregate size should strongly vary with total concentration.  Here, we illustrate why, notwithstanding the variation of $\mu$ with concentration, $n_*$ is nearly independent of $\Phi$ and almost entirely determined by the form of aggregation energetics, $\epsilon(n)$.  Following this, we summarize the effects of dispersity (i.e. finite width of the aggregation peaks in Fig.~\ref{fig: SLA_generic}B) on aggregation thermodynamics, which is necessary to account for the (weak) concentration dependence of the optimal aggregate size.

{\bf Two-state aggregation:} As a first approximation, consider the two-state aggregation model deep into the aggregation regime, i.e. well above the CAC ($\Phi\gg\phi^*$).  The optimal aggregate size derives from the condition $\frac{ d \rho_n}{d n}\big|_{n_*} =0$, or 
\begin{equation}
    \epsilon (n_*) + n_* \epsilon'(n_*)- \mu = 0 ,
    \label{eq: deltamu}
\end{equation}
where $\epsilon'=\frac{ d \epsilon}{d n}$. Using the fact that $\phi_1( \Phi \gg \phi_*) \simeq \phi_* (\Phi/\phi_*)^{1/n_*} = e^{ \epsilon_*}(\Phi/n_*)^{1/n_*}$ from eq. (\ref{eq: massaction}) in the previous section, this transforms to the condition for the optimal (peak) aggregate size
\begin{equation}
\epsilon'(n_*) = \frac{\kt}{n_*^2} \ln\big(\Phi/n_*\big) \ \ \ \ {\rm (two-state)}.
\label{eq:optimalSize}
\end{equation}
From eq. (\ref{eq:optimalSize}) we may draw two key conclusions.  First, in the limit of large target size $n_* \gg 1$, the optimal size corresponds to a minimum of $\epsilon(n)$.  That is, since $\epsilon'(n_*) \to 0$, $n_* \to n_{\rm T}$ and the aggregate peak is selected by minimizing per sub unit aggregation energy, independent (to a first approximation) of concentration. Second, the right-hand side of eq.~(\ref{eq:optimalSize}), which is proportional to the translational free energy of a dilute concentration of $n_*$-mers, is negative, and hence $\epsilon'(n_*)<0$.  Combining this with eq. (\ref{eq: deltamu}), we find the inequality,
\begin{equation}
\mu < \epsilon(n_*) .
\label{eq: mubound}
\end{equation}
This last condition shows that the equilibrium chemical potential approaches from below, but never quite reaches, the interaction energy of the optimal aggregate $\epsilon(n_*)$ (excepting the unphysical limit $\Phi/n_* \to 1$).

Finally, the fact that $n_*$ corresponds to a \emph{maximum} in the aggregate size distribution, suggests the following condition from $\frac{d^2 \rho_n}{dn^2}<0$
\begin{align}
\epsilon''(n_*) > -\frac{2\epsilon'(n_*)}{n_*} >0  . 
\label{eq:nStarMaximum}
\end{align}
As the righthand side goes to zero as $\sim n_*^{-3}$, the aggregation energetics must be convex in the vicinity of the optimal size. Strictly speaking however, a stronger condition than convexity alone is needed to justify the neglect of aggregation number fluctuations in the 2-state approximation, as discussed next.  

{\bf Gaussian approximation:}  We now consider the effect of {\it convexity} of the aggregation energetics, characterized by the second-derivative of $\epsilon(n)$ at the peak aggregate size.  As above, we restrict our analysis to the case of a single, well-defined minimum in $\epsilon(n)$ occurring at a finite target size $n_{\rm T}>1$.  Close to the minimal-energy size, the energetics have the form
\begin{equation}
\epsilon (n) \simeq \epsilon_{\rm T} + \frac{ \epsilon''_{\rm T}}{2}(n-n_{\rm T})^2 ,
\end{equation}
where $\epsilon_{\rm T} <0$ and $\epsilon''_{\rm T}>0$ respectively characterize the minimum energy and convexity of the target aggregate, as illustrated in the harmonic approximation in Fig.~\ref{fig: SLA_generic}A.  Physically, $\epsilon''_{\rm T}$, which we call the {\it convexity}, quantifies (twice) the energetic cost (in $\kt$) to alter the aggregate number from its target by $\pm 1$.  In the following section, we will describe the physical effects that control convexity in different models of self-limiting assembly.  Here, we see that the concentration-dependence of the mean (or peak) self-limiting size, as well as the size-dispersity, are controlled by a single combination of $\epsilon''_{\rm T}$ {\bf and} $n_{\rm T}$.

The effect of finite convexity is to allow fluctuations in $n$ around the peak size $n_*$.  When $\epsilon''_{\rm T}$ and $n_{\rm T}$ are sufficiently large, the aggregate distribution follows a Gaussian,
\begin{equation}
\rho_n(n \gg 1) \simeq e^{n_*\beta( \mu - \epsilon_*)} e^{-\frac{(n-n_*)^2}{2 \langle \Delta n^2 \rangle}} ,
\end{equation}
where $\langle \Delta n^2 \rangle$ characterizes the variance of aggregate sizes relative to $n_*$.  Assuming that the Gaussian distribution of aggregates is well separated from the monomer peak, the size fluctuations around $n_*$ may be summed in $\phi_n = n \rho_n$, yielding the same mass-action formula as eq. (\ref{eq: massaction}), but with a redefined CAC,
\begin{equation}
\phi_* \approx \frac{e^{\beta \epsilon_*}}{ \big( n_* \sqrt{ 2 \pi  \langle \Delta n^2 \rangle} \big)^{1/n_*}} ,  \ \ \ \ {\rm (Gaussian)}.
\label{eq: cac_guass}
\end{equation}
Compared to the two-state approximation, $\phi_*$, is depressed by a factor proportional to $ \langle \Delta n^2 \rangle^{1/2n_*}$ due to the comparative increase in the number of aggregates states and associated entropy.  Likewise, well above the CAC, the monomer population is depressed (relative to the  2-state approximation) by the same factor.  Combining this effect into the chemical potential with the peak aggregate condition in eq. (\ref{eq: deltamu}) gives the following prediction for peak (mean) aggregate
\begin{equation}
    n_* \simeq n_{\rm T} \bigg( 1 + \frac{\kt}{n_{\rm T}^3 \epsilon_{\rm T}'' } \ln \Big[ \frac{\Phi}{n_{\rm T} \sqrt{ 2 \pi \langle \Delta n^2 \rangle} } \Big] \bigg),  \ \ {\rm (Gaussian)} 
    \label{eq: n_gauss}
\end{equation}
where we are considering only the leading correction to $n_* - n_{\rm T}$.  In this same limit, aggregate dispersity becomes,
\begin{equation}
    \frac{\langle \Delta n^2 \rangle^{1/2} }{n_{\rm T}}\simeq \frac{1}{ \big( n_{\rm T}^3 \beta \epsilon_{\rm T}'' \big)^{1/2} },  \ \ \ \ {\rm (Gaussian)} ,
    \label{eq: disp}
\end{equation}
where again averages are taken with respect to aggregate distribution $\rho_n$ (i.e. number average).

The results of the Gaussian approximation in eqs. (\ref{eq: n_gauss}) and (\ref{eq: disp}) highlight two physical effects of convexity.  First, in eq. (\ref{eq: n_gauss}), the mean aggregate size always falls {\it slightly below} the minimal-energy target size (i.e. because $\Phi < 1$, and hence, the logarithmic factor is always negative).  This ``sub-optimal'' aggregate size derives from the (translational) entropic preference for smaller-$n$ aggregates, and hence, this weak depression of $n_*$ decreases with increasing supersaturation as the translation free energy of aggregates $(\kt /n_{\rm T})\ln(\Phi/n_{\rm T})$ tends toward zero. Second, the relative shift of mean aggregation number $(n_*-n_{\rm T})/n_{\rm T}$ {\it and} the relative size variance $\langle \Delta n^2 \rangle^{1/2}/n_{\rm T}$ decrease with the reciprocal of $\epsilon_{\rm T}'' n_{\rm T}^3/\kt$.  Hence, corrections from size variations become small, in relative terms, either for sharp minima, when $\epsilon_{\rm T}'' \gg \kt$, or for larger aggregate number, when $n_{\rm T} \gg (\kt/ \epsilon_{\rm T}'')^{1/3}$.  While at first glance, this might suggest a generic tendencies toward monodisperse aggregation in the large $n_{\rm T}$ limit, SLA models described in Sec.~\ref{sec: models} show convexity to be a {\it decreasing} function of $n_{\rm T}$.  Hence, as it turns out, the decrease of relative size fluctuations with target size becomes non-trivially dependent on the geometric sensitivity of aggregation energy.

\textbf{Self-limitation without minima:} Before  considering landscapes with more complex equilibria, we briefly note that it is possible to construct functional forms of $\epsilon(n)$ which exhibit self-limitation without local minima.  As an example, consider a variation of the general type of energetics described in Sec~\ref{sec: unlimited}, 
\begin{equation}
    \epsilon(n) = -\epsilon_0 + \Delta(n)
    \label{eq: asymp1}
\end{equation}
where $\Delta(n)$ is a monotonically decreasing, and convex, function of $n$ so that the minimal energy per particle occurs for infinite aggregates, i.e. $\epsilon(n \to \infty) = - \epsilon_0$.  The condition for a maximum in $\rho_n$ in eq. (\ref{eq: deltamu}) gives
\begin{equation}
  \mu+\epsilon_0 \equiv - (\Delta \mu)_\infty=  \Delta(n_*) +n_* \Delta'(n_*)
    \label{eq: nomin}
\end{equation}
Because the chemical potential is bounded from above by $ - \epsilon_0$ (from eq. (\ref{eq: mubound})) and hence $(\Delta \mu)_\infty > 0$, the conditions for finite optimal aggregate size can be satisfied (at some accessible 
$\mu$) provided that $\Delta(n)$ decreases {\it faster} than $1/n$, or more specifically, that
\begin{equation}
\frac{\Delta'(n)}{\Delta(n)} < - \frac{1}{n},
\label{eq: asymp2}
\end{equation}
for some range of finite $n$.  For example, for any model that approaches the bulk energy as a power-law $\Delta(n) =\Delta_0/n^\gamma$ where $\gamma>1$, it is straightforward to show the peak size obeys $n_* = \big[(\gamma-1) \Delta_0/(\Delta \mu)_\infty\big]^{1/\beta}$, which increases continuously with concentration as $\Delta \mu_\infty \to 0$ from above.  

While the previous argument shows that it is possible to construct mathematical examples of minima-free energy densities that result in finite-$n$ peaks in $\rho_n$, physical cases of SLA fall outside of this category. For example, generic physical grounds suggest that aggregates possess a boundary, or surface, at which the assembly energy is different (generally higher) than in the interior, as described for cases of short-ranged cohesive interactions in Sec. ~\ref{sec: unlimited}. In the limit of large $n$, the asymptotic contribution to the energy density from this boundary, $n^{-1/d}$ where $d \geq 1$, will dominate over other possible terms falling off faster than $n^{-1}$ (such as the $\Delta(n)$ term in  eqs. (\ref{eq: asymp1}) and (\ref{eq: asymp2})), leading to unlimited assembly. Hence, we exclude such anomalous cases, and focus the discussion on situations where self-limitation is directly associated with well-defined minima of $\epsilon(n)$.

\begin{figure*}
\centering
\includegraphics[width=1\linewidth]{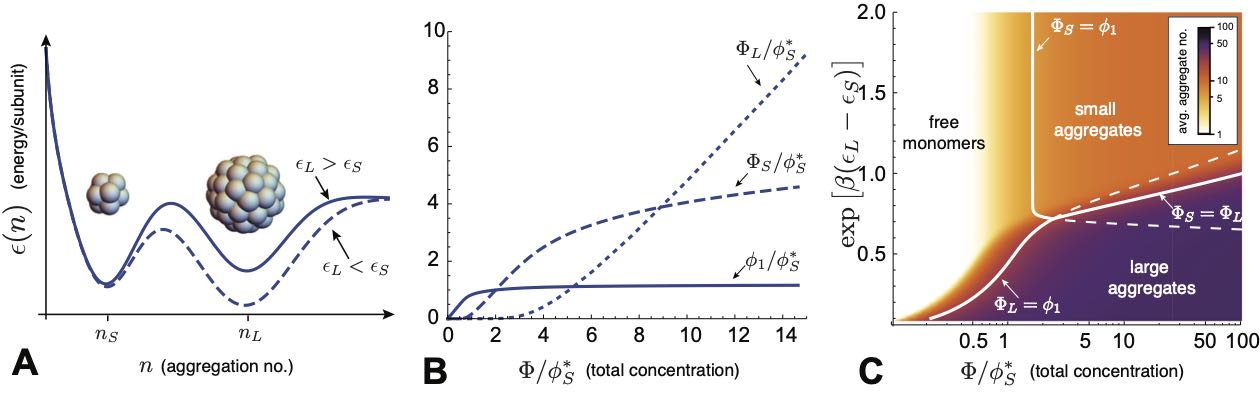}
\caption{\footnotesize (A) A schematic plot of the aggregation energetics for a case with two local minima corresponding to {\it small} and {\it large} finite aggregates, containing $n_{\rm S}$ and $n_{\rm L}$ respective subunits in their target sizes.  The solid curve shows a case where small aggregates are the global minimum of $\epsilon(n)$, while the dashed curve shows a case where the larger aggregate is the global minimum.  The latter case can lead to secondary CAC transitions between small and large aggregates, as shown in (B), which plots the monomer, small aggregate and large aggregate populations as functions of total concentration for large aggregates that are slightly energetically more favorable than small ones ($e^{\beta(\epsilon_{\rm L}-\epsilon_{\rm S})}=0.8$).  In this case, small aggregates dominate at intermediate concentrations, but ultimately are overtaken by a second population of large aggregates above a second CAC.  (C) An assembly state map for the two finite aggregate model, as a function of the (per subunit) energy difference between small and large aggrates and total concentration.  The color scale shows the mean aggregate size, while the solid lines indicate boundaries between  states dominated by monomers, $n_{\rm S}$-mers, and $n_{\rm L}$-mers.  Dashed lines indicate where subdominant aggregates reach the monomer concentration.  Plots in (B) and (C) show results of the two finite aggregate model for $n_{\rm S}=10$ and $n_{\rm L}=50$.  }
\label{fig: 2state}
\end{figure*}

\subsubsection{Competing states of aggregation}
\label{sec:SecondCMC}

Secs. \ref{sec:CAC} and \ref{sec:maxClusterSize} above describe the simplest case of self-limiting assembly:  a concentration-controlled crossover, or ``pseudo-transition", from a monomer-dominated state to a state dominated by aggregates of one finite size. The finite aggregate size corresponds to a single minimum in $\epsilon(n)$, and the transition occurs at a single CAC.  In this section, we overview the thermodynamics of cases in which assembly is characterized by multiple local minima, or instead, by transitions between self-limiting and unlimited aggregation states.  In these cases, the aggregation thermodynamics can exhibit a more complex dependence on concentration, corresponding to {\it secondary CACs} between different aggregation states.  While concentration-dependent transitions between different aggregates are commonly attributed to interactions between aggregates~\cite{Israelachvili2011}, it is less widely appreciated that they can also occur in ideal aggregation models, which strictly neglect inter-aggregate interactions.  As we review in Sec. \ref{sec: amphiphiles}, the possibility of an ``ideal" secondary CAC was first considered in the context of transitions between spherical and cylindrical surfactant micelles~\cite{Porte1984,May2001}.  In this section, we describe this behavior as a generic consequence of the translational entropy preferences for smaller aggregate sizes, and as such, ideal secondary CACs can occur in a much broader class of SLA models.

{\bf Two finite aggregate states:} We first describe a simple model with only 2 states of finite aggregates in equilibrium with a population of free monomers, for simplicity ignoring number fluctuations around these three states.  We consider two states of {\it small} and {\it large} aggregates, corresponding to two well-defined local minima of $\epsilon(n)$, at $n_{\rm S}$ and $n_{\rm L}> n_{\rm S}$ subunits, respectively, as shown schematically in Fig.~\ref{fig: 2state}A.  In this case, aggregation is controlled by not only the difference in the respective energy minima $\epsilon_{\rm S}$ and $\epsilon_{\rm L}$, but also the difference in the aggregation number.  To understand aggregation in the presence of multiple minima, it is convenient to define the nominal CACs corresponding to either aggregate state, from eq. (\ref{eq: cac}),
\begin{equation}
\phi^*_{\nu} \equiv \Big[n_\nu e^{-  \beta n_\nu  \epsilon_\nu} \Big] ^{-1/(n_\nu -1) } \approx \frac{ e^{\beta \epsilon_{\nu} }}{ n_\nu^{1/n_\nu} } \  \ {\rm for \ } \nu = S, L 
\end{equation}
These are concentrations at which aggregates of either type would overtake free monomers, were it not for the additional equilibrium between the populations of small and large aggregates.  In terms of these   concentration scales, the law of mass action takes the form
\begin{equation} 
\Phi = \phi_1 + \phi^*_{\rm S} \Big( \frac{ \phi_1}{ \phi^*_{\rm S} } \Big)^{n_{\rm S}} +  \phi^*_{\rm L} \Big( \frac{ \phi_1}{ \phi^*_{\rm L} } \Big)^{n_{\rm L}},
\label{eq: binaryagg}
\end{equation}
where the last two terms represent the respective populations of subunits in $n_{\rm S}$-mers  and $n_{\rm L}$-mers , which we denote as $\Phi_{\rm S}$ and $\Phi_{\rm L}$.  As above for the case of a single minimum in $\epsilon(n)$, in the limit of high concentration, aggregation always proceeds towards the state with the lowest energy.  However, in this case, there are two possible thermodynamic scenarios for the concentration dependence, depending on the relative energy difference between small and large aggregates: 

{\it i) $\epsilon_{\rm S} < \epsilon_{\rm L}$}:  Because $n_{\rm S}< n_{\rm L}$, in this case it is always true that $\phi^*_{\rm S} < \phi^*_{\rm L}$, which means that as concentration increases, $\phi_1 \to \phi^*_{\rm S}$ before reaching $\phi^*_{\rm L}$.   Above the threshold where $\phi_1 \approx \Phi_{\rm S}$ , monomers remain effectively ``buffered'' at $\phi_1 \approx \phi^*_{\rm S}$, and it is straightforward to show that $\Phi_{\rm L} \ll \Phi_{\rm S}$~\footnote{The assumptions of ideal aggregation require that $\Phi$ must remain below unity, a condition that requires $\phi_1 \ll \phi^*_{\rm L}$ for case $(i)$.}. Thus, when the smaller aggregate has a lower per subunit aggregation energy,  aggregation proceeds as if there was only a single target state with a CAC at $\phi^*_{\rm S}$ and never yields a significant number of large aggregates.

{\it ii) $\epsilon_{\rm S} > \epsilon_{\rm L}$}:  When the large aggregates are energetically favored, there are two possibilities.  Consider first the case of large energy differences, such that $e^{\beta(\epsilon_{\rm L}-\epsilon_{\rm S})} <\frac{n_{\rm L}^{1/n_{\rm L}} }{n_{\rm S}^{1/n_{\rm S}}}$. For this first regime $\phi^*_{\rm L} < \phi^*_{\rm S}$, and there is only a single CAC at the (lower) critical concentration for large aggregates. For the second regime, when the energy difference between large and small aggregates is smaller, in the range $1>e^{\beta (\epsilon_{\rm L}-\epsilon_{\rm S})} >\frac{n_{\rm L}^{1/n_{\rm L}} }{n_{\rm S}^{1/n_{\rm S}}}$, the order of the CACs reverses: $\phi^*_{\rm S} < \phi^*_{\rm L}$, and leads to two CACs.  As shown shown in Fig.~\ref{fig: 2state}B, for this case upon increasing concentration from the dilute limit, the concentration reaches a {\it first CAC} at $\Phi \approx \phi^*_{\rm S}$, with a transition to a state dominated by small aggregates (i.e. $\Phi_{\rm S}\gg \phi_1,\Phi_{\rm L}$).  This state persists until reaching a {\it second CAC} at $\Phi \approx \phi^{**}$, defined by a crossover in dominant aggregation state to $\Phi_{\rm L}>\Phi_{\rm S}$.  The concentration threshold condition can be estimated by solving for the monomer concentration at which $\Phi_{\rm S} (\phi_1)= \Phi_{\rm L}(\phi_1) \gg \phi_1$ from eq. (\ref{eq: binaryagg}) 
\begin{equation}
\phi^{**}\approx \phi_{\rm L}^* \Big(\frac{\phi_{\rm L}^*}{\phi_{\rm S}^*} \Big)^{n_{\rm S}/(n_{\rm L}-n_{\rm S})},
\end{equation}
which is larger than the ``bare'' value $\phi_{\rm L}^*$ owing to the depletion of free monomers by small aggregates, and corresponds to a total concentration $\Phi^{**}\approx 2 \phi^{**}$.  The high-concentration regime above the second CAC is dominated by minimal-energy large aggregates, but maintains a sizeable amount of {\it small} aggregates (i.e. $\Phi_{\rm L} \gg \Phi_{\rm S} \gg \phi_1)$, approximately buffered at the second CAC concentration. 

To summarize this 2-aggregate model, it is possible to have {\it two} pseudo-critical transitions (as in Fig.~\ref{fig: 2state}B), first from disassembled monomers to small aggregates and then from small to large aggregates, provided the energy difference between aggregates is sufficiently small, in the window
\begin{equation}
    0<\epsilon_{\rm S}-\epsilon_{\rm L}<\frac{\kt}{n_{\rm S}}\ln n_{\rm S} - \frac{\kt}{n_{\rm L}}\ln n_{\rm L}.
\end{equation}
This is consistent with the secondary CAC behavior shown in the assembly state diagram in Fig.~\ref{fig: 2state}C, calculated for the case of $n_{\rm S}=10$ and $n_{\rm L}=50$.

The physical origin of this double CAC behavior can be traced to a competition between the higher cohesive  energy of large aggregates pitted against the higher (per subunit) translational entropy of smaller aggregates.  This can be cast in terms of the chemical equilibrium between large and small aggregates, which requires 
\begin{equation}
\mu = \epsilon_{\rm S} + \frac{\kt}{n_{\rm S}} \ln \rho_{\rm S}=\epsilon_{\rm L} + \frac{\kt}{n_{\rm L}} \ln \rho_{\rm L} .
\end{equation}
Energetically favorable large aggregates, $\epsilon_{\rm L}<\epsilon_{\rm S}$, require aggregate concentrations to adjust to maintain a suitably higher translational entropy of {\it small} aggregates, namely $\frac{1}{n_{\rm S}} \ln \rho_{\rm S}<\frac{1}{n_{\rm L}} \ln \rho_{\rm L}$, specifically 
\begin{equation}
\rho_{\rm S}/\rho_{\rm L} = \frac{e^{n_{\rm S}\beta (\epsilon_{\rm L} - \epsilon_{\rm S})}}{ \rho_{\rm L}^{(n_{\rm L}-n_{\rm S})/n_{\rm L}}} 
\end{equation}
This condition shows that the larger entropy of smaller aggregates requires that $\rho_{\rm S}/\rho_{\rm L} > 1$ provided that the concentration of large aggregates remains sufficiently small, below $\rho_{\rm L} < \rho_{\rm L}^{**}=\Big[e^{\beta(\epsilon_{\rm L}-\epsilon_{\rm S})} \Big]^{n_{\rm S} n_{\rm L} /(n_{\rm L}-n_{\rm S})}$.  Hence, when $n_{\rm S}$ and the differential in aggregation energy are small enough, small aggregates remain more populous than large aggregates up to total concentrations that exceed the first CAC to the small aggregate state, until the second CAC.

This simplified model illustrates a generic conclusion.  Even if an aggregate state does not correspond to the {\it global} minimum of $\epsilon(n)$, it may exhibit an entropically-stabilized window of thermodynamic dominance at intermediate concentrations, provided its target size is sufficiently small and its energy is sufficiently close to the global minimum.  Next, we illustrate this entropic stabilization of finite (compact) aggregates in models for which the competing states are {\it unlimited}.

{\bf Finite and unlimited aggregates:} While polymorphic assembly into multiple finite-number aggregates occurs in some natural and biomimetic systems~\cite{Wingfield1995,Lutomski2018,Sun2007} and may be desirable for nanomaterials applications, cases in which aggregates {\it change dimensionality} are more common.  That is, aggregate structures that remain finite in at least one or more spatial directions, but undergo essentially unlimited growth in other directions.  The most common examples are amphiphillic assemblies, which can either form spherical micelles (finite in all directions), cylindrical micelles (finite in two spatial dimensions, unlimited in one), or lamellar/layered assemblies (finite in one dimension, unlimited in two).   In Sec. \ref{sec: amphiphiles} below we describe the molecular ingredients that lead to polymorphic transitions between aggregate dimensionality based on a model of surfactant assembly~\cite{May2001, MagnusBergstrom2016}.  In this section, we illustrate how the principles of secondary CAC behavior apply to models which can exhibit states of finite aggregation number (e.g. spherical) that can transition to states of 1D aggregation or a bulk (unlimited) morphology.  

\begin{figure}
    \centering
    \includegraphics[width=0.95\columnwidth]{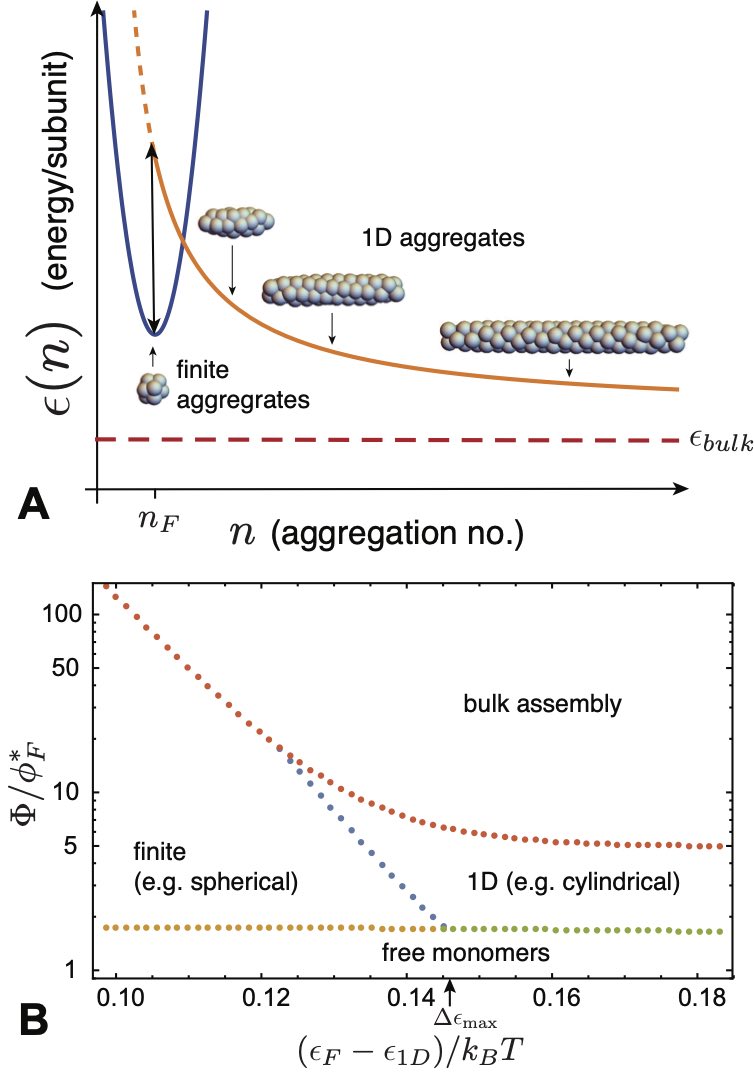}
    \caption{ 
        (A) A schematic plot of polymorphic aggregation energetics with three competing branches of assembly:  finite aggregates with a local minimum at $n_{\rm F}$, 1D aggregates, and bulk aggregates ($n \to \infty$ energy shown as horizontal dashed line).  In this case, the infinite 1D aggregate has a lower per subunit energy than finite aggregates, and there is a barrier (in total energy) $\delta$ that separates these states at $n = n_{\rm F}$, i.e. the double arrow in (A) corresponds to $\delta/n_{\rm F}$.  (B) Phase diagram for concentration-dependent size selection. The dominant aggregation state is shown for a system with coexistence among finite aggregates with $n_{\rm F}=100$ subunits, separated by an energy gap $\epsilon_{\rm F}-\epsilon_{\rm 1D}$ and a barrier of $\delta$, eq.~\eqref{eq: gap}, to 1D aggregates. There is an additional per subunit energy gap of $\epsilon_{\rm 1D} - \epsilon_{\rm bulk}=0.0005 \kt$ between 1D and bulk aggregates. The horizontal axis gives the energy gap between spheres and cylinders, and the vertical axis gives the total concentration relative to the CAC for finite aggregates $\phi^*_{\rm F} \simeq e^{\beta \epsilon_{\rm F}}/n_{\rm F}$. The boundaries between monomers, finite and 1D aggregates are determined by crossovers in the most populous aggregate type from eq. (\ref{eq: totdim}), while the point of bulk saturation is determined by the point when $\mu = \epsilon_{\rm bulk}$. In this example, the energy per subunit in finite aggregates is fixed at $\epsilon_{\rm F}=-10~ \kt$ and the endcap energy of spherocylinders is $\Delta_0=20~\kt$. The maximum energy gap for which 2nd CAC behavior occurs ($\Delta \epsilon_{\rm max}\approx 0.14~\kt$, eq.~\ref{eq:deltaStar}) is indicated on the x-axis.     
    }
\label{fig: DCMCphaseDiagram}
\end{figure}

Since our primary interest is to describe conditions where ideal aggregation gives rise to concentration-dependent transitions in morphology, we consider a minimal description of a generic model including finite and unlimited aggregation states.  As summarized in Fig.\ref{fig: DCMCphaseDiagram}A, this model considers three disconnected ``branches'' of assembly:  

{\it 1) Finite (compact) aggregates,} with respective target size and energy, $n_{\rm F}$ and $\epsilon_{\rm F}$

{\it 2) 1D aggregates,} with energy per subunit 
$$\epsilon (n) = \epsilon_{\rm 1D} + \Delta_0/n $$
where $\Delta_0>0$ characterizes the cost of finite ``endcaps'' and $\epsilon_{\rm 1D}<0$ is the limiting $n\to \infty$ per subunit assembly energy in this morphology

{\it 3) Bulk aggregates,} with energy density $\epsilon_\text{bulk}$. Here, we consider only one macroscopic aggregate ($n \to \infty$) with negligible boundary energy.

Based on the foregoing analysis of the 2-finite assembly state model, it can be anticipated that secondary CAC behavior from finite to 1D aggregation takes places when the $n\to \infty$ energy density of 1D aggregates is lower than finite aggregates, but the energy gap is sufficiently small that the translational entropy associated with the compact aggregates can stabilize a window of $n_{\rm F}$-mer aggregates.  Likewise, when the energy density of the bulk state falls below these dimensionally limited  states, we anticipate an upper limit to concentration (i.e. saturation) which can maintain equilibrium with dispersed aggregates.

With this in mind, we consider a simplified law of mass action for subunit populations
\begin{equation}
    \Phi = \phi_1 + \Phi_{\rm F} + \Phi_{\rm 1D} + \Phi_\text{bulk},
    \label{eq: totdim}
\end{equation}
where the terms respectively describe the populations of free monomers, subunits in a single finite aggregate size, 1D aggregates of various size, and bulk aggregation. The population of subunits in finite aggregates is given by (neglecting number fluctuations)
\begin{equation}
    \Phi_{\rm F} = n_{\rm F} \big( \phi_1 e^{-\beta \epsilon_{\rm F}}  \big)^{n_{\rm F}},
\end{equation}
while the 1D aggregate population is given by
\begin{multline}
    \Phi_{\rm 1D} \simeq \sum_{n=n_{\rm F}}^{\infty} n e^{- \beta \Delta_0} \big( \phi_1  e^{- \beta \epsilon_{\rm 1D}}   \big)^n \\  \simeq n_{\rm F} e^{- \beta \Delta_0} \frac{ \big( \phi_1 e^{- \beta \epsilon_{\rm 1D}}  \big)^{n_{\rm F}} }{\big(1- \phi_1 e^{- \beta \epsilon_{\rm 1D}}  \big)}.
\end{multline}
In this final term, we made the additional assumptions that 1D assembly is not favorable below some aggregate size close to $n_{\rm F} > 1$, and that the monomer concentration remains well below the value $e^{ \beta \epsilon_{\rm 1D}}$ where $\Phi_{\rm 1D}$ diverges\footnote{\label{fn:six} Specifically, we assume $n_{\rm F}>e^{ \beta \delta}$. For $n_{\rm F}<e^{ \beta \delta}$, $\Phi_{\rm 1D} \simeq e^{-\beta \Delta} \big( \phi_1 e^{-\beta \epsilon_{\rm 1D}}  \big)^{n_{\rm F}+1} \big(1- \phi_1 e^{-\beta \epsilon_{\rm 1D}}  \big)^{-2}$, $\phi_1^{**} \approx e^{\beta \epsilon_{\rm 1D}} \big(1-e^{-\beta \delta/2}/n_{\rm F}^{1/2} \big)$.}.  
The results are not qualitatively sensitive to these approximations. 

Based on these forms, it is straightforward to find the free monomer concentration $\phi_1^{**}$ where finite aggregates and 1D aggregates are equally populous, i.e. $\Phi_{\rm F} (\phi_1^{**})= \Phi_{\rm 1D}(\phi_1^{**})$,
\begin{equation}
    \phi_1^{**}=e^{\beta \epsilon_{\rm 1D}} \big(1-e^{-\beta \delta} \big),
    \label{eq:phiStarStarCyl}
\end{equation}
where
\begin{equation}
    \delta =   \Delta_0+n_{\rm F} (\epsilon_{\rm 1D} - \epsilon_{\rm F})
    \label{eq: gap}
\end{equation}
is the energy difference, or ``barrier", between an $n_{\rm F}$-mer and a 1D aggregate of the same size (see Fig.~\ref{fig: DCMCphaseDiagram}A).  As described above, a stable aggregate population requires at least a local minimum in the energy and hence a barrier necessarily separates aggregation states associated with distinct local maxima in population.  Critically, the size of this barrier determines the window of secondary CAC transition behavior, as follows.

First, note that $\delta > 0$ implies that the energy of forming two ``end caps'' on the 1D aggregate exceeds that of the target $n_{\rm F}$-mer.  Second, the existence of a second CAC requires that this concentration exceeds the primary CAC to a $n_{\rm F}$-mer dominated state, that is, the condition $\phi_1^{**} > \phi_{\rm F}^{*} \simeq e^{\beta \epsilon_{\rm F}}/n_{\rm F}^{1/n_{\rm F}}$.  This gives an upper limit to the energy gap between $n_{\rm F}$-mer aggregation and 1D assembly for second CAC behavior, $\epsilon_{\rm F}-\epsilon_{\rm 1D} < \depsMax$, with 
\begin{align}
\label{eq:deltaStar}
\Delta \epsilon_{\rm max} \approx \frac{\kt}{n_{\rm F}} \ln n_{\rm F} +\kt \ln \big(1-e^{-\delta} \big) . 
\end{align}
, the stability window of the $n_{\rm F}$-mer state expands, in terms of $\Delta \epsilon_\text{max}$, both with {\it decreasing} target size and {\it increasing} energy barrier to 1D aggregation.  For energy gaps larger than this limiting condition, the intermediate $n_{\rm F}$-mer state disappears~\footnote{  
Note that eq.~(\ref{eq:deltaStar}) is an implicit relation for $\depsMax$, since $\delta$ is a function of $(\epsilon_{\rm 1D} - \epsilon_{\rm F})$. However, in the limit $\delta \gg 1$, the maximum gap is approximately $\depsMax\approx\left(\epsilon_{\rm 1D}+\kt \ln n_{\rm F}\right)/{n_{\rm F}}$. Similarly, in the limit $n_{\rm F}<e^{\beta \delta}$, $\depsMax \approx \left(\Delta_0-2 \kt \right)/n_{\rm F}$ (see previous footnote~\ref{fn:six}).}.  

Last, note that monomers reach chemical equilibrium with the bulk state at a concentration $\phi_1 = e^{\beta \epsilonBulk}$, which sets an additional condition for second CAC behavior, $\phi_1^{**} < e^{\beta \epsilonBulk}$.  This gives the upper limit to the energy gap between 1D and bulk assembly for second CAC behavior before saturation,
\begin{equation}
 \epsilon_{\rm 1D} - \epsilonBulk < -\kt \ln \big(1-e^{-\beta \delta} \big)    
\end{equation}
This condition shows that concentration range of the 1D aggregate state diminishes with increasing energy barrier between compact and 1D aggregates.

An example assembly state diagram for this model under conditions $\epsilon_{\rm F} > \epsilon_{\rm 1D} > \epsilonBulk$ is shown in Fig.~\ref{fig: DCMCphaseDiagram}B.  The concentration dependent state of aggregation is plotted versus energy gap between $n_{\rm F}$-mers and 1D aggregates for variable $\epsilon_{\rm F}$ with fixed $n_{\rm F}$, $\Delta$, $\epsilon_{\rm 1D}$, and $\epsilon_\text{bulk}$.  Under these conditions, when the aggregation energy of (compact) $n_{\rm F}$-mers is  larger than, but sufficiently close to, that of (infinite) 1D aggregation, the system undergoes a sequence of concentration-driven transitions:  first from monomers to finite-aggregations; then to 1D aggregates; and finally, to bulk (unlimited) assembly. In the following section, we revisit the possibility of multi-CAC behavior in the context of polymorphism of surfactant aggregates, focusing on its microscopic origin for this effect in terms of an underlying molecular model of aggregation.

\section{Mechanisms and models of self-limiting assembly}

\label{sec: models}

In Sec. \ref{sec: thermo}, we overviewed the basic thermodynamics of ideal aggregation, and described some of the generic ingredients and outcomes of finite-size equilibria.  We showed that the key ingredient is a per subunit {\it aggregation free energy} which has one or more local minima as a function of aggregation number (for {\it finite-number} aggregates), or as a function of the size of one or more spatial dimensions of the aggregate (for {\it spatially finite} aggregates like quasi-cylindrical or planar structures).  In this section, we review four broad mechanisms of self-limiting assembly, and a physical example of each mechanism.  In each case, we first focus on the physical ingredients that give rise to the self-limiting aggregation energetics, and then illustrate some of the implications of the generic phenomonology overviewed in Sec. ~\ref{sec: thermo}.  

To organize the discussion, we divide these mechanisms into two broadly-delineated classes:  {\it self-closing assembly} and {\it open-boundary assembly}.  These classes are distinguished by the presence or absence of an open boundary and gradients in intra-aggregate stress in the target assembly.  Self-closing assembly describes aggregates in which the subunits, by and large, share the same ``cohesive environment'' of neighboring subunits, and further adopt a common shape in the target assembly.  In contrast, self-limitation of open-boundary assemblies requires {\it gradients} of the inter-subunit forces throughout the aggregate.  Within the discussion, we highlight the distinct outcomes and potential ``tradeoffs'' between these different mechanisms in terms of size selection.

We divide the remainder of Sec. \ref{sec: models} into two main parts, focusing respectively on self-closing and open-boundary assemblies.  Following the introduction of each broader class of SLA, we further subdivide each into two subclasses, representing four basic physical mechanisms of SLA.  For each of the four basic mechanisms, we briefly overview the applicable physical systems and then introduce a simple example model that captures the emergence of self-limited assembly.

\subsection{Self-closing assembly}

We define {\it self-closing assembly} (SCA) as class of self-limiting assembly that achieves a finite target size, or finite target dimension, due to {\it anisotropic binding} between neighbors that leads to a preferred rotation of neighbor bonds.  Such interactions generically arise when subunits are tapered or wedged-shaped, such that cohesive bonding leads to a relative rotation of the axes of neighbor units (see Figs.~\ref{fig: classes}A and \ref{fig: capsules}A).  In combination with the relative displacement of subunit centers, this relative rotation, when built up over multiple subunits leads to a preferred intra-assembly curvature (along one or more principal directions). In the simplest case, this can be visualized as 1D ``loops" of subunits, whose preferred curvature radius $R_\text{close}$ leads the structure to close upon itself. 

\begin{figure}
\centering
\includegraphics[width=1\linewidth]{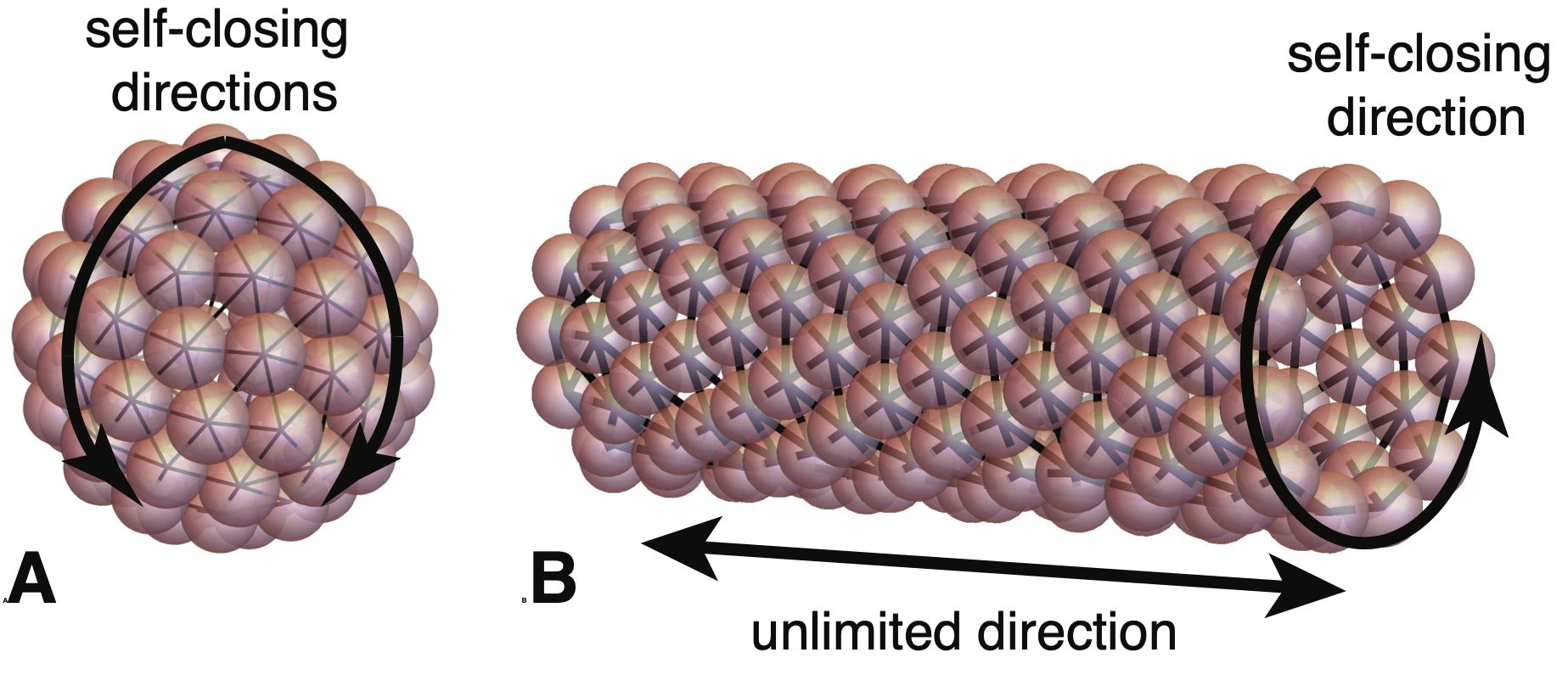}
\caption{\footnotesize (A) Schematic of spherical (shell) assembly with two independent self-closing directions of assembly.  (B) Schematic of tubule assembly with one self-closing direction (circumferential) and one unlimited direction (axial) of assembly.  }
\label{fig: capandtube}
\end{figure}

We include in the SCA class structures that close upon themselves in all directions of assembly and thus achieve a \emph{finite number} of subunits, such as the spherical shell in Fig.~\ref{fig: capandtube}A, as well as structures that close in one or more directions but remain unlimited in others, such as the tubule in Fig.~\ref{fig: capandtube}B. In the latter case, structures have an unlimited number of subunits but achieve a \emph{finite size} in the self-closing direction(s). For example, the tubule is unlimited in the axial direction, but has a well-defined radius and corresponding number of subunits in the circumferential direction. Notably, the underlying principles for size selection remain the same as for finite number; namely, the self-limiting size $W$ of a self-closing direction is determined by the minimum of the energy per subunit with respect $W$.  This can be readily understood by considering a system with nearly all of its $N$ subunits assembled in aggregates of unlimited number, and correspondingly a negligible fraction of free monomers. To first approximation, the energetic costs of free edges and the translational entropy  of the unlimited aggregates can be neglected, and the concentration of free monomers can be assumed a small contribution to the total free energy, $F \approx N \epsilon(W)$. Hence, the thermodynamic equilibrium at fixed $N$ corresponds to the selection of the energy-minimizing size $W_*$, corresponding to $\partial_W \epsilon|_{W_*}=0$.  We describe similar considerations for self-limiting, open-boundary assemblies in Sec. \ref{sec: accumulant} below.

Strictly speaking, for SCA it need not be necessary to identify a continuously loop of bonds along the self-closing direction(s), nor that target curvatures are strictly uniform. We only require that there are one or more periodic directions on a representative 2D surface of the aggregate (e.g.,\ the surface spanned by the subunit centers).   Moreover, the preferred curvature does not need to select a perfectly commensurate number of subunits per cycle, since physical subunits generically possess some flexibility of shape and cohesion (bonding) that permits fluctuation in the inter-unit rotation. In the simplest case, e.g. with fluid-like intra-assembly order, a SCA may accommodate such strains through uniform deformation of subunits and their bonds.  However, certain physical examples introduce extra geometric constraints (e.g. solid-like, spherical shells) which require at least some variable intra-assembly strains.  Notwithstanding the possibility of such gradients, we categorize assembly as self-closing if its target size is selected through the curvature radius $R_\text{close}$, as opposed to the accumulation of stress-gradients, which is described as a distinct class of self-limiting assembly below.

Physical examples of SCA can be divided, roughly, into two groups according to the ratio of subunit size, characterized by some thickness $d$, and the target curvature radius $R_\text{close}$.  When $R_\text{close}/d \gg 1$ the target number of units per cycle is proportionately large.  This case describes {\it tubules, shells and capsules}.  The second case, $R_\text{close} \approx d$ describes assemblies whose curvature (and thus self-limited size) is most often selected and regulated by the molecular dimension itself, which is characteristic of {\it amphiphiles} and their micellar  aggregates.

\subsubsection{Shells, capsules, and tubules}

\label{sec: capsules}

We first review the case of tubule or shell-like assemblies.  Examples of these are common in biology, where ``tapered'' protein subunits select a preferred radius of assembly curvature~\cite{Oosawa1975}.  Quasi-cylindrical (tubular) examples include microtubules~\cite{Nogales2000, Cheng2012} and the bacterial flagella
~\cite{Namba1997}, while quasi-spherical (shell and capsule) examples include clathrin cages \cite{Kirchhausen2014,Mettlen2018,Bucher2018,Giani2017}, viral capsids \cite{Zlotnick2011,Mateu2013,Hagan2014,Perlmutter2015,Bruinsma2015,Hagan2016,Twarock2018,Zandi2020}, bacterial microcompartments \cite{Kerfeld2010,Schmid2006,Iancu2007,Rae2013,Bobik2015,Chowdhury2014,Kerfeld2016}, and other protein-shell organelles  \cite{Nott2015,Zaslavsky2018,Pfeifer2012,Sutter2008}.  Recently, engineered tapered or patchy colloids have also drawn interest for their ability to realize synthetic analogs of self-closing tubules and shells~\cite{Li2011, Morphew2017}, although many if not most such realizations to date are properly categorized as analogs to micelles where curvatures are comparable to colloidal dimension.  Whatever the underlying subunit structure, the existence of 1D curvature only ensures equilibrium self-limitation along one of the two assembly dimensions in the tubular constructs, while the preferred positive Gaussian curvature of shells and capsules leads to equilibrium states with finite subunit {\it number}.

\begin{figure}
\centering
\includegraphics[width=1\linewidth]{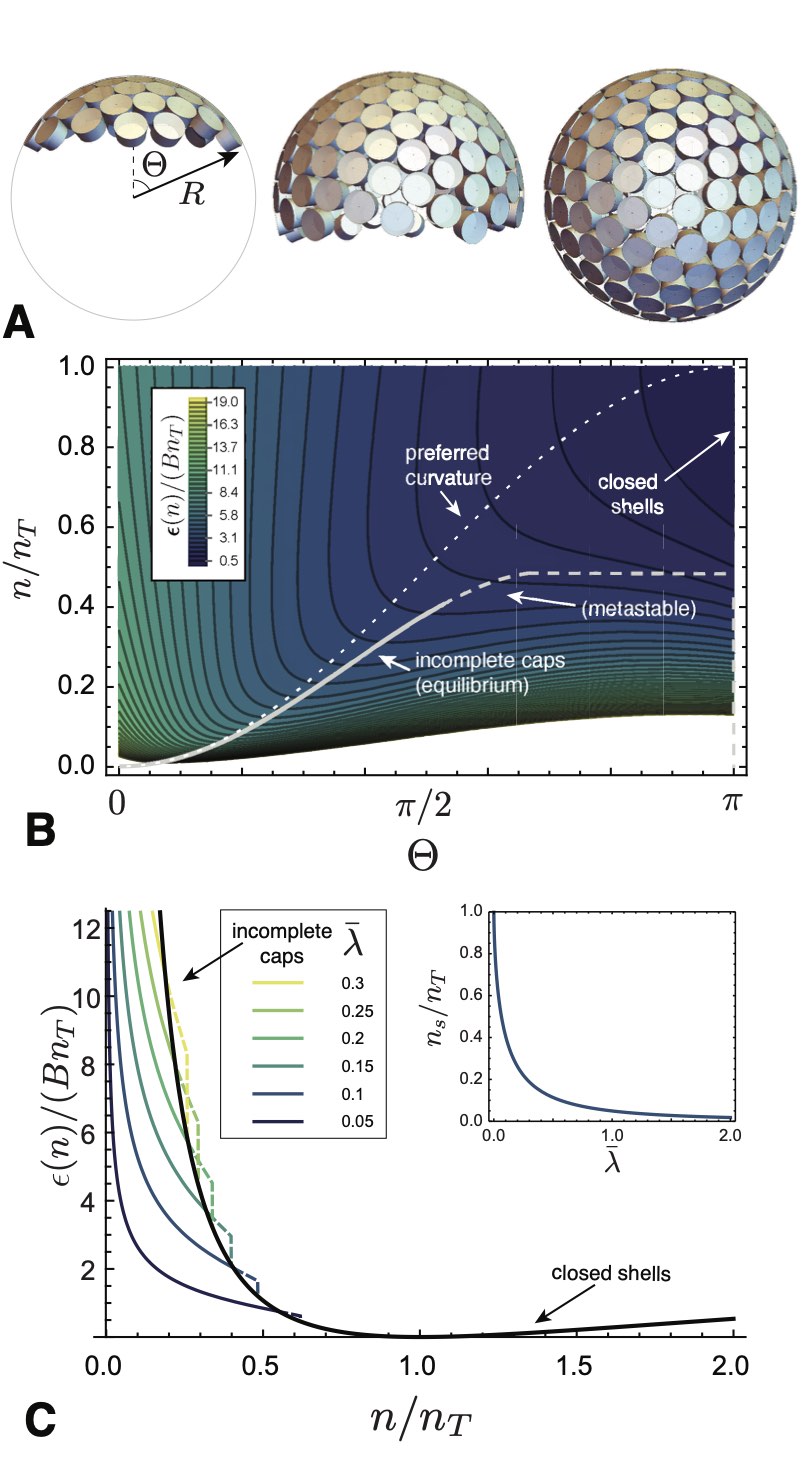}
\caption{\footnotesize (A) Schematic geometry of a ``fluid capsule'' of tapered subunits, which assumes partial shell geometries of spherical caps with aperture angle $0<\Theta\leq \pi$.  (B) A contour map of the energy density landscape of the capsule model as a function of aperture angle and the ratio of subunits $n$ to the preferred number in the ideal closed shell $n_{\rm T}$.  Low (high) values of $\epsilon(n)$ appear as purple (red).  The dotted line shows a partial shell with the target curvature, while the solid line indicates the minimal energy cap, whose curvature radius is slightly compressed by the line tension of the boundary.  Beyond a threshold cap size $n = n_{\rm S} \simeq 0.4 n_{\rm T}$, this open cap becomes unstable to preclosure, and the minimal energy branch runs along $\Theta \to \pi$.  This landscape corresponds to a dimensionless line tension $\bar{\lambda} =0.1$.  (C) Plots of the minimal energy branches of assembly:  incomplete caps are shown as colored curves (metastable portions are dashed), and the closed shell ($\Theta \to \pi$) is shown as a black curve.  The inset shows the size $n_{\rm S}$ corresponding to the preclosure, or ``snap'', transition between stable open caps and closed shells as a function of line tension.}
\label{fig: capsules}
\end{figure}

To illustrate the self-limiting thermodynamics of shells and capsules, where target curvature radii are much larger subunit dimensions, consider the following minimal model. Spherical fluid capsules, shown schematically in Fig.~\ref{fig: capsules}A, are composed of subunits with nominal area $a_0$ and a tapered shape that favors a preferred (target) spherical curvature radius $R_{\rm T}$.  Here, we restrict the analysis to cases in which the curvature preference sufficiently disfavors locally anisotropic curvature \cite{Lazaro2018} to limit incomplete assembly to cap-like aggregation states with positive Gaussian curvature. Competition between incomplete assemblies with positive and zero Gaussian curvature was recently considered in Ref. \cite{Mendoza2020}. We assume the cap covers an axisymmetric ``cap'' domain of radius $R$ from its pole up to the {\it aperture angle} $\Theta$ (see Fig.~\ref{fig: capsules}A).  While a closed capsule with a preferred curvature $R_{\rm T}$ has a target aggregation number $n_{\rm T} = 4 \pi R_{\rm T}^2 / a_0$, a capsule may realize a different aggregation number $n = 2 \pi R^2 (1- \cos \Theta)$, provided that either it is open (i.e. $\Theta \neq \pi$) or is deformed from its preferred taper (i.e. $R \neq R_{\rm T}$).

Taking the simplest possible model, we assume that the intra-shell order is fluid-like, such that the only elastic penalty derives from bending deformations away from target curvature, which we consider via a membrane-like bending energy
\begin{equation}
    E_\text{bend} = \frac{B}{2} \int dA \Big(\frac{1}{R} - \frac{1}{R_{\rm T}} \Big)^2,
\end{equation}
where $B$ is a bending modulus and the area integration is carried out over the incomplete shell.  Additionally, we consider the line energy associated with open edge for an incomplete cap,
\begin{equation}
    E_\text{open} = 2 \pi R \sin \Theta ~ \lambda ,
\end{equation}
where $\lambda$ is the energy per unit length of the exposed edge, associated with the fewer cohesive bonds as well as a difference in solvation of subunits at the edge.  The aggregation energy as a function of $n$ then takes the form
\begin{multline}
    \epsilon(n, \Theta) =-\epsilon_{\rm T}+ \frac{2 \pi B}{ n_{\rm T}} \Big( \sqrt{\frac{n_{\rm T}(1- \cos \Theta) }{2 n}  } -1 \Big)^2 \\ + \frac{ \sqrt{ 2 \pi a_0} \lambda}{ n^{1/2}} \frac{ \sin \Theta}{ \sqrt{1- \cos \Theta}}
    \label{eq: capsule}
\end{multline}
where $-\epsilon_{\rm T}$ is the ``bare'' aggregation energy for subunits in the bulk of undeformed capsules.

It is easy to see that the form of $\epsilon(n, \Theta)$ in eq. (\ref{eq: capsule}) has a global minimum for closed capsules of the target size (i.e. $n=n_{\rm T}$ and $\Theta = \pi$).  Nevertheless, the combination of bending elasticity and the edge energy of incomplete shells influences assembly for $n \neq n_{\rm T}$.   This can be seen by plotting the landscape of assembly energetics in the $n/n_{\rm T}$ - $\Theta$ plane, as shown in Fig.~\ref{fig: capsules}B.  For very small aggregate sizes $n/n_{\rm T} \ll 1$, caps lock into the preferred curvature, $R \to R_{\rm T}$, described by the condition $\cos \Theta \to 1- 2 n/n_{\rm T}$, shown as a dotted line in Fig.~\ref{fig: capsules}B.  As $n$ increases, the edge energy favors compression of the open caps to smaller curvature radii $R<R_{\rm T}$.  The amount of this shape compression grows up to a critical aggregation number $n_{\rm S}$, beyond which the minimal energy capsule ``snaps'' discontinuously to a closed shell of suboptimal size, that is $\Theta \to \pi$ for $n_{\rm S}<n<n_0$.  

Hence, the minimal  aggregration vs. $n$ generically exhibits two branches (Fig.~\ref{fig: capsules}C):  an open cap for $n<n_{\rm S}$ and an (edge-free) closed shell for $n_{\rm S}<n<n_{\rm T}$.  The transition between these two branches can be understood in terms of ``nucleation'' of an open pore in an over-curved shell, where $n_{\rm T}-n_{\rm S}$ corresponds to the size of the ``critical nucleus''.  The inset of Fig.\ref{fig: capsules}C shows that the critical preclosure size $n_{\rm S}$ generically decreases with an increasing (dimensionless) ratio of edge energy and bending stiffness, $\bar{\lambda} \equiv ( a_0 \nT /8 \pi)^{1/2} \lambda/B$.  

Notwithstanding the generic existence of a transition between open-cap and preclosed  branches of the energy landscape, the transition does not lead to stable partial shells (i.e. a minima in $\epsilon(n)$ for $n \neq n_{\rm T}$).  Hence, while the open cap branch and its transition to the closed shell may have implications for assembly pathways and kinetics, the equilibrium distribution of self-limiting capsules is independent of the edge energy and generically governed by the energetics of the closed-shell branch.  This fact has further generic consequences for the concentration dependence and dispersity of aggregate size, both of which are governed by the product of the convexity and cube of the target size, $ \epsilon''(n_{\rm T}) n_{\rm T}^3/ \kt$, according to eqs. (\ref{eq: n_gauss}) and (\ref{eq: disp}).  The $\Theta \to \pi$ limit of (\ref{eq: capsule}) shows that $\epsilon''(n_{\rm T}) = B \pi/n_{\rm T}^3$.  Hence, the decrease of convexity with target assembly size precisely cancels that entropic factor of $n_{\rm T}^3$, such that the {\it relative} shift in mean aggregate size and {\it relative} dispersity, $(n_* - n_{\rm T})/n_{\rm T}$ and $\langle \Delta n^2 \rangle^{1/2}/n_{\rm T}$, respectively, are limited only by the ratio of bending modulus to thermal energy, $B/\kt$, and independent of self-closing target size.   \footnote{This continues until the asymptotic limit of zero spontaneous curvature ($n_{\rm T}\to\infty$), at which point the free energy per subunit is independent of size and the size distribution becomes an unlimited exponential \cite{Helfrich1986} as shown for 1D assemblies in section~\ref{sec: unlimited}.}
Therefore,  to  regulate the self-limiting size of self-closing assemblies in {\it absolute} terms, the rigidity of subunits and their angular interactions must {\it grow with target size}.

The predicted growth of size fluctuations with target size would seem to contradict observations of the best studied example of self-closing shells, icosahedral virus capsids.  At conditions of optimal assembly, size polydispersity of viruses is remarkably small. In fact, this high degree of monodispersity has been exploited by using 3D crystalline arrays of virus capsids for optical applications requiring precise spatial periodicity~\cite{Young2008, Steinmetz2011, Minten2011, Dang2011, Judd2014, Park2015, Malyutin2015, Delalande2016, Rother2016, Chen2016, Brillault2017}.  While high-precision measurements of size-dispersity in capsids are challenging, electron microscopy structures that were obtained without the assumption of icosahedral symmetry show that as many as 40\% of alphavirus nucleocapsid core particles exhibit defects \cite{Wang2015,Wang2018}, and hence some dispersity in shape.  Notably some non-icosahedral structures, like immature HIV caspids, exhibit variations in number of subunits ($\sim1000$ GAG protein subunits) that are on the order of the mean capsid size \cite{Briggs2009}, although such effects may also be attributed to assembly kinetics \cite{Dharmavaram2019}.  More recently, size distributions of hepatitis B virus (HBV) capsids (and capsids of other viruses) have been achieved near or at single-subunit precision using resistive pulse sensing ~ \cite{Zhou2011,Zhou2018}, mass spectrometry ~ \cite{Uetrecht2011}, and charge detection mass spectrometry  ~\cite{Pierson2014,Pierson2016,Lutomski2018}. Although metastable defective capsids are observed \cite{Pierson2016,Lutomski2018}, these measurements show that at long times (potentially corresponding to a near equilibrium state) the population is dominated by icosahedral capsids with the native size \footnote{In fact HBV is dimorphic. Both in vitro and in vivo HBV capsid assembly yields mostly 120 protein dimer capsids with $T=4$ icosahedral symmetry in the Caspar Klug nomenclature (see section~\ref{sec: AAs}), but also a few percent of $T=3$ icosahedral capsids. However, size fluctuations around the dominant $T=4$ population were shown to be insignificant at long times.}.

\begin{figure}
\centering
\includegraphics[width=1\linewidth]{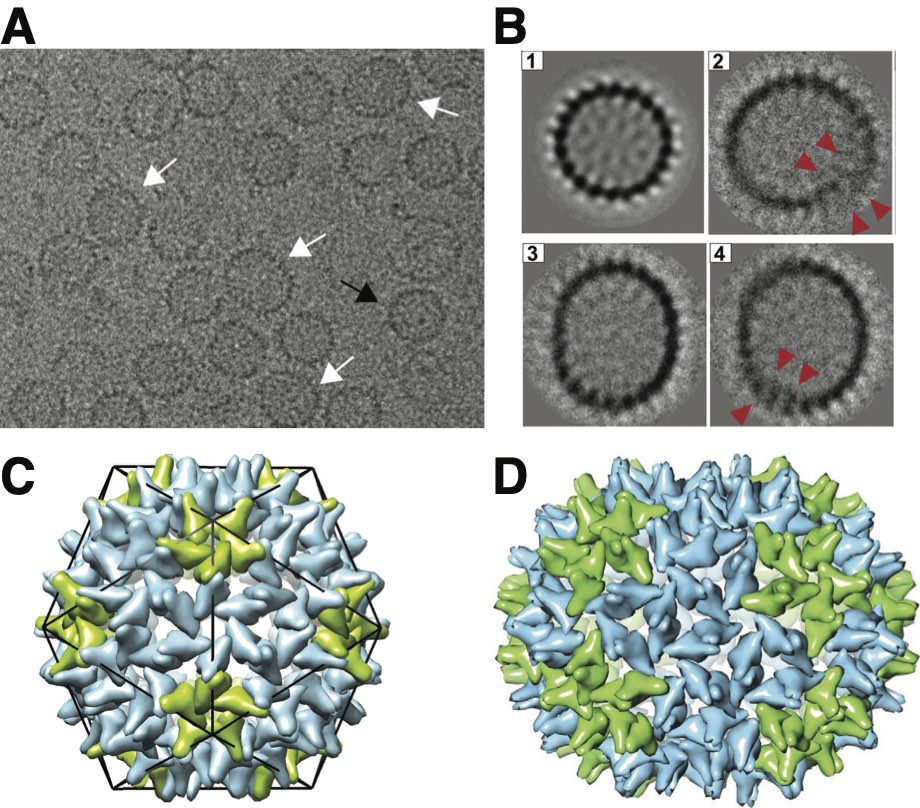}
\caption{\footnotesize Examples of well-formed and defective woodchuck hepatitis B virus (WHV) capsids, adapted from Ref. ~\cite{Pierson2016}. (A) Portion of a representative cryo-EM image of empty WHV capsids assembled in vitro in the absence of RNA (from capsid proteins with the C-terminal RNA binding  deleted, wCp149).  The population of capsids is structurally heterogeneous; the black arrow indicates an example of a $T=4$ icosahedral capsid, while the white arrows indicate examples of defective capsids. (B) 2D class averages of (1) icosahedral and (2-4) defective particles. Red arrows in (2) and (4) indicate locations where the capsid shell has overgrown and overlaps itself. (C) Schematic model of a $T=4$ icosahedral HBV capsid, with the monomers forming the 12 five-fold vertices colored green, and others colored blue.  (D) Hypothetical model of the structure of a capsid that is non-icosahedral but closed, with an elongated structure containing 150 protein dimers (the icosahedral capsid has 120 dimers). Here the green dimers are in pentamers or extend between pentamers and hexamers, and the blue dimers are in hexamers or extend between two hexamers.
Images reprinted from ~\cite{Pierson2016}.}
\label{fig: capsids}
\end{figure}

To place these measurements in the context of the results for the fluid shell model, we note that bending moduli for virus capsids have been estimated from the force-displacement curves measured in nanoindentation experiments in which virus capsids are compressed using an AFM tip~\footnote{It should be noted that estimating elastic moduli from the force-displacement curve is sensitive to the value chosen for thickness of the capsid shell, and the relationship between the atomic structure and the effective mechanical thickness remains at least somewhat obscure \cite{May2011,May2011a}.}.  Estimated bending modulus values vary from 10-200 $\kt$, but typical values fall in the higher end of the range, 100-200 $\kt$. (For comparison, bending moduli of lipid bilayer membranes are typically in the range of $10-20$ $\kt$.)
Using the result from above that $\langle \Delta n^2 \rangle^{1/2}\approx n_0 /\sqrt{\pi B}$, with $n_0=120$ and an estimate of $B\thicksim 60 \kt$ for HBV\footnote{The bending modulus is calculated from the 3D Youngs modulus $E=0.26$ GPa obtained from nanoindentation measurements in Roos et al.\cite{Roos2010}, and assuming a thin shell model so that the 2D Youngs modulus and bending modulus are respectively given by $Y=E t$ and $B = Y t^3/(12 (1 - \nu^2))$, with $t=2.1$ nm the effective shell thickness \cite{Roos2010,Wynne1999} and $\nu=0.4$ the Poisson's ratio \cite{Uetrecht2008,Roos2009}.} gives a root mean squared size fluctuation of about 9 dimers, considerably larger than the long-time experimental estimates.

This discrepancy highlights two physical ingredients neglected in the model described above: the discrete subunit size and the fact that most virus capsids, as well as most other protein shells, are crystalline rather than fluid. As discussed below in section~\ref{sec: AAs}, the arrangement of proteins within icosahedral capsids can be mapped onto a triangular net. However, tiling a spherical topology with a triangular lattice requires the formation of 12 five-fold sites, often consider as ``defects'' in a hexagonal packing. A number of equilibrium calculations have shown that the elastic energy of the defects themselves and inter-defect elastic interactions significantly affect the energy landscape of such shells  \cite{Chen2007,Bruinsma2003,Zandi2004,Li2018,Li2019,Mendoza2020}. These effects are reflected in local minima in the aggregate energy at certain `magic numbers' of subunits~\cite{Zandi2004}.  Notably, these minima correspond to shells with high degrees of symmetry, with the in-plane bonding energies corresponding to shells with icosahedral symmetry. Thus, when including the energetics of this bond-ordering a size fluctuation of even one subunit can incur a significant energy cost ($\gtrsim 10 \kt$), since it requires disruption of the low-energy geometry, for example, through the introduction of a pair of 5- and 7-fold defects. In fact, the metastable structures observed in HBV capsids are typically found at discrete intervals corresponding to deviations of multiple subunits from the native capsid size \cite{Pierson2014,Pierson2016,Lutomski2018}, suggesting that typical fluctuations correspond to insertion or deletion of multi-subunit oligomers (e.g. hexamers of the capsid protein) which would minimize disruptions to the capsid symmetry. For example, Fig.~\ref{fig: capsids}A,B show cryo-EM images and examples of 2D class averages respectively of in vitro assembly products of woodchuck hepatitis B virus (WHV) capsid proteins, which exhibit heterogeneous structures including icosahedral capsids, elongated closed shells, and shells with overlapping edges.  Fig.~\ref{fig: capsids}C shows a hypothetical model of a non-icosahedral closed shell, in which insertion of additional hexamers leads to a prolate structure.

Computational models of icosahedral assembly have also identified ensembles of defective capsules in which additional hexamers were added to the icosahedral shell \cite{Nguyen2009,Elrad2010}, although these models and the corresponding defective structures differ from the HBV system.   Hence, for more realistic models that incorporate both bending and ``bond-network" elasticity, we expect that these corrugations in the energy landscape (i.e. due to communicability with icosahedral symmetry)  versus $n$ will be superposed on the smooth landscapes illustrated in Fig.~\ref{fig: capsules} for the fluid shell model, which may account for size fluctuations to be restricted for a limited set of low-energy values of $n$ at or near to high-symmetry, or magic number capsomer arrangements.

\subsubsection{Amphiphillic aggregates}

\begin{figure}
\centering
\includegraphics[width=0.975\linewidth]{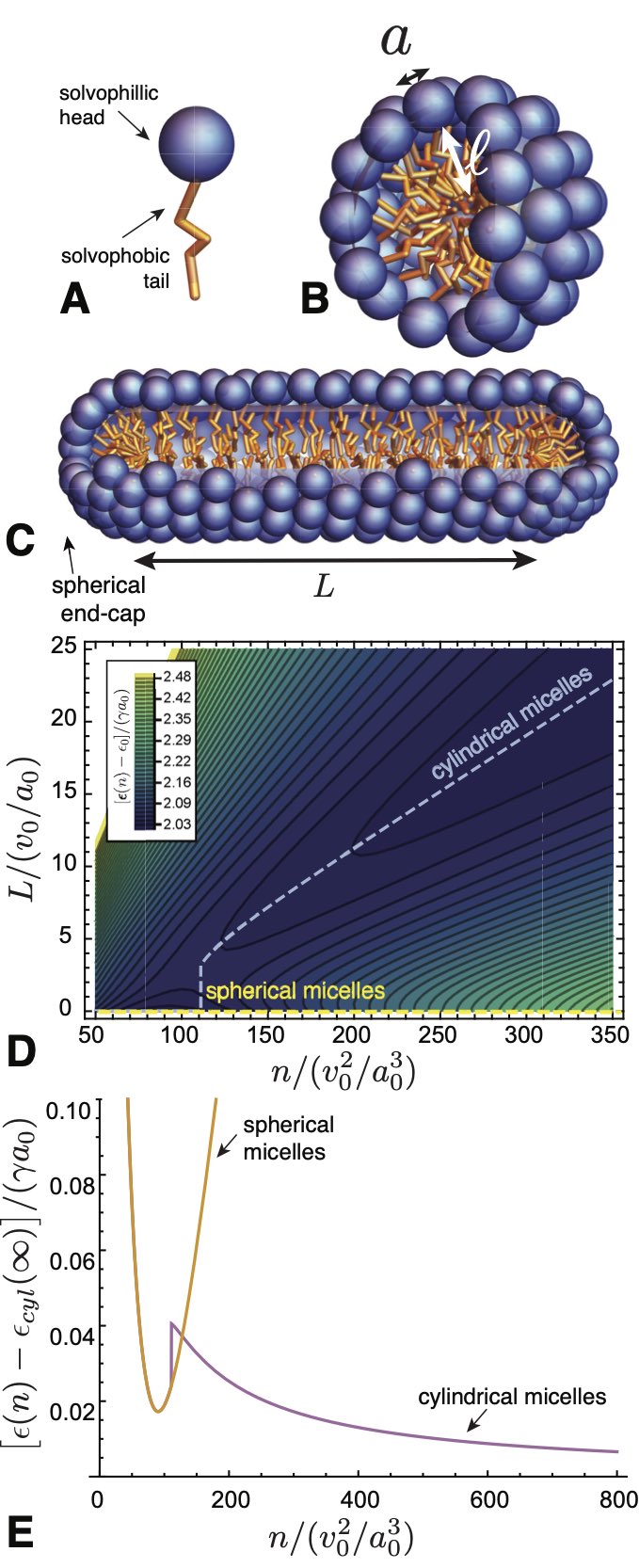}
\caption{\footnotesize (A) A cartoon model of a (single-tailed) surfactant molecule.  (B) A schematic of a spherical micelle with the thickness of the solvophobic core ($\ell$) and the area per head group ($a$) highlighted.  A wedge-like portion of the micelle is cutaway to illustrate the interior packing of tails.  (C) A spherocylinder model of a worm-like micelle, which has a cylindrical portion of  length $L$ capped by two hemispherical micelles of equal radius.  (D) An energy landscape for spherocylinders described by the model in eq. (\ref{eq: amphi}) with $P^{-1} = 2.25$ and $\bar{k} = 0.1$.  The dashed orange line shows the ($L=0$) spherical micelle branch and the dashed pink line shows the ($L=0$) worm-like micelle branch, whose corresponding energy as a function of reduced aggregation number is plotted in (E).}
\label{fig: surfactants}
\end{figure}

\label{sec: amphiphiles}

Arguably the most common and well-studied class of self-limiting assemblies is {\it amphiphiles}.  In the broadest sense, these refer to subunits with chemically dissimilar ends, which consequently favor distinct solvent environments.  For example, lipids and surfactants possess oily hydrocarbon tails attached to a polar or charged head group~\cite{Israelachvili2011}, which imparts a respective hydrophobic and hydrophillic character to either end of the same molecule (e.g. schematic in Fig.~\ref{fig: surfactants}A).  Dispersing such amphiphiles in a solvent that has higher affinity to one end of the molecule generically drives them to form aggregates that partially hide, or sequester, the solvophobic portions while maintaining exposure of the solvophillic portions.  Examples of such aggregrates, spherical and cylindrical micelles, and bilayer sheets, are shown schematically in Fig. ~\ref{fig: surfactants}B-C.  Notably these structures curve upon themselves, but do so on a lengthscale that is limited by, and comparable to, the size of the amphiphile itself, e.g. the molecular tail length in  Fig. ~\ref{fig: surfactants}A.  The tendency to exclude unfavorable solvent from the ``core'' of the aggregate, in combination with the packing constraints of filling this region with the solvophobic portions, requires each amphiphillic subunit to ``span'' the entire thickness of the aggregate, which is fundamental to their self-limiting assembly.

In this section we describe a simple model to capture the self-limiting assembling of amphiphiles, and highlight, in particular, how thermodynamic considerations of changes in {\it aggregate thickness} shape the preferred aggregate curvature, but also give rise to polymorphism in the dimensionality of aggregates (e.g. spheres, cylinders, membranes).  For illustration, we review a model for the thermodynamics of surfactant aggregation, of the type shown in Fig.~\ref{fig: surfactants}A, capturing central ingredients of the well-known packing model developed by Israelachvili and coworkers~\cite{Israelachvili1976, Israelachvili2011}.  While this model aims to capture molecular elements of low-molecular weight surfactants and lipids, the essential thermodynamic features  carry over to other amphiphillic assemblies, such as block copolymers in selective solvents~\cite{Halperin1992, Leibler1983, Zhang1996, Jain2003}.  The thermodynamics of amphiphile aggregation incorporates three ingredients:  (i) the thermodynamics of area per solvophillic head group; (ii) the thermodynamics of molecular extension; and (iii) the constraint of uniform density in the solvophobic core, which links the first two elements.  

A simple model to describe the head-group energetics considers the per subunit energy to form aggregates with an area $a$ at the solvent/core interface~\cite{Israelachvili1976, Israelachvili2011}.  The generic tendency to ``hide" amphiphiles from the solvent is parameterized by a surface energy cost $\gamma a$, where $\gamma>0$ favors dense lateral packing of head groups.  Competing against this lateral compression is the cost of inter-unit repulsive interactions, which in the simplest case are described by the two-body term in the virial series, giving a per subunit energy $A_2/a$, where $A_2>0$~\cite{Tanford1974}.  These two terms can be combined into a single form
\begin{equation}
    \epsilon_{\rm int} = \gamma\Big(a + \frac{a_0^2}{a} \Big),
    \label{eq: epint}
\end{equation}
where $a_0 = \sqrt{A_2/\gamma}$ is the {\it optimal head group area}.  

In combination with the tendency to achieve optimal head group area are additional thermodynamics of tail packing in the core, and the costs to extend its length, $\ell$.  There are various models proposed for this effect, including a finite-maximum extension~\cite{Israelachvili1976, Israelachvili2011} or instead treating the core as melt of flexible polymers~\cite{Dill1980, Ben-Shaul1984, Nagarajan1991, Nagarajan2002}.  Here, we adopt a simplified model used by May and Ben-Shaul for the free energy of tail length $\ell$ that spans from the solvent/core interface into the center of the aggregate
\begin{equation}
\epsilon_{\rm stretch} = \frac{k}{2} \big( \ell - \ell_0\big)^2 ,
\end{equation}
where $k$ is an elastic constant for intra-subunit stretch and $\ell_0$ is a preferred length, which parameterizes the free energy cost of deformations from a preferred conformation state of the short tail.   Here, we consider $k$ and $\ell_0$ as a minimal description of the extensional thermodynamics, and like $\gamma$ and $a_0$, these parameters can be varied through a combination of subunit structure and physical-chemical conditions, e.g. temperature and solvent properties.

Extensional energetics are linked by packing constraints associated with occupying the core with a fixed density of solvophobic portions of the subunits~\cite{Israelachvili1976, Israelachvili2011}. These constraints vary with the dimensionality of the limited directions in the aggregate: $d_{\rm L}=3$, spherical micelles; $d_{\rm L}=2$, cylindrical micelles; and $d_{\rm L}=1$ planar bilayers~\footnote{Note that $d_{\rm L}$ refers to the number of \emph{limited} directions, while we use $d$ to refer to the dimensionality of the {\it unlimited} directions, e.g. $d=1$ and $d=2$ for cylindrical and lamellar aggregates.}.  By considering the ratio between the core volume and interfacial area of an aggregate of thickness (radius) $\ell$, it is straightforward to show that the (solvophobic) volume per subunit satisfies
\begin{equation}
    v_0 = \frac{a \ell}{d_{\rm L}} .
\end{equation}
As aggregates change their shape and number, uniform density requires adjustment of $a$ and $\ell$ to maintain constant $v_0$.  Using this constraint, we rewrite the assembly energy in terms of a single {\it dimensionless thickness},
\begin{equation}
    r \equiv \frac{\ell}{ (v_0/a_0)} = d_{\rm L}\frac{a_0}{a} ,
\end{equation}
giving
\begin{equation}
    \frac{\epsilon(r,d_{\rm L})}{ \gamma a_0} = \Big(\frac{d_{\rm L}}{r}+\frac{r}{d_{\rm L}} \Big) + \frac{\bar{k}}{2} \big( r - P^{-1}\big)^2 + \epsilon_0 ,
\label{eq: amphi}
\end{equation}
where $\bar{k} = k v_0^2/ (\gamma a_0^3)$ is a scaled stretch modulus of the tail, and $\epsilon_0$ parameterizes the negative energetic gain to assemble.   The parameter $P$ was introduced by Isrealachvili as the {\it packing parameter}, a measure of the commensurability of the preferred shape with accessible aggregate geometries,
\begin{equation}
    P \equiv \frac{a_0 \ell_0}{ v_0} .
\end{equation}
Written in this way, it is straightforward to understand how area and stretch thermodynamics compete to determine the optimal aggregate morphology.  While area terms favor a thickness $r=d_{\rm L}$,  stretch thermodynamics favor a thickness $r=1/P$.  Only for particular preferred headgroup areas and subunit lengths do these two values coincide, i.e. when $P = 1/d_{\rm L}$; otherwise there is at least some shape frustration between these terms.  As a heuristic, we therefore expect aggregation to favor the dimensionality $d_{\rm L}$ closest to $1/P$, which corresponds to the ``tapered'' geometry that most closely approximates the favored areal and thickness packing at uniform density.  

\begin{figure*}
\centering
\includegraphics[width=0.95\linewidth]{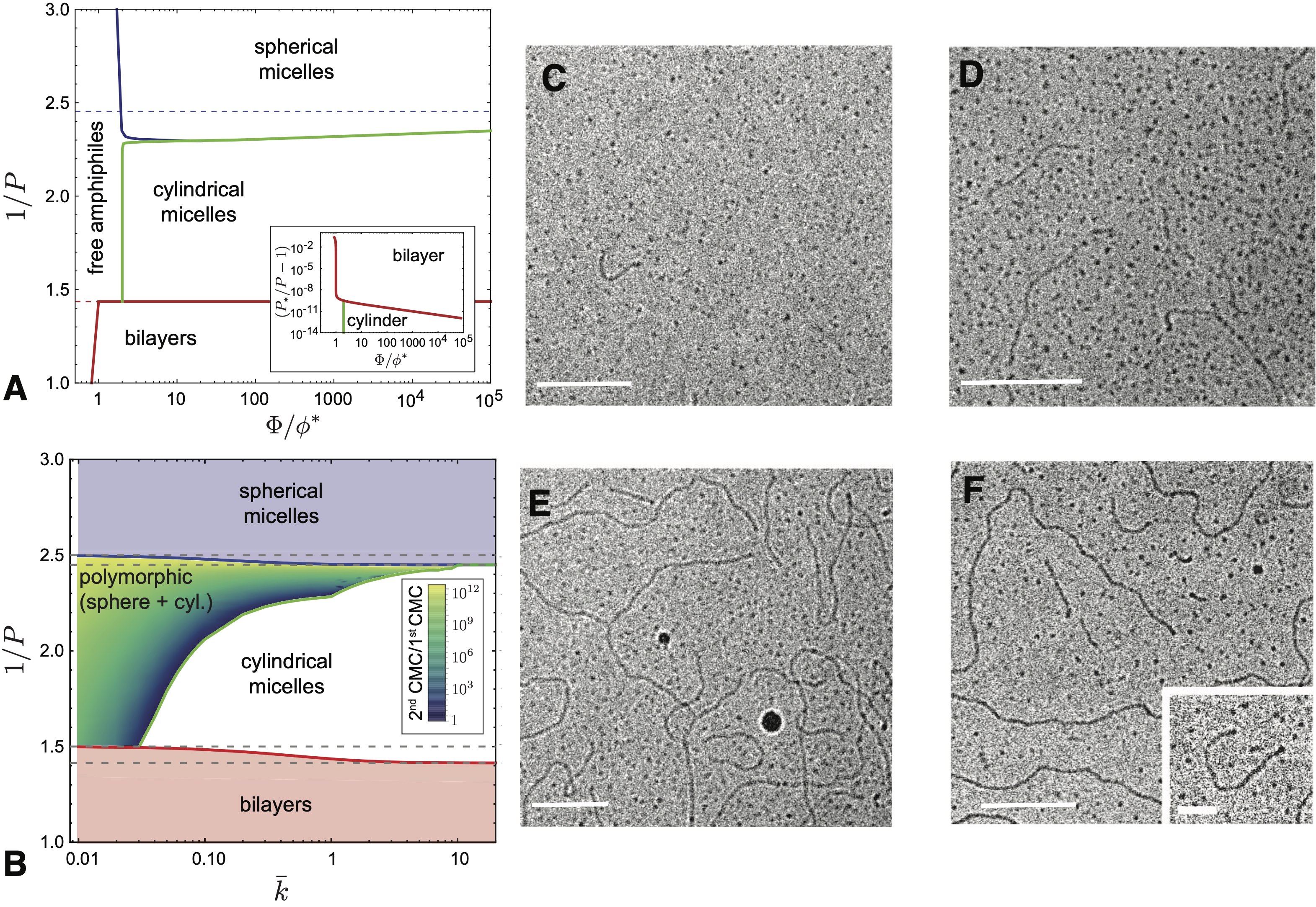}
\caption{\footnotesize (A) An assembly state phase diagram for the amphiphillic aggregate model in eq. (\ref{eq: amphi}) for  dimensionless stiffness $\bar{k} =1$ and $v_0^2/a_0^3 = 10$, where $\phi_*$ is the nominal CMC for cylinders.  Solid lines mark the boundaries between the most populous aggregate type, and dashed lines indicate the boundaries in the infinite $\Phi$ limit.  The inset shows a zoom in near the boundary between cylinders and bilayers, illustrating an extremely narrow window of secondary CMC behavior due to the large mean (finite) size of cylinders.  (B) Summarizes the polymorphic assembly of the amphiphile model in the plane of stiffness $\bar{k}$ and inverse packing parameter $P^{-1}$, for $v_0^2/a_0^3 = 10$.  Regimes of single CMC behavior are shown as solid red, white, and blue for bilayers, cylinders, and spheres, respectively.  The regime of polymorphic concentration-driven sphere-to-spherocylinder transitions is colored on a blue-green scale according to the ratio of second CMC (spheres to spherocylinders) to the first CMC (monomers to spheres).  (C-F) show cryo-transmission electron micrographs of micelles formed by dimeric (gemini) surfactancts, at $25^\circ$ C at increasing weight percent:  (C) 0.26\%; (D) 0.5\%;  (E) 0.62\%; and (F) 0.74\% (scale bars equal 100 nm).  The coexistence of cylindrical micelles of increasing total length for (D-F), and absence of lengths intermediate to spheres and the shortest cylinders is consistent with a concentration-dependent second-CMC.  The inset of (F) shows the bulbous ends of a cylindrical micelle (scale bar, 25 nm), consistent with an energy barrier between spherical and (long) cylinders due to a mismatch in preferred radius.  (C-F) are adapted from \cite{Bernheim-Groswasser2000}.  }
\label{fig:phasecp}
\end{figure*}

A more complete picture of the polymorphism of aggregates is given by considering assembly landscapes that allow for transitions of micellar dimension.  Fig.~\ref{fig: surfactants}C shows the structure of a ``wormlike" micelle, modeled as a spherocylinder composed of a length $L$ of cylindrical micelle capped by two equal-radius hemispherical micelle caps.  As $L$ increases, the fraction of the aggregate in the $d_{\rm L}=2$ (vs. $d_{\rm L}=3$) packing increases, and thus consideration of the energy as a function of both $n$ and $L$ illustrates the landscape of aggregates intermediate to a uniformly cylindrical or spherical geometry.   An example landscape is shown in Fig. ~\ref{fig: surfactants}D, for a packing parameter $P^{-1}=2.25$ intermediate to spheres and cylinders, which exhibits two branches of local minima.  In the spherical branch, $L=0$ and changes in number are accommodated purely through changes in micelle radius.  The second, cylindrical, branch appears only above a threshold aggregate number ($n \simeq 111 v_0^2/a_0^3$), beyond which further subunit addition is accommodated through increasing the length.  The per subunit aggregation energies for these two branches are shown in Fig. ~\ref{fig: surfactants}E.  Notably, an energy barrier separates the convex minimum of the spherical branch from the cylindrical branch, which asymptotically approaches the global minimum of $\epsilon$ at $n\to \infty$ via the $1/n$ falloff characteristic of 1D assembly.  The origin of this ``barrier'' between spherical and cylindrical aggregates derives from the fact that the preferred radii of these two micelle types are different, and hence the confinement energy of the ``endcaps'' on the spherocylindrical micelle exceeds that of a minimal-energy (larger radius) sphere.  We note that this spherocylinder geometry will overestimate the cost of endcaps relative to a more realistic model that includes, say, variation of the radial thickness along the micelle~\cite{May2001}.

As described in Sec.~\ref{sec:SecondCMC}, the existence of multiple low-energy branches of aggregation leads to a rich phase behavior, which can be analyzed according to the law of mass action eq.~\eqref{eq: lma}.   As summarized in Appendix~\ref{sec:SurfAppendix}, we analyze the polymorphic assembly of the model described by eq.~(\ref{eq: amphi}), making several simplifying assumptions.  Specifically, we adopt the continuum limit for $n$, calculate the optimal spherocylinder radius for each length $L$ by minimizing the per-molecule energy eq.~\eqref{eq: amphi} as a function of $r$ (i.e. neglecting radius fluctuations), and numerically calculate the mass fractions of spheres, spherocylinders, and membranes as functions of $k$, $P$, $\Phi$ from eq.~\eqref{eq: lma}.

Fig.~\ref{fig:phasecp}A shows the phase diagram as a function of the packing parameter and total concentration $\Phi$ for a dimensionless stretching stiffness $\bar{k}=1$. Analogous to the phase diagram of the generic model in Fig.~\ref{fig: DCMCphaseDiagram}B, with increasing packing parameter the system undergoes dimensional transitions from spheres to spherocylindrical assemblies and finally to bulk lamellar aggregates (corresponding to packing dimensionalities of $d_{\rm L}=3,\, 2,\, 1$). The infinite concentration limits of the phase boundaries  (indicated by dashed lines in Fig.~\ref{fig:phasecp}A) correspond to the packing parameter values where the bulk free energy per subunit of two aggregate geometries are equal: $\epsilon_3= \epsilon_2$ at the sphere/spherocylinder boundary, and $\epsilon_{2}=\epsilon_{1}$ at the spherocylinder/lamella boundary, with $\epsilon_{d_{\rm L}}$ the optimal energy per subunit in the aggregate interior for aggregates with dimensionality $d_{\rm L}$. Notably, for values of the packing parameter near each of these dimensional transitions, there is a concentration-dependent dimensional transition. I.e., for packing parameter values close to the infinite-concentration value for the spherocylinder-sphere transition $2.45\gtrsim P^{-1} \gtrsim 2.3$, spheres are favored at low concentrations, with a transition to spherocylinders occurring above a threshold concentration that diverges exponentially as the packing parameter approaches the transition value, i.e. $P^{-1} \to 2.45$. An analogous behavior occurs at the spherocylinder/lamella transition (Fig.~\ref{fig:phasecp} inset), but the range of $P^{-1}$ values is exceptionally narrow due to the very large (yet finite) mean size of spherocylinder aggregates (see below).

In Fig.~\ref{fig:phasecp}B, we show an overview of the polymorphic assembly in terms of the two control parameters of the amphiphile aggregation model, $P^{-1}$ and $k$. Boundaries in the $P^{-1}-\bar{k}$ space are shown for infinite-concentration sphere/spherocylinder and spherocylinder/lamella transitions, indicated by blue and red lines respectively, with gray dashed lines corresponding to the values of these transitions in the extensionally ``floppy'' ($\bar{k}\rightarrow 0$) and ``stiff'' ($\bar{k}\rightarrow\infty$) limits~\footnote{The transition values between $d_{\rm L}$ and $d_1+1$ are $P^{-1}=(d(d-1))^{1/2}$ and $P^{-1}=d_{\rm L}-1/2$ in the respectively stiff ($\bar{k}\rightarrow\infty$) and floppy ($\bar{k}\rightarrow 0$) limits.}.
We also show the region in $P^{-1}-\bar{k}$ space for which there is a concentration-driven sphere/spherocylinder transition (a secondary CMC), with color indicating the width in concentration space of the transition.  More specifically, the color scale indicates the ratio of pseudo-critical concentrations, $\Phi_{\rm cyl}/\Phi_{\rm sph}$ where spherocylinders or spheres, respectively, become the most populous subunit state (i.e. greater than 50\%) at a given value of stiffness and packing parameter. 

The emergence of the polymorphic, concentration-driven transition between spherical and spherocylindrical micelles captured in Fig.~\ref{fig:phasecp} can be understood in terms of the three ingredients of the secondary CAC behavior encoded in eq. (\ref{eq:deltaStar}):  the energy gap between minimal-energy spheres and infinite cylinders $\Delta \epsilon = \epsilon_3-\epsilon_2$, the finite aggregation number in spherical micelles $n_{\rm sph}$, and an energy barrier $\delta$ separating the spherical aggregates from spherocylinders (as in Fig.~\ref{fig: surfactants}D).  Below the blue curve in Fig.~\ref{fig:phasecp}B, where $\Delta \epsilon>0$, the physical origin of the intermediate concentration state of spheres is the higher (per subunit) entropy associated with their fewer subunits.  This window of second CMC behavior widens in $\Phi$ as the gap between infinite cylinders and spheres vanishes to zero, which happens as the inverse packing parameter increases and approaches the blue curve in Fig.~\ref{fig:phasecp}B.  Likewise, from eq. (\ref{eq: amphi}) it is straightforward to see that the gap between $d_{\rm L}=3$ and $d_{\rm L}=2$ vanishes as $\bar{k}\to 0$, since the is no obstacle to achieving the optimal head group packing ($a \to a_0$)  for any $d_{\rm L}$ in the absence of extensional stiffness.  Hence, the ratio of the second CMC (spheres to spherocylinders) relative to the first (monomers to spheres) grows large in both of these regimes.  

The regime of second CMC behavior is restricted to lower values of extensional stiffness, and disappears above a threshold value of $\bar{k}$, due to its effect on the energy barrier between spheres and spherocylinder micelles.  In the limit of $k \to 0$, the thickness of spheres and cylinders is determined purely by head group packing, and hence $d_{\rm L}=3$ and $d_{\rm L}=2$ micelles have different radii, implying a finite frustration cost for the hemispherical endcaps of the spherocylindrical micelles.  As summarized in eq. (\ref{eq:deltaStar}), the window of second CMC behavior is widened with increasing energy barrier between compact and 1D assemblies.  With increasing $\bar{k}$, this barrier diminishes, ultimately vanishing the in the $\bar{k} \to \infty$ limit, because the high stiffness requires the micelle thickness to maintain $\ell = \ell_0$ independent of dimensionality.  Hence, as is the case for the ``ladder model'' of cylindrical micelle thermodynamics \cite{Missel1980}, in the absence of an energy barrier between spheres and elongated cylinders, there is only a single CMC to a state where mean aggregation number continuously increases with $\Phi$.

Similar arguments apply to the spherocylinder/lamella transition, except that spherocylindrical aggregates have a very small translational entropy due to their large mean size, and thus are stabilized by entropy over a vanishingly narrow region of parameter space.

Evidence for the secondary CMC transition between spherical and cylindrical micelles has been reported for range of surfactant systems; see \cite{MagnusBergstrom2016} for a review.  Many experiments report an indirect signature of an inflection point, and secondary upturn, in the mean aggregation number as a function of concentration.  For example, the convex dependence of viscosity on concentration for certain ionic surfactants was interpreted by Porte in coworkers as a secondary CMC, and  motivated their model with an energetic gap between spherical and cylindrical states for second CMC behavior \cite{Porte1984}.  Elsewhere, experimental imaging of the concentration-dependence micelle morphologies has been used to probe second CMC behavior.  For example, Figs.~\ref{fig:phasecp} C-F show cryo-electron microscopy (cryo-EM) images of micelles formed by dimeric (gemini) surfactants~\cite{Bernheim-Groswasser2000}.  Above the first CMC, but still below a second threshold  concentration, cryo-EM shows monodisperse spherical micelles  Fig.~\ref{fig:phasecp}.  With increased concentration (Fig.~\ref{fig:phasecp}D-F), cryo-EM shows the appearance of cylindrical, or worm-like, micelles.  While these grow in length with concentration, they also retain coexistence with a population of spherical micelles. This, along with an observed gap in micelle sizes intermediate to spheres and the shortest cylinders, is consistent with the secondary CMC transition induced by an energetic gap between spheres and cylinders.  As described above, and consistent with the apparently ``bulbous'' ends of worm-like micelles  (Fig.~\ref{fig:phasecp}F), such an energetic gap is a natural consequence of the mismatch between radii of spherical and cylindrical micelle packings.

To conclude the overview of amphiphillic aggregation, we briefly return to the question of {\it convexity} of the energetics described by (\ref{eq: epint}) - (\ref{eq: amphi}) for spherical micelles ($d_{\rm L} =3$).  It is straightforward to consider two simple limits to esimtate the dependence of convexity on target size.  When $\bar{k} \ll 1$, energetics are dominated by headgroup area terms and $\epsilon_{\rm T}'' (\bar{k} \ll 1) \propto \gamma v_0^{2/3} /n_{\rm T}^{7/3}$.  Whereas, the opposite limit is controlled by length elasticity, and $\epsilon_{\rm T}'' (\bar{k} \gg 1) \propto k v_0^{2/3} /n_{\rm T}^{4/3}$.  Notably, as with the fluid capsule model in the previous section, convexity generically decreases with target size, however, not nearly as quickly.  In particular, for the amphiphile model the product $n_{\rm T}^3 \epsilon_{\rm T}''/ kt$ is always {\it increasing} with $n_{\rm T}$ (i.e. as $n_{\rm T}^{2/3}$ and $n_{\rm T}^{5/3}$ for small and large $\bar{k}$, respectively).  Hence, according to eq. (\ref{eq: disp}) the relative number fluctuations of micelles {\it decrease} with mean size.  This is in marked distinction with the predictions of the bending elasticity of the fluid shell model, for which $\epsilon_{\rm T}'' \propto n_{\rm T}^{-3}$ and size fluctuations grow in proportion to mean size.  The origin of this relatively sharper minima of aggregation energetics, and correspondingly tighter control of aggregate size, can be traced to the fact the the closure radius is on the scale of the subunit thickness (i.e. $R_{\rm close} \approx t$), or more specifically, the additional considerations of tail packing constant density in the micellar cores, which are absent for the membrane bending elasticity of fluid capsules.

\subsection{Self-limited, open-boundary assembly}
\label{sec:OBA}

Here we describe a class of self-limiting assembly characterized by an {\it open boundary}, a surface that separates the aggregate interior from the solution of freely associating subunits.  Unlike the SCA described above, in open-boundary assemblies (OBA) this free boundary is maintained in the target self-limiting states. Therefore, the target state has a finite surface energy associated with loss of short-range cohesion or differences in solvation at its exterior.  

As described in Sec.~\ref{sec: capsules}, a finite boundary energy alone generically favors unlimited aggregates. Hence, to be self-limited, OBA structures require additional interaction terms that grow with aggregate dimensions, and thus balance the generic tendency to minimize the boundary to interior ratio.  We define this additional (non-surface) energetics as the {\it excess energy}, and its essential feature is a regime of {\it super-extensive growth}, meaning the total excess energy increases with size faster than the number of subunits $n$.  Below we describe two example mechanisms that generate such a form of excess energy, but the key underlying feature is the existence of {\it gradients in stress} throughout the aggregate.  Whereas SCA can realize finite target dimensions with uniform subunit shape and packing, in OBA long-range, gradient patterns of intra-aggregate stress are required for self-limitation and ultimately dictate the range of possible self-limiting sizes.  

Before introducing these two physical mechanisms, which will illustrate the microscopic origins of excess energy accumulation, we begin with a generalized description of the thermodynamics of OBA.

\begin{figure}
\centering
\includegraphics[width=0.95\linewidth]{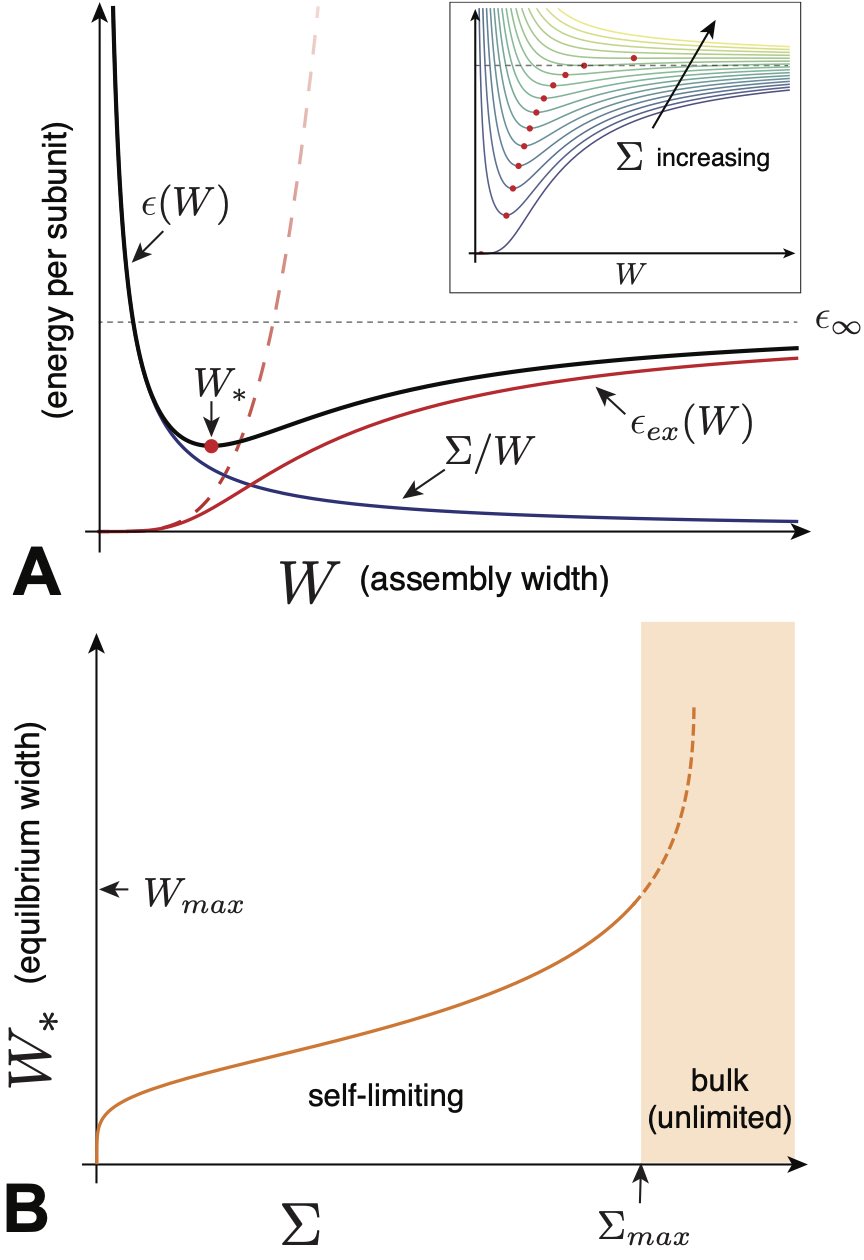}
\caption{\footnotesize (A) A schematic plot of the aggregation energetics (per subunit) for self-limiting open boundary assembly, where the blue, red, and black curves respectively show the surface energy density, excess energy, and total energy density. The dashed curve shows power-law growth of excess energy at small size.  The red point highlights the equilibrium self-limiting width $W_*$, and the inset shows the variation of this equilibrium width with increasing surface energy $\Sigma$, assuming fixed $\epsilon_\text{ex}(W)$.  (B) A schematic plot of the equilibrium width as a function of $\Sigma$, where the dashed portion indicates the possibility of a finite-width branch that is metastable relative to the bulk state $W\to \infty$.  The boundary between equilibrium self-limiting and bulk states is marked by a maximum self-limiting size $W_\text{max}$, denoted as the {\it escape size}. } 
\label{fig: openboundary}
\end{figure}

\subsubsection{Limits of self-limitation}
\label{sec: accumulant}

To describe the aggregation energetics of OBA, we consider a structure with an open boundary that can grow in $D$ possible spatial dimensions (directions). We let $d_{\rm L}$ of these directions be potentially self-limiting, while in the remaining $d=D-d_{\rm L}$ directions the structure undergoes unlimited growth.   Denoting the limited and unlimited dimensions of the structure as $W$ and $L$ respectively, the scaling of aggregate volume gives $n \propto W^{d_{\rm L}} L^{d}$. In the limit that $L \gg W$, the amount of the open boundary then grows as $A_{\rm b} \sim W^{d_{\rm L}-1} L^{d}$.~\footnote{For $D=2$, assemblies are sheet-like and the boundary corresponds to a 1D edge due to fewer (lateral) cohesive bonds, while for $D=3$, the boundary corresponds to the entire 2D surface surrounding the aggregate.} For example, consider subunits that can bind in all three spatial dimensions, i.e. $D=3$. Quasi-cylindrical aggregates of such subunits accrue a surface energy cost deriving from their full boundaries, which are limited in two transverse spatial dimensions ($d_{\rm L}=2$) and unlimited in the axial direction ($d=1$); whereas finite-thickness planar aggregates are limited along the normal direction ($d_{\rm L}=1$), but unlimited in the two in-plane directions ($d=2$).  Notably, such structures are self-limiting spatially, but do not necessarily have a finite or even well-defined peak aggregation number~\footnote{Indeed, cylindrical aggregates have an exponential length distribution, while planar aggregates (e.g. plates and membranes) will correspond to a bulk state according to the analysis of Sec.~\ref{sec: unlimited}.}.  Nevertheless, equilibrium in these cases (e.g. finite-thickness ``filaments'' and ``slabs") derives from the optimal energy per subunit with respect to the self-limiting dimension(s) of the assembly.

Because surface subunits typically have fewer cohesive bonds with neighbor subunits and potentially more unfavorable contacts with surrounding solvent, an open boundary generically accrues a surface energy cost proportional to $A_{\rm b} $.  Parameterizing this cost by the boundary energy $\Sigma$ and the ``bare'' aggregation energy for interior subunits $-\epsilon_0$, we may write a generic form for the per-subunit aggregation energy as a function of the self-limiting dimension $W$:
\begin{equation}
    \epsilon(W) = -\epsilon_0 + \frac{\Sigma}{W} +\epsilon_\text{ex}(W).
    \label{eq: excess}
\end{equation}
For simplicity, we absorbed a geometric factor associated with the dimensionsality of the boundary into the definition of $\Sigma$.  The first two terms describe the short-range cohesive interactions, with a constant bulk energy gain and a surface energy penalty, while the final term defines the {\it excess energy} relative to the short-range model.  An example energy of this form is shown in Fig.~\ref{fig: openboundary}A. As we illustrate below, such an excess energy arises from effects such as long-range inter-subunit repulsions or elastic stresses that increase with aggregate size.  In self-limiting assemblies, these ``cumulative'' effects give rise to an excess energy (per subunit) that increases monotonically in size; hence, $\epsilon_\text{ex}(W)$ captures energetic effects that grow {\it super-extensively} with aggregate size.  In any physical system, this super-extensive energy growth will only persist up to some threshold assembly size, crossing over from convex (e.g. power-law) growth at small sizes to some asymptotically saturating energy density as $W \to \infty$.  The large-$W$ saturation of excess energy can occur for a variety of reasons. For example, long-range repulsions may be screened beyond some length scale. Alternatively, above a threshold excess energy cost, it will become energetically favorable for subunits to reorganize or deform to avoid further excess energy accumulation.  Notwithstanding its microscopic origin, the effect of this saturating excess energy is to renormalize the per subunit energetics from its bare value to $-\epsilon_0+\epsilon_\infty$.

In OBA, self-limitation follows directly from the balance between the accumulating cost of $\epsilon_\text{ex}(W)$ and the generic decrease of surface energy with increasing $W$.  It is then straightforward to show that equilibrium assemblies satisfy the following ``equation of state'' that links the finite equilibrium size $W_*$ to the surface energy,
\begin{equation}
    \Sigma = W_*^2 \epsilon'_\text{ex} (W_*) , 
    \label{eq: Sigeos}
\end{equation}
with $ \epsilon'_\text{ex} = \partial_W \epsilon_{\rm ex}$.  Stability criteria additionally require that $\epsilon''_\text{ex}>0$, but the basic results of the competition between surface energy and excess energy accumulation are shown schematically in Fig.~\ref{fig: openboundary}.  Since the surface energy always drives assembly toward larger sizes, the equilibrium finite size $W_*$ generically increases with $\Sigma$ given a fixed form of $\epsilon_\text{ex}(W)$.  Self-limitation can arise in two ways:  either the equilibrium width can increase {\it continuously} with $\Sigma$ to the bulk state (i.e. $W_* \to \infty$), or as illustrated in Fig.~\ref{fig: openboundary}B, self-limitation will persist only up to a {\it maximal} finite size before a discontinuous transition to bulk assembly occurs.  In the latter case, the energy density of the finite state eventually increases with surface energy, to the point where the energy densities of the finite and bulk (unlimited) states become equal; i.e. $ \epsilon(W \to \infty)=-\epsilon_0+\epsilon_\infty$. For surface energies above this {\it maximal} value, the bulk (unlimited) state is favored. Hence, such systems can be characterized by a maximal self-limiting size, $W_\text{max}$, and a maximal surface energy $\Sigma_\text{max}$ below which equilibrium structures are finite (e.g. Fig.~\ref{fig: openboundary}B).

For a general OBA, it is then useful to consider the following question:  What are the range of possible self-limited equilibrium states that a given system can exhibit?  As described in Sec.~\ref{sec:maxClusterSize}, provided that the concentration is well above the aggregation threshold (CAC), the mean aggregate size is determined by the minimum of the per-subunit energy $\epsilon(n)$. Thus, the answer to this generic question depends only on the excess energy and its accumulation with width~\footnote{Eq.~\ref{eq: n_gauss} shows a small concentration-dependent shift of the optimal size $n^*$ below the size corresponding to the minimum of $\epsilon(n)$; the optimal size approaches the energy-minimizing size as $\Phi \to \infty$.}. 

\begin{figure}
\centering
\includegraphics[width=0.95\linewidth]{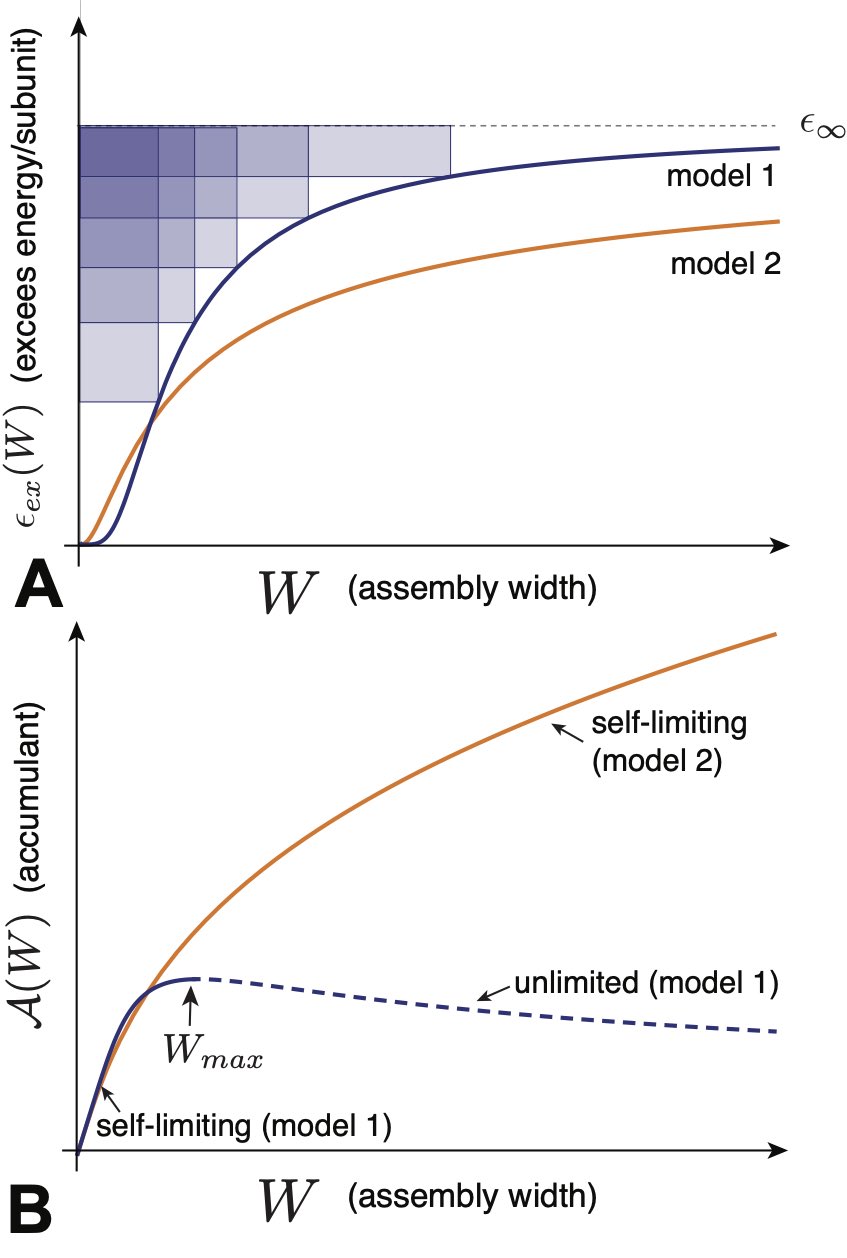}
\caption{\footnotesize (A) Schematic plots of excess energy versus width for two different models of OBA: model 1 (blue); and model 2 (orange).  Both crossover from power-law growth at small size to saturating excess energy at large size, but do so with different functional forms (notably, the asymptotic approach to $\epsilon_\infty$ is slower in model 2 compared to model 1).  The area of the blue rectangles  graphically illustrates the definition of the {\it accumulant} in eq. (\ref{eq: accumulant}). (B) Accumulant ${\cal A}(W)$ as a function of width for the two models shown in (A).  Regions of increasing ${\cal A}(W)$ correspond to possible ranges of equilibrium self-limiting sizes for a given form of $\epsilon_\text{ex}(W)$.  That is, values of $W$ for which ${\cal A}'(W)>0$ correspond to energy minima of $\epsilon(W)$, which are lower in energy than the bulk state $\epsilon(W\to\infty)$, for a particular value of the surface energy $\Sigma$ give by eq. (\ref{eq: Sigeos}).  Hence, model 1 shows an upper limit for maximal self-limiting size $W_\text{max}$, while model 2 exhibits stable self-limiting equilibria at all sizes.}
\label{fig: accumulant}
\end{figure}

To see this, we reformulate the condition for equilibrium of the self-limiting state relative to bulk assembly, $\epsilon(W_*) < \epsilon(W \to \infty)$, in terms of the surplus of energy in the bulk relative to the finite state, 
\begin{align}
    \Delta \epsilon (W_*) \equiv & \epsilon (W \to \infty) - \left(\frac{\Sigma}{W_*} + \epsilon_\text{ex}(W_*) - \epsilon_0\right)  \\
    = & \frac{\partial }{\partial W_*} \Big(W_* \big[\epsilon_{\infty} -\epsilon_\text{ex}(W_*) \big] \Big) ,
    \label{eq: stablefinite}
\end{align}
where we used the equation of state linking stable size to surface energy in eq. (\ref{eq: Sigeos}).  The condition that $\Delta \epsilon (W_*) >0$ is required for  equilibrium finite (self-limited) states can then be simply formulated in terms of the first integral of surplus bulk energy
\begin{equation}
    {\cal A}(W) \equiv  W \big[\epsilon(W \to \infty) -\epsilon(W) \Big] = W \big[\epsilon_{\infty} -\epsilon_\text{ex}(W) \Big] - \Sigma, 
        \label{eq: accumulant}
\end{equation}
so the equilibrium of finite structures relative to bulk corresponds to the condition
\begin{equation}
    \Big( \frac{ \partial {\cal A} }{\partial W} \Big)_{\Sigma} > 0.
    \label{eq: increase}
\end{equation}
We refer to the function ${\cal A}(W)$ as the {\it accumulant}, and note that graphically it corresponds to the area of the rectangular regions of the plot of $\epsilon_\text{ex}(W)$ versus $W$ highlighted in Fig.~\ref{fig: accumulant}A, for the model (1) described by the blue curve. Notably, because the fixed-$\Sigma$ partial derivative in eq. (\ref{eq: increase}) is independent of surface energy, for a given form of excess energy the accumulant may be constructed for any value of surface energy, and analyzed as a function of $W$ to consider potential finite-size equilibria at {\it all values} of $\Sigma$.

According to eq. (\ref{eq: increase}), finite-$W$ equilibria correspond to the range of {\it increasing} ${\cal A}(W)$.  Fig. \ref{fig: accumulant} illustrates two models of excess energy, both of which are characterized by crossovers from power-law growth at small-$W$ to asymptotic saturation to finite values of $\epsilon_\text{ex}(W \to \infty)=\epsilon_{\infty}$, but do so via different functional dependencies on the finite size $W$. 

For model 1 (blue), the range of increasing ${\cal A}(W)$ extends only up to a maximum, with corresponding width $W_\text{max}$, indicating that a first-order transition between finite and bulk states occurs at $W_\text{max}$ and the corresponding value of $\Sigma_\text{max}$~\footnote{Defining the accumulant in terms of the $\Sigma = 0$ energetics, i.e. ${\cal A}(W) \equiv  W \big[\epsilon_{\infty} -\epsilon_\text{ex}(W) \Big]$, as in Fig.~\ref{fig: accumulant}B, it can be easily shown that $\Sigma_\text{max} = {\cal A}(W_\text{max})$.}.  For model 2 (orange), the monotonically increasing range of the accumulant extends to $W \to \infty$.  This indicates that model 2 supports equilibrium self-limited states at all values of $\Sigma$, and only reaches bulk assembly in the limit of infinite surface cost.   Hence, even if the excess energy saturates in the $W \to \infty$ limit, self-limitation may still, in principle, extend to all possible size ranges, depending on the nature of the asymptotic approach to the bulk energy.

Given the form of the accumulant defined in eq. (\ref{eq: accumulant}) and the condition eq. (\ref{eq: increase}) for its increase, the possibility for self-limited states that extend continuously up to the bulk state (i.e. $\lim_{\Sigma\to \infty} W_* \to \infty$) can be deduced from the asymptotic form of the residual energy $\Delta \epsilon(W) =\epsilon_\text{ex}(\infty) -\epsilon_\text{ex}(W) $ as $W \to \infty$.  Following an argument made by \cite{LeRoy2018, Terzi2020}, we assume this residual vanishes as a power law $\epsilon_\text{ex}(\infty) -\epsilon_\text{ex}(W) \sim W^{-\nu}$.  It is then straightforward to show that when $\nu > 1$, ${\cal A}(W)$ decreases as $W \to \infty$. Such cases correspond to the first-order type self-limitation exhibited by model 1 in Fig.~\ref{fig: accumulant}.  Alternatively, when $0<\nu<1$, indicating a slower saturation of excess energy, the accumulant continues to increase as $W \to \infty$, indicating the existence of self-limited equilibria extending up to the bulk state.   The case of $\nu =1$ is marginal, and can exhibit either first- or second-order like behavior.  Below, we describe models in the context of geometrically frustrated assemblies that can illustrate both types of behavior, either a {\it continuous} or {\it discontinuous} transition between the finite and bulk states depending on the mechanisms underlying the accumulation of excess energy.

\subsubsection{Short-range attractions, long-range repulsions}
\label{sec:SALR}

As described in the previous section, self-limitation in OBA requires superextensive growth of the excess energy.  In this first class of examples, the accumulation of excess energy derives from long-range interactions between subunits, specifically, interactions characterized by {\it short-range attraction and long-range repulsion} (SALR)~\cite{Groenewold2001, Sciortino2004}.  Models of such systems usually consider isotropic pair-potentials $u(r)$ that can be split into two parts,
\begin{equation}
    u(r) = u_\text{SA}(r) + u_\text{LR}(r),
\end{equation}
where the short-ranged potential describes cohesive interactions $u_\text{SA}(r)<0$ between neighboring subunits, which act on scales comparable to the subunit hard-core diameter $d$; i.e., $u_\text{SA}(r \gg d) \simeq 0$.  Outside of this cohesive range, the potential is dominated by a long-range repulsion $u_\text{LR}(r)>0$ that extends over sizes much larger than single particles.  Even when the repulsive interactions between neighboring subunits are much weaker than the cohesion, the fact that repulsive interactions extend far beyond the first shell of neighbors can lead to superextensive growth of repulsive energy with increasing assembly size.  A model of this form has been applied to explain finite-sized aggregate formation in a broad range of systems~\cite{Dinsmore2011}, including protein complexes~\cite{Stradner2004, Cardinaux2007, Fodera2013}, charged nanoparticles~\cite{Nguyen2007}, colloidal particles in low-dielectric solvents~\cite{Sedgwick2004, VanSchooneveld2009}, dipolar mesophases~\cite{Seul1995}, and even models of ``nuclear matter''~\cite{Caplan2017}.

\begin{figure*}
\centering
\includegraphics[width=1\linewidth]{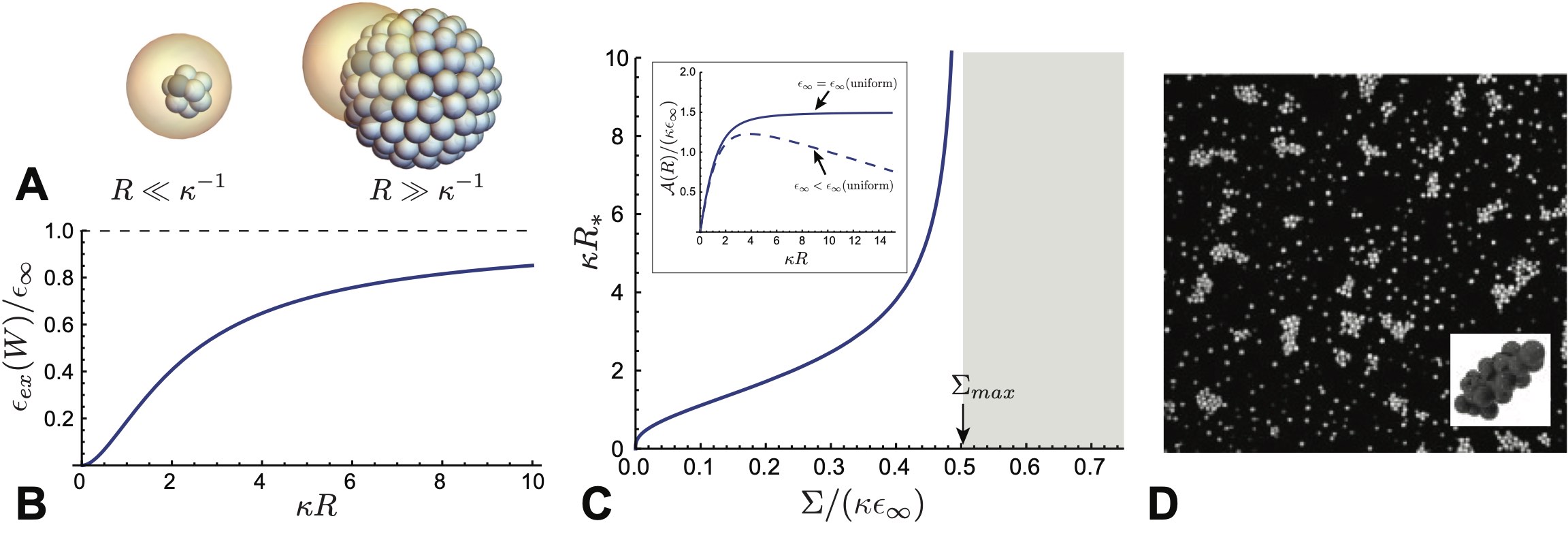}
\caption{\footnotesize (A) Schematics of aggregates of particles interacting via short-range (i.e. contact) interactions and long-range repulsions, with a screening length $\kappa^{-1}$, which can be large or small compared to the aggregate size $R$. (B) Excess energy of spherical aggregates versus radius, showing a crossover from power-law growth for $R \ll\kappa^{-1}$ to an asymptotically saturating bulk energy for $R \gg \kappa^{-1}$.  (C) Plot of the equilibrium finite radius $R_*$ of the SALR model versus surface energy $\Sigma$, showing a continuous divergence to the bulk at a finite $\Sigma=\Sigma_\text{max}$. The second-order-like transition from finite-to-bulk states predicted by this model (solid curve) is consistent with the monotonically increasing form of the accumulant plotted in the inset, where this result assumes the bulk state has uniform density.  Assuming instead, that the bulk state has a lower energy, non-uniform density (e.g. a periodic aggregate morphology~\cite{Zhang2019}), the transition will be first-order, as illustrated by the non-monotonic accumulant shown as a dashed line.  (D) shows a confocal microscope image of clusters of charged colloidal particles (radius 660 nm) adapted from \cite{Stradner2004}.  In these experiments short-range attractions are induced by depletion forces generated by inert polymers in suspension, while electrostatic repulsions are maintained at relative long range due to the low-concentration of mobile ions in host organic solvent~\cite{Stradner2004}.}
\label{fig: SALR}
\end{figure*}

To illustrate the mechanism of self-limitation in this class of systems, we consider the following specific model: Short-range cohesive interactions lead to $-u_0$ per neighbor contact, and aggregates maintain an (approximately) uniform density $\rho_0 \approx d^{-3}$.  For repulsive interactions, we assume that subunits are isotropic and repel according to a screened (Yukawa) repulsion,
\begin{equation}
    u_\text{LR}(r) = \frac{q^2}{r}e^{-\kappa r}.
\end{equation}
Here $q$ is the electrostatic charge per subunit, assumed to be fixed, and $\kappa$ is the screening length, which arises from Debye-H\"uckel screening by mobile ions in solution, and truncates the far-field repulsions for $r \gg \kappa^{-1}$.  For simplicity, we illustrate a simplified version of the theory by Groenowold and Kegel (GK)~\cite{Groenewold2001}.  In our presentation, we consider the case where charge per subunit is fixed, a point which we revisit below.  Our primary purpose is to illuminate a model with the minimal ingredients for a self-limiting equilibrium, and due its  finite screening length, the long-range Yukawa potential provides a convenient example.  For purposes of illustration, we consider spherical aggregates with radius $R$, whose interaction (free) energy can be described by
\begin{equation}
    \epsilon(R) =- u_0 \frac{\langle z \rangle}{2} + \frac{ 3 \Sigma}{R} + \epsilon_\text{ex}(R) \
\end{equation}
where $ \langle z \rangle$ is the mean number of neighbor contacts in the bulk of the aggregate, and $\Sigma \approx u_0/d^2$ is the surface energy associated with fewer (short-ranged) cohesive contacts at the boundary.  Here, the excess energy derives directly from the sum of long-range, pairwise repulsions in the aggregate volume ${\cal V}$~\footnote{The form of the self-energy can be readily calculated using Greens theorem, where the total repulsive energy can be written as $\int_{\cal V} d^3 {\bf r}~ q \rho ({\bf r}) \phi ({\bf r})$ with a  potential that satisfies the linearized (Debye-H\"uckel) equation, $(\nabla^2 + \kappa^2) \phi ({\bf r}) = 4 \pi q \rho ({\bf r}) $.  Solving this for spherically symmetric aggregates, $\rho(r\leq R) = \rho_0$ and $\rho(r>R)$, yields the explicit form of screened electrostatic energy.},
\begin{multline}
  \epsilon_\text{ex}(R) = \frac{q^2 \rho_0}{2{\cal V} } \int_{\cal V} d^3 {\bf r}~ d^3 {\bf r}' u_{LR}(|{\bf r}- {\bf r}'|)  \\ =  \frac{q^2 \rho_0}{2 \kappa^2} \Big\{ 1 - \frac{3(1+\kappa R)}{2(\kappa R)^3}(1+e^{- 2\kappa R}) \big[\kappa R - \tanh (\kappa R) \big] \Big\}.
  \label{eq: SALRex}
\end{multline}
The behavior encoded in $\epsilon_\text{ex} (R)$ is plotted in Fig.~\ref{fig: SALR}B, and can be understood physically by considering the asymptotic limits of small and large aggregate size relative to the screening-length of the repulsive interactions, as shown schematically in Fig.~\ref{fig: SALR}A.  For small aggregates, 
\begin{equation}
\epsilon_\text{ex} (R \ll \kappa^{-1}) \simeq \frac{q^2 \rho_0}{5}R^2, 
\end{equation}
which derives from the fact that, when repulsions extend over the entire aggregate, the per-subunit cost of the Coulomb self-energy of an aggregate of is roughly $Q^2/(nR) \propto R^2$, where the total aggregate charge $Q = q n \propto R^3$. In the opposite regime, where aggregate sizes far exceed the screening length, 
\begin{equation}
\epsilon_\text{ex} (R \gg \kappa^{-1}) \simeq
\epsilon_{\infty} \Big( 1- \frac{3}{2 \kappa R} + \frac{3}{2 (\kappa R)^3} \Big), 
\label{eq: SALRlarge}
\end{equation}
where $\epsilon_{\infty} \equiv q^2 \rho_0 \kappa^{-2}/2$. This leading term $R \to \infty$ derives from the fact that each subunit in the bulk of the aggregate experiences repulsive interactions with  roughly $\rho_0 \kappa^{-3}$ other subunits within a screening length, while the first correction accounts for the surface layer of thickness $\kappa^{-1}$ with fewer neighbors within the screening length.  Notably, the subleading $1/R^3$ term can be associated with a square curvature cost (per unit area) for deforming the boundary shape from planar, which alters the distribution of repulsive particles near the free surface of the aggregate.

Fig.~\ref{fig: SALR}C shows the predicted equation of state for the self-limiting radius $R*$ as a function of surface energy, $\Sigma(R_*) = R_*^2 \epsilon'_\text{ex}(R_*)/3$.  For small $\Sigma$, the balance between the ``charging energy" of aggregates and surface energy leads to a growth $R_* \sim \Sigma^{1/3}$, which proceeds until the optimal aggregate size grows beyond the screening length.  In the large aggregate regime, the asymptotic approach to $\epsilon_{\infty}$ leads to an aggregate size that diverges continuously at a critical surface energy $\Sigma_\text{max} = \kappa \epsilon_\infty/2$.  The origin of the second-order-like transition to the bulk in this model can be traced to the $R \gg \kappa^{-1}$ form of the excess energy in eq. (\ref{eq: SALRlarge}).  The leading correction to $\epsilon_\text{ex} (R \to \infty) = \epsilon_\infty$ goes as $-1/R$, that is, in the notation of the foregoing Sec.~\ref{sec: accumulant}, $\nu =1$.  Hence, this leading correction at large $R$ behaves like a negative contribution to the surface energy, $\Sigma_\text{eff} = \Sigma-\Sigma_\text{max} $, due to the reduced electrostatic repulsion within a screening length of the surface.  When $\Sigma< \Sigma_\text{max}$, the effective surface energy is {\it negative}, and the aggregate equilibrium derives from the balance between the $-1/R$ drive to create more surface and the sub-leading $+1/R^3$ term, to give
\begin{equation}
    R_* \sim \sqrt{ \frac{ \epsilon_\infty \kappa^{-3} }{\Sigma_\text{max} -\Sigma} } ,
\end{equation}
which diverges (continuously) as $\Sigma \to \Sigma_\text{max}$.  This prediction, that there is a maximal cohesive surface energy for aggregates, but that their equilibrium self-limiting size extends to arbitrarily large values, is consistent with the plot in Fig.~\ref{fig: SALR}C(inset), which shows the accumulant of spherical aggregates to be monotonically increasing over the full range of $R$.

The result that self-limitation can extend up to arbitrarily large sizes (i.e. $W_\text{max} \to \infty$) for this simplified model may be surprising, as it implies that the thermodynamics are sensitive to the finite size of aggregates over much larger size ranges than the finite interaction range, $\kappa^{-1}$.  That is, a subunit whose interactions extend only $\kappa^{-1}$ can still ``sense'' that it should join an aggregate with a radius smaller than $R_* \gg \kappa^{-1}$, but not a larger one. The resolution of this puzzle is that the physical term that restrains aggregate growth in this $R_* \gg \kappa^{-1}$ regime derives from the square curvature cost of the free boundary (contributing as $+1/R^3$ to the excess energy).  Although repulsions are short-ranged compared to large aggregates (i.e. $R \gg \kappa^{-1}$), they are sufficiently non-local to sense the {\it curvature} of the boundary, and thereby, the global radius of the aggregate.  That is, repulsions in this regime give rise to precisely the type of square-curvature energetics that selects for finite sizes in self-closing assemblies in Sec. \ref{sec: capsules}; i.e., with a preferred boundary curvature that vanishes as $\Sigma \to \Sigma_\text{max}$~\footnote{Consistent with this result, performing an analogous calculation for a planar slab geometry shows that the self-limited size diverges exponentially above a threshold value of $\Sigma$, due to the absence of such a curvature term.}.

While self-limited assembly due to a competition between cohesive boundary costs and accumulation of long-range repulsion is generic, the specific features of this simplified model, including continuously diverging finite size for $R_* \gg \kappa^{-1}$, require several caveats.  Foremost, the convexity of the assembly energetics decreases with aggregate size as $\epsilon''_* \sim n_*^{-3}$ for $R_* \gg \kappa^{-1}$ for this model, similar to the case of self-closing shells in Sec. ~\ref{sec: capsules}. Thus, according to eq.~(\ref{eq: disp}), fluctuations in $n$ grow with self-limiting size, and diverge at the threshold as $\langle (\Delta n)^2 \rangle^{1/2} \propto n_* \sim (\Sigma_c- \Sigma)^{-3/2}$.  Hence, for all practical purposes the self-limitation to finite and well-defined sizes will not persist to arbitrarily large structures.  

Beyond this, the model is oversimplified and fails to incorporate a number of physical ingredients, which can influence the form of excess energy accumulation.  For example, more realistic models would include:  a finite compressibilty of the short-range cohesive forces, which allows for density variation with aggregate size and position within the aggregate; and for charged subunits in solvent, the effects of dielectric contrast between aggregates and the solvent, as well as variable degrees of charge condensation/dissociation as aggregates vary in size.  Indeed, when initial applications of this GK model aimed to understand the aggregation of lysosyme proteins in low-salt aqueous solutions, comparison to scattering measurements of the aggregate size suggested that variable ionization of subunits could not be neglected~\cite{Zaccarelli2007}.  While to a first approximation, it was argued that subunit charging is independent of aggregation number~\cite{Groenewold2001}, the charge per subunit and the screening length, which varies with counterion concentration, both vary with \textit{total concentration} of ionizable subunit species~\cite{Cardinaux2007}.  Hence, accounting quantitatively for the concentration dependence of aggregation for these electrostatic systems requires consideration of subunit charge in a self-consistent fashion, leading potentially to aggregation-dependent renormalization of repulisive interactions~\cite{Nguyen2007}, not to mention additional effects associated with non-spherical aggregate shapes~\cite{Sciortino2005}.  Notably, the non-spherical nature of aggregates of charged particles is evident in confocal images of a colloidal analog to the aggregation of nanoscopic proteins~\cite{Sedgwick2004}, as shown Fig. ~\ref{fig: SALR}D.

Finally, even beyond specific physical considerations of electrostatic SALR systems, the prediction of the simplified Yukawa model of diverging finite aggregate size assumes a spatially uniform bulk state, whereas models of systems with short-range attractions and long-range repulsions have been shown to form periodically-modulated aggregate phases, e.g. stripes, layers and spheres, at high concentration~\cite{Seul1995, Sear1999, Sciortino2004, Zhuang2016}.  That is, at sufficiently high densities, the uniform density bulk state is unstable to lower free energy periodic bulk states such that $\epsilon_\infty({\rm periodic}) = \epsilon_\infty({\rm uniform})$.  If the stability of equilibrium self-limitation is reanalyzed in terms of a competition with a lower-energy, non-uniform bulk state, then the maximal size range of self-limiting equilibrium becomes finite (see e.g. the accumulant plot in Fig.~\ref{fig: SALR}D).  The full thermodynamics of the transition between a dilute phase of self-limiting aggregates (above the CAC) to the long-range ordered bulk aggregate phases requires considerations beyond the ideal aggregation thermodynamics considered here.

\subsubsection{Geometrically frustrated assembly (GFA)} 

\label{sec: GFA}

\begin{figure}
\centering
\includegraphics[width=0.9\linewidth]{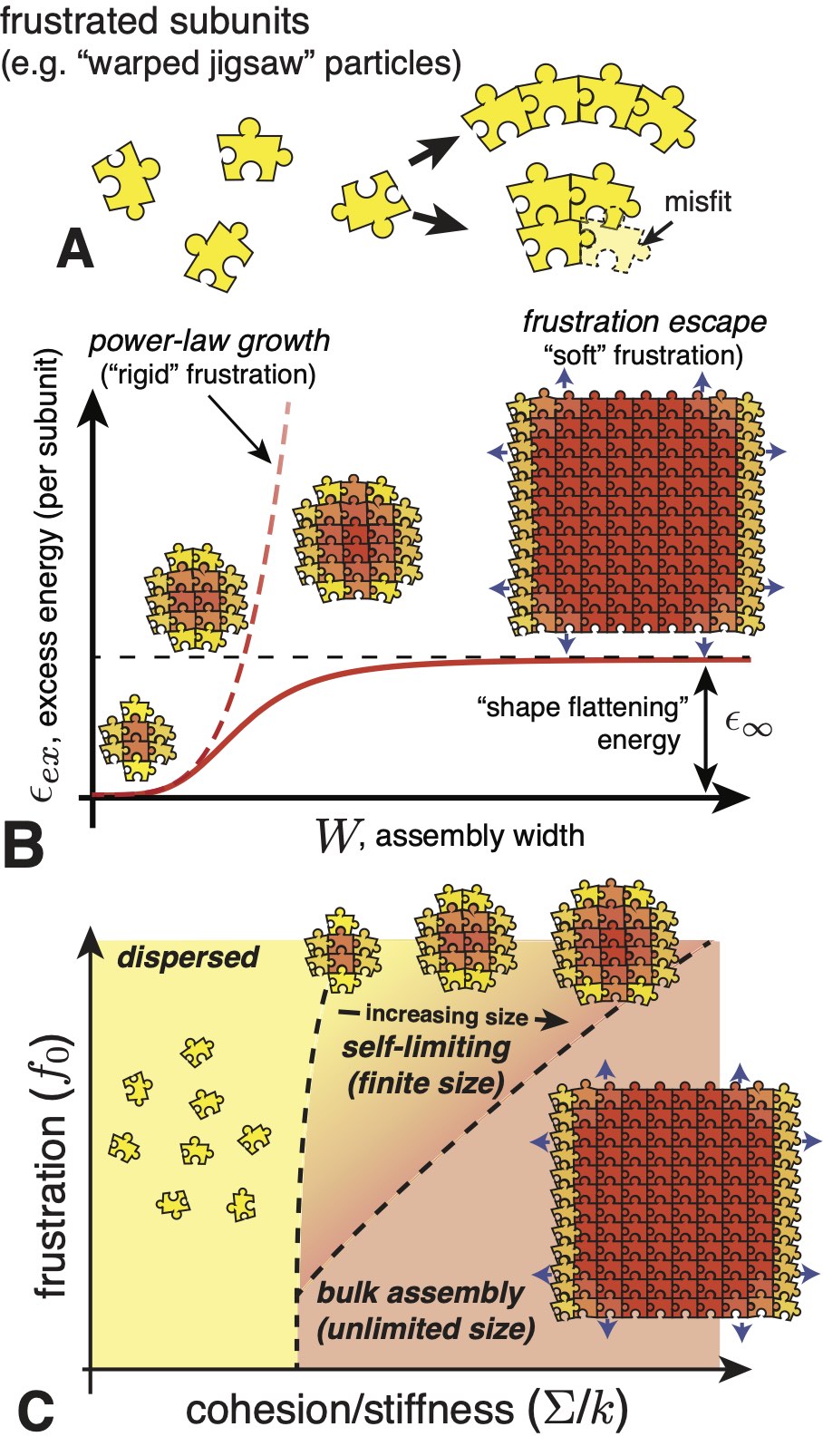}
\caption{\footnotesize (A) A heuristic ``warped jigsaw'' model for GFA, after ~\cite{Grason2017}, in which directional interactions promote curvature along rows of a locally preferred 2D ``lattice''.  Assembly in both bonding directions leads to ``misfits'', as in the tetrameric aggregate.  (B) A schematic plot of the excess energy (per subunit) for the warped jigsaw model, due to the superextensive build-up of elastic costs of misfits.  Intra-assembly strains are illustrated via particle color, from unstrained (yellow) to highly-strained (red).  The excess energy shows the characteristic crossover from power-law growth at small aggregates to an asymptotic approach to a strained bulk state (in this case envisioned as ``shape flattening'' of jigsaw particles).  (C) A schematic phase diagram for a generic model GFA in eq. (\ref{eq: epex2}), considered (at fixed concentration and temperature) as a function of the ratio of surface energy to elastic stiffness and a measure of the ``strength'' of frustration, $f_0$.  }
\label{fig: GFA}
\end{figure}

The notion of {\it geometric frustration} (GF) originally emerged in the context of low-temperature, condensed matter systems (e.g. magnetic materials, spin models)~\cite{Vannimenus1977}.  It refers to the impossibility of propagating an energetically-preferred arrangement throughout space, due to global geometric constraints~\cite{Kleman1989, Sadoc2006}.  For bulk, infinite systems, GF leads to a rich phenomenology: extensive arrays of topological defects thread through the highly-degenerate bulk ground states that populate a rough energy landscape.  

Recently has it been recognized that GF gives rise to new behaviors in self-assembling materials~\cite{Grason2016}, deriving from two key features.  First, the constituent subunits (e.g. polymers, colloids, proteins) are relatively ``soft" and held together by weak, non-covalent forces. Second, assemblies need not reach bulk states, and thus have additional degrees of freedom associated with the (potentially) finite-size and shape of the assembled domain.  Unlike bulk or rigid systems where GF must be resolved by defects~\cite{Sadoc2006}, in soft assemblies it can be tolerated, at least over some size range, by smooth gradients in the subunit shapes and packings.  As an illustration, see the schematic of ``warped jigsaw'' particles in Fig.~\ref{fig: GFA}A, where the tapering of the particle shape favors curvature along one row of the lattice assembly~\cite{Grason2017}.  Provided particles or their interactions are sufficiently deformable, aggregates can accommodate frustration through strain gradients, leading to arrangements that are more (less) relaxed near to (far from) the free boundary of the aggregate.  

This {\it self-organization of long-range stress gradients} is the defining characteristic of geometrically frustrated assembly (GFA), giving rise to the form of accumulating excess energy that can limit assembly size (Sec.~\ref{sec: accumulant}).  The balance between the surface energy and the (superextensive) cost of GF can select equilibrium domain sizes that are finite, and in principle, arbitrarily larger than the subunits themselves.

To date, GFA has been implicated in the emergent structures of various soft matter systems, including self-twisting protein bundles~\cite{Aggeli2001, Turner2003, Grason2007, Hall2016, Brown2014, Cameron2018,Yang2010}, twisted molecular crystals~\cite{Haddad2019, Li2020}, chiral smectics~\cite{Hough2007, Matsumoto2009} and membranes~\cite{Ghafouri2005, Selinger2004, Armon2014, Zhang2019, Gibaud2012, Sharma2014,Sakhardande2017,Kang2017}, particle-coated droplets~\cite{Bausch2003, Irvine2010, Meng2014, Yu2016}, curved protein shells~\cite{Zandi2004, Li2018}, and phase-separated lipid vesicles~\cite{Schneider2005}.  Consideration of GF in these systems has primarily stemmed from experimental observations of assemblies that (A) terminate at finite size and/or (B) exhibit ``defect-ordered" morphologies.  Continuum models that consider the interplay between the elastic costs of ``misfit'' and domain formation have been developed to address a range of distinct frustration mechanisms, including frustration of 2D liquid crystalline/crystalline order on non-Euclidean manifolds~\cite{Nelson1987, Bowick2009}, metric and orientational frustration of chiral fibers~\cite{Brown2014,Grason2015,Haddad2019}, shape-frustration in stacking assemblies of curved layers~\cite{DiDonna2003, Achard2005, Matsumoto2009}, chirality frustration in crystalline~\cite{Ghafouri2005, Armon2014} and liquid crystalline membranes~\cite{Selinger2001, Kang2017}, and the assembly of non-tiling polygonal particles~\cite{Lenz2017}.

We first describe a heuristic model that highlights the common thermodynamic features of these apparently diverse realizations of GFA, which are encoded in a simplified form of the excess energy,
\begin{equation}
\label{eq: epex2}
\epsilon_\text{ex} (W) \approx \frac{k}{2}  \delta^2(f, W)  + \frac{C}{2}(s-s_0)^2 .
\end{equation}
Here, $k$ is a generalized elastic parameter for straining inter-element packing, as measured by a generalized (mean) strain $\delta(f,W)$, which itself varies with the finite size $W$ of the domain and a parameter $f$ that measures the ``strength of frustration''. To be clear, local strains vary with position throughout the assembly, as we describe for a specific example below, but for simplicity we focus here on how the magnitude of strain varies with size and frustration.  While frustration mechanisms vary considerably among distinct GFA systems, they all exhibit power-law growth of strain with domain size $W$.  This can be modeled heuristically as 
\begin{equation}
    \delta \approx f(s) W^\eta
\end{equation}
where $\eta$ and $f$ vary for different cases of GFA. For example, the orientational strains in 2D liquid crystal domains grow linearly with domain size ($\eta=1$), whereas positional strains in frustrated 2D crystals grow quadratically ($\eta=2$) ~\cite{Grason2016, Niv2018}.  The definition of ``frustration strength'' $f$ depends on the specific GF mechanism; however, it can generally be expressed as a function $f(s)$ of the local shape of inter-subunit packing (e.g. inter-subunit bend or twist), which we denote generically with the shape parameter $s$.  In many cases, the frustration strength can be expressed as a simple power law of shape,
\begin{equation}
f(s) \approx s^\mu
\label{eq: fs}
\end{equation}
where $\mu$ is a positive exponent.  For the example of a crystalline cap on a spherical surface~\cite{Grason2016}, $f$ corresponds to the Gaussian curvature, which is the square of the 1D curvature in this geometry and thus corresponds to a shape parameter with $\mu=2$.  In the form of eq. (\ref{eq: fs}), the strength of frustration generically vanishes, intuitively, in the limit that the shape ``flattens" to $s \to 0$.  The second term in eq. (\ref{eq: epex2}) describes generic costs for deformations away from an ideal, misfitting shape with $s=s_0\neq 0$, which incur elastic penalties described by the ``shape modulus'' $C$.  

This basic form of eq. (\ref{eq: epex2}) implies a generic size dependence for the excess energy shown schematically in Fig.~\ref{fig: GFA}B. For small sizes, assemblies retain their preferred, misfitting shape ($s \simeq s_0$) leading to a power-law growth of excess energy, $\epsilon_\text{ex} (W \to 0) \simeq k f_0^2 W^{2\eta}/2$, where $f_0=f(s_0) = s_0^\mu$ is the frustration strength of the preferred shape.  If power-law growth of $\epsilon_\text{ex} (W)$ extended to all size scales, then the compromise between costs of GF and surface energy would select a finite equilibrium size $W_* \thicksim f_0^{-2/(2\eta-1)}$ for any surface energy $\Sigma$.  However, in any physical system, the excess energy can only accumulate up to some maximal size scale, beyond which the assemblies {\it escape frustration} through one of a number of competing morphological ``modes''~\cite{Hall2016, Hall2017}.  These include, for example, the formation of {\it topological defects} that screen far-field frustration stresses~\cite{Grason2012, Li2019}, as well as {\it shape flattening}, which refers to the smooth deformation of an incompatible (i.e. misfit) shape to a  uniformly strained,  compatible one~\cite{Grason2020}.  Because subunits and their interactions are generically soft, the excess energy required to escape frustration must be finite.  Thus, as shown in Fig. ~\ref{fig: GFA}B, $\epsilon_\text{ex} (W)$ generically crosses over from power-law accumulation at small $W$ to an asymptotic approach to this finite energy $\epsilon_\text{ex} (W \to \infty) = \epsilon_{\infty}$.  In the minimal description of eq. (\ref{eq: epex2}), the cost of shape flattening is simply $\epsilon_{\infty} =C s_0^2 /2$ per subunit.

\begin{figure*}
\centering
\includegraphics[width=0.95\linewidth]{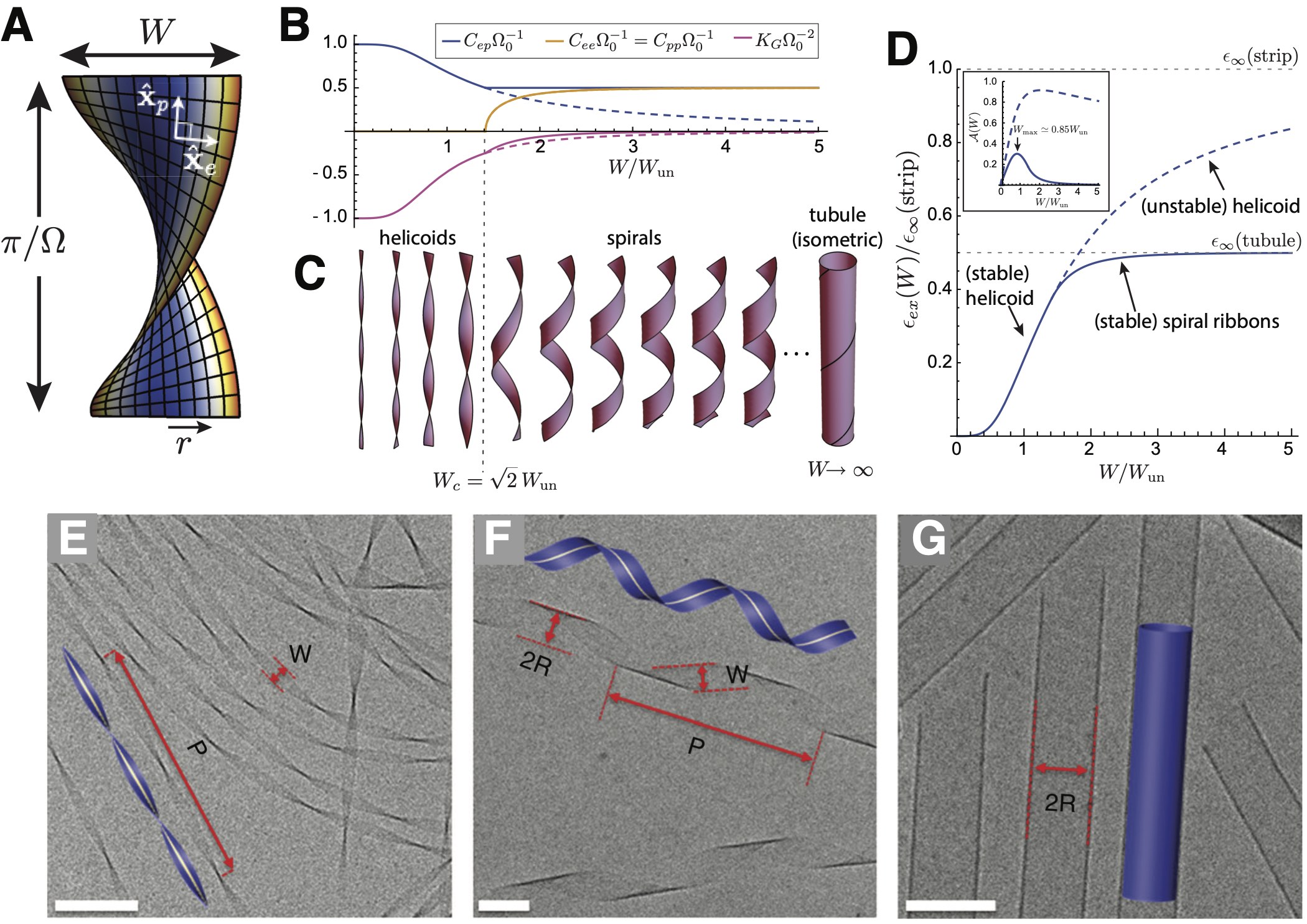}
\caption{\footnotesize (A) A schematic of a chiral, crystalline ribbon, where colors indicate the local extensional strains required by negative Gaussian curvature, from low (blue) to high (red).  (B) Plots of the equilibrium shape relaxation of the narrow ribbon model of \cite{Ghafouri2005} as a function of increasing ribbon width, for curvature components $C_{ij}$ and Gaussian curvature $\KG$, which is approximately uniform over the ribbon.  The solid branches show the minimal energy configurations, while the dashed line indicates the (unstable) helicoidal equilibrium.  Widths are scaled by the characteristic length $W_{\rm un}$, defined in Eq. (\ref{eq: unwinding}).  (C) shows schematics of the shape equilibrium, in particular the shape transition from helicoids to spiral ribbons at a critical width $\Wc$.  (D) and (E) plot the respective excess energies for helicoidal (dashed) and helicoid-spiral shape branches (solid).  While both shape modes predict an asymptotically relaxed frustration energy and a finite self-limiting width, spiral ribbons expel $\KG$ at much faster rate, leading to a narrower range of self-limitation. (E-G) show transmission electron microscopy images of ribbons assembled from chiral bola-amphiphiles adapted from ~\cite{Zhang2019}.  The panels illustrate morphologies evolving with assembly time (scale bars 100 nm): in (E), helicoids are observed after 24 hours; in (F) spiral ribbons are observed after one week; and finally, in (G), closed tubules are observed after 5 months. }
\label{fig: ribbons}
\end{figure*}

Comparing power-law accumulation at small sizes to the asymptotic shape-flattened energy, one expects a crossover between these regimes at a characteristic ``flattening'' size $W_{\rm flat} \approx \big[(C/k)/ s_0^{2(\mu-1)}\big]^{1/\eta}$.  This size scale defines the maximum size for which assemblies will tolerate the accumulating frustration cost, beyond which deformation to an unfrustrated shape ($s \to 0$) becomes energetically favorable. Intuitively, this length scale also sets a bound on the {\it escape size}, $W_\text{max} \leq W_{\rm flat}$, since for $W \gg W_{\rm flat}$ the bulk energy is simply renormalized by the cost of flattening.  In general, the escape size is set by the lowest-energy mode of relaxing frustration in the bulk state, which may be also involve ``Wigner lattice'' states of defects that neutralize the long-range cost of frustration~\cite{Li2019}.  Understanding the practical limit of self-limitation then requires considering all possible competing modes of relaxing frustration, and determining which of these has the lowest energy for a particular regime of assembly.  Notably, this distinction between power-law growth of energy density with $W$ vs. asymptotic approach to a constant value has been proposed, by Meiri and Efrati, to classify distinct regimes of GF itself, as   respectively ``cumulative'' and ``non-cumulative'' frustration~\cite{Meiri2021}.  The analysis of the accumulant in Sec.~\ref{sec: accumulant} translates that criteria into directly into their thermodynamics consequences for equlibrium SLA.

Notwithstanding which mode facilitates escape to the bulk, the heuristic picture of GFA implies a common phase diagram (Fig. ~\ref{fig: GFA}C), spanned by the bare frustration strength $f_0$ on one axis, and the ratio of cohesion to intra-assembly stiffness $\Sigma/k$ on the other (say for fixed subunit concentration and temperature).  Above a critical frustration strength, there is a regime of self-limiting aggregates between the dispersed state (below the CAC) and the bulk state. Within the self-limited regime, the equilibrium domain size increases with $\Sigma/k$ but decreases with frustration strength $f_0$.

As a specific example of the origin and implications of GFA, we briefly discuss a well-studied model for crystalline ribbons frustrated by chirality~\cite{Ghafouri2005, Selinger2004, Armon2014}.  This model has been developed to understand the polymorphic assembly of chiral surfactant bilayers that adopt a variety of quasi-1D structures~\cite{Oda1999, Selinger2001, Ziserman2011, Zhang2019}; i.e., they form 2D sheets that are much narrower in width than length, $W \ll L$.  Ghafouri and Bruinsma first described this mechanism of frustration~\cite{Ghafouri2005}, posing it as a competition of the bending cost of a chiral (anisotropic) membrane against the elastic costs of in-plane stretching for a 2D crystal with Gaussian curvature.  More recently, this model has been extended~\cite{Armon2014, Grossman2016} to describe experimental observations of bola-amphiphile ribbons by Sharon and coworkers~\cite{Zhang2019}. 

For the purposes of this discussion, we present here only a simplified picture of the frustration and its effects in chiral ribbons, and provide a more detailed summary in Appendix ~\ref{sec: ribbon}.  As shown in Fig.~\ref{fig: ribbons}A, we consider assemblies of ribbons of width $W$ and much longer (unlimited) length $L \gg W$.  For the case of sufficiently narrow ribbons, it can be shown that the favored morphology is that of a {\it helicoid}:  a strip whose width axis twists around a straight central ribbon at a uniform rate $\Omega$.  In this regime, the excess energy of the helicoidal chiral ribbon takes the following form (see Appendix ~\ref{sec: ribbon}),
\begin{equation}
    \epsilon_\text{ex}  (W)= \frac{Y a_0 }{1440} \Omega_0^4 W^4 + B a_0 (\Omega-\Omega_0)^2 ,
    \label{eq: helicoid}
\end{equation}
where $a_0$ is the area per subunit in the membrane, and $Y$ and $B$ are the respective in-plane stretching and out-of-plane bending moduli for the assembled membrane.  Here $\Omega$ plays the role of a shape parameter in the heuristic model introduced above, and the second term has the form of a {\it chiral bending energy} that favors uniform twist $\Omega_0$~\cite{Helfrich1988, Ghafouri2005}. The first term derives from the elastic cost of in-plane deformation of the 2D lattice of the ribbon, and describes the accumulating elastic cost of frustration.  It is intuitive to understand the growth of stretching energy with ribbon width by considering the contour length per helical turn of the longitudinal strip of the ribbon at a distance $r$ from the center, $(2 \pi/\Omega) \sqrt{1+(\Omega r)^2 }$.  Due to the resistance to shearing and stretching of local elements in the 2D solid, this motif leads to strains that grow with relative longitudinal extension compared to the midline, roughly $\approx (\Omega r)^2$ to lowest order, and hence, it generates an elastic energy density of $\sim Y (\Omega^2 W^2)^2$.  It can be shown that the underlying source for stress gradients in such an assembly derives from the incompatibility of a 2D planar metric with the non-zero (negative) Gaussian curvature of the membrane, $\KG \simeq -\Omega^2$~\cite{Ghafouri2005, Armon2014}.  In the language of the heuristic picture introduced above, we can identify the strength of frustration with $\KG$, and hence $\eta=\mu=2$, due to the quartic growth of frustration cost with shape misfit $\Omega$ and width $W$.

An instructive, albeit oversimplified, analysis of the size-dependence of frustration in chiral ribbons considers only the relaxation of helicoidal twist for variable width.   For small widths, the ribbon adopts a preferred twist, $\Omega_*( W \to 0) = \Omega_0$,  due to the vanishing cost of frustration as $W \to 0$, and $\epsilon_\text{ex} \approx Y \Omega_0^4 W^4$.  For large widths $W \to \infty$, the balance between stretching and bending favors unwinding of the pitch, $\Omega_* \simeq  \Omega_0 (W_{\rm un}/ W)^{4/3}$, where 
\begin{equation}
    W_{\rm un} \equiv \Big( \frac{720 B}{Y \Omega_0^2 }\Big)^{1/4} ,
\label{eq: unwinding}
    \end{equation}
defines a characteristic {\it unwinding size}.  That is, for $W \ll W_{\rm un}$ the ribbon retains roughly the twist preferred by chirality and the power-law ($\sim W^4$) accumulation in excess energy, while for much larger ribbons, the prohibitive cost of frustration causes the ribbon to unwind, expelling Gaussian curvature.  Fig. ~\ref{fig: ribbons}D (dashed curve) shows the characteristic crossover in the excess energy of helicoidal ribbons, with an asymptotic flattening of the excess energy characterized by the exponent $\nu =4/3$, which implies self-limiting widths are not stable in the unwinding region (see section~\ref{sec: accumulant}).  In terms of the accumulant analysis (inset Fig. ~\ref{fig: ribbons}D), the escape of frustration by helicoid unwinding would imply a maximum self-limiting size of $W_{\rm max} ({\rm helicoid}) \simeq 2.1 W_{\rm un}$.

How do we understand physical parameters that determine the range of possible self-limiting widths?  Here, we note that it arises from the combination of two physical lengths.  One of these derives from the ratio of bending to stretching moduli, $\sqrt{B/Y} \equiv t $, which is typically of order of the thickness an elastic membrane.  In other words, in can be expected that $t$ is of order of the molecular ($\approx {\rm nm}$) size of the constituent amphiphiles.  The second length scale is the preferred pitch $P_0= 2 \pi/|\Omega_0|$, which derives from the chiral preference for local skew packing of neighbor amphiphiles \cite{Zhang2019}.  Unlike $t$, the size range of $P_0$ is mesoscopic, on the order of 100s of nm.  With these definitions, we see that the unwinding size scale, and hence the maximum self-limiting width, are the geometric mean of these two length scales $W_{\rm un} \simeq 11 \sqrt{ t P_0}$, one molecular and one mesoscopic.  It can then be understood that frustration can limit the size of helicoids up to a size range intermediate to these molecular and mesoscopic sizes, that is, ribbons of order of 10s of nm in width.

As we describe in more detail in Appendix~\ref{sec: ribbon}, frustration escape in chiral ribbons is more complex than suggested by considering only helicoidal shapes.  As chiral ribbons grow beyond a critical width $\Wc = \sqrt{2} W_{\rm un}$ (Fig.~\ref{fig: ribbons}B-C), they become mechanically unstable to a new class of shape equilibria, {\it spiral ribbons}~\cite{Ghafouri2005, Armon2014}.  With increasing width, this class of shapes approaches an isometric strip that is wound helically around a cylinder.  Hence, unlike the helicoid, this more complex shape relaxation allows the assembly to retain some residual chiral twist while at the same time expelling the Gaussian curvature that includes in-plane stresses.  Therefore, these spiral shapes facilitate  a more ``efficient'' escape of frustration than helicoidal unwinding (see Fig.~\ref{fig: ribbons}D).

Notwithstanding the quantitative frustration relaxation for large widths, it can be shown that the mechanical instability does not change the basic conclusion, that the scale of the self-limiting width is set by the geometric mean of the $t$ and $P_0$.  That is, according to the accumulant analysis of the spiral ribbon branch (inset of Fig.~\ref{fig: ribbons}D), $W_\mathrm{max} \simeq 0.85 W_{\rm un}$, and the instability only reduces the numerical prefactor of the unwinding scale. This is consistent with a recent experimental study of ribbon morphologies of bola-amphiphiles~\cite{Zhang2019} (Figs.~\ref{fig: ribbons}E-G).  In this system, the molecular size suggests $t \approx 3-4 \ {\rm nm}$, while skewed packing of chiral neighbors in the crystal leads to much larger pitches of $200 \ {\rm nm}$.  The observation of helicoidal ribbons up to $\approx 50 \ {\rm nm}$ (consistent with the geometric mean of molecular size and helical pitch) demonstrates that frustration stress can propagate far beyond molecular dimensions.  Notably, when ribbons grow beyond this size, they do not grow to infinitely wide and untwisted sheets, but instead, transition to a second mechanism of self-limitation, forming closed tubules of finite diameter proportional to the pitch.

The preceding discussion has neglected relaxation of frustration by defects.  For 2D solid chiral ribbons this can be justified because the critical width for defect formation in helicoids~\footnote{This can be conjectured to be $\propto |\Omega_0|^{-1}$ based on standard arguments of Gaussian curvature ``screening'', see e.g. \cite{Bowick2009}.} far exceeds the transition to a shape-flattening state, whose lower energy is facilitated by the soft bending modes of thin solids.  This conclusion does not hold in general, even when the same morphologies arise, if the underlying  mechanism of frustration differs.  For example, chiral ribbons with only liquid-crystalline (LC) in-plane order (e.g. hexatic) would be described by the same shape (bending) elasticity, but with a frustration cost that arises from {\it orientational strains} from Gaussian curvature~\cite{Vitelli2004, Mbanga2012}.  Such angular strains grow with a weaker power-law $\sim \KG W$ than positional strain~\cite{Niv2018}, and hence, are an example of $\eta=1$ strain growth.  This ``softer'' growth of frustration energetics implies a shape flattening transition that takes place at a much larger size, proportional to the mesoscopic pitch $|\Omega_0|^{-1}$.  This is a size scale at which disclinations may also be expected to lower the ribbon energy, implying that frustration escape for LC ribbons likely falls into a different class than solid ribbons, one that may mix both smooth (shape flattening) and singular (defect-mediated) modes.

\section{Kinetic pathways towards self-limiting equilibrium}

\label{sec: kinetics}

As the primary focus of this review is the equilibrium ingredients and thermodynamics of SLA, the foregoing sections have not considered the non-equilibrium pathways by which such systems, starting from an out-of-equilibrium initial condition, may arrive at a self-limited (equilibrium) distribution.  However,  the kinetics of assembly can have significant influence over the size distributions and resulting morphologies formed in experimental systems of SLA systems, since they are necessarily limited to observations at finite times. Hence, interpretation of practically all experiments must allow for the possibility of non-equilibrium effects. In this section, we provide a basic introduction to a few of the key principles for understanding and modeling kinetics of SLA.  Rather than an exhaustive review, our purpose is largely to illustrate how the language and formalism introduced in the foregoing discussions of equilibrium SLA ``translates" to the basic conceptual and theoretical frameworks used for analyzing these same systems out of equilibrium.  A reader will find a far more thorough and general description of the interplay between kinetics and assembly products, in both bulk and SLA processes, in several reviews  \cite{Sear2007,Agarwal2014,Whitelam2015,Hagan2014}.

Theory, computation, and experiments have shown that the ability of a self-assembling system to approach equilibrium within practical timescales depends on a delicate balance between thermodynamic and kinetic effects~\cite{Zlotnick1999,Endres2002,Ceres2002a,Zlotnick2003,Whitesides2002,Ceres2002,Hagan2006,Jack2007,Rapaport2008,Whitelam2009,Nguyen2007,Wilber2007,Wilber2009,Klotsa2011,Grant2011,Grant2012,Hagan2011,Cheng2012,Hagan2014,Whitelam2015}. In a broad variety of systems, yields are nonmonotonic with the strength of cohesive interactions that drive assembly. Optimal assembly occurs when the cohesive free energy is on the order of $5-10\kt$ per subunit-subunit contact depending on the initial subunit concentration and valency of subunit contacts. To understand this behavior, we require a consideration of the kinetics of assembly.

To begin, we consider a typical bulk (in vitro) assembly kinetics experiment, such as, for example, solution assembly of reconstituted proteins into multi-unit structures.  To prepare an initial state, the system is equilibrated under conditions that do not lead to assembly; i.e., the total subunit concentration is below the CAC ($\Phi<\phiStar$) and thus the equilibrium aggregate distribution is peaked at monomers (see Fig.~\ref{fig: SLA_generic}). In practice, this is accomplished by setting a low subunit concentration, a solution $p$H and salt concentration that ensure very weak subunit-subunit attractions, or sufficiently high temperature that the translational entropy of unassembled subunits dominates over inter-subunit attractions.  The system is then rapidly quenched into a condition that favors assembly ($\Phi > \phiStar$), either by increasing the subunit concentration $\Phi$, or decreasing the CAC above the current concentration, by changing temperature or physico-chemical changes that increase subunit-subunit interactions (e.g. changes to $p$H and salt concentration). That is, the new equilibrium state (post-quench) corresponds to a population of aggregates coexisting with monomers as described in Sec. \ref{sec: thermo}. 
However,  the pre-quench distribution is out-of-equilibrium, consisting predominantly of unassembled monomers, and as such gradients in the system free energy will drive assembly toward lower free energy states with assembled clusters.   Note that this scenario  can equally well describe experiments of both unlimited (e.g. bulk crystals) and self-limited (e.g. capsules) assembly.

The study of assembly kinetics is concerned with the timescale required to approach the equilibrium state, as well as the structures and lifetimes of long-lived metastable states which may occur along the way. To understand the influence of kinetics on practical applications or biological function, these timescales must be compared to those that are accessible to an experimenter or a biological organism.

\subsection{Classical nucleation theory and assembly timescales}
A useful starting point to understand the dependence of assembly timescales on control parameters is given by the framework of classical nucleation theory; see e.g. \cite{Becker1935,Binder1976,Oxtoby1992,Zandi2006,Hagan2010,Hagan2014,Agarwal2014,Whitelam2015,DeYoreo2003} and  \cite{Oxtoby1992} for a review. In this approach, the assembly of a cluster is broken into two phases: \textit{nucleation} and \textit{growth} (often called `elongation' in the context of finite structures). Nucleation refers to the process of overcoming a free energy barrier to form a small but relatively stable aggregate, while the second phase describes growth of such an aggregate to its final (optimal) size. We will see below that for most assembly reactions to be productive (i.e. observable on experimentally realistic timescales),  nucleation must be the rate limiting process, and thus the assembly timescale can be estimated by calculating the nucleation rate.

 \subsubsection{Nucleation kinetics}
Consider a general form of aggregation free energy for aggregates that describes assembly driven by short-range cohesive interactions, with possible additional terms to describe higher-order effects such as those that give rise to SLA:
\begin{align}
\epsilon(n) = -\eMin + \frac{\Delta_0} {n^{1/d}} + \epsilon_{\rm ex}(n)
\label{eq:epsilonCNT}
\end{align}
where $\eMin$ is the bulk energy per subunit in the aggregate with optimal size, $\Delta_0$ accounts for surface energy (as in Eq.~\eqref{eq:epsilonUnlimited}) and following Sec.~\ref{sec: accumulant} we define $\epsilon_{\rm ex}(n)$ as the excess energy relative to the bulk and surface effects, including effects that are super-extensive in $n$~\footnote{Here, the notion of excess energy is shifted by an unimportant constant relative to eq. (\ref{eq: excess}).}.  Note that this description may apply equally to either self-limiting (i.e. with a minimum at a finite $n= n_{\rm T}$ and $\epsilon_{\rm T} = -\eMin$ from Sec.~\ref{sec:self-limiting} ) or unlimited (i.e. where $ \epsilon(n) $ is minimal for $n \to \infty$ and $\epsilon_{\rm T} = -\eMin$ per Sec.~\ref{sec: unlimited} ) assembly.  We begin the discussion by considering nucleation for the simplest case $\epsilon_{\rm ex}(n) =0$, i.e. unlimited assembly. We will see shortly that this analysis also qualitatively applies to most SLA examples, since nucleation occurs at small sizes, where the size dependence of the excess energy, $\epsilon_{\rm ex}(n)$\, is much smaller than that of the surface terms, whose large values at small sizes constitute the generic origin of the nucleation barrier.

As noted in section~\ref{sec: unlimited}, small aggregates generically have a smaller cohesive energy than large aggregates, because the fraction of subunits at the aggregate surface with unsatisfied interactions (accounted for by the second term $\Delta_0$ in Eq.~\ref{eq:epsilonUnlimited}) decreases with aggregate size.  In contrast, the bulk energy ($\eMin$) dominates over surface terms for large aggregates. We define the `critical nucleus size' $\nCrit$ as the crossover between these two regimes. Below $\nCrit$ disassembly is favored over assembly because the bulk cohesive energy driving assembly is outcompeted by this unfavorable surface energy and the greater translational entropy of unassembled monomers. Above the critical nucleus size, on the other hand, the bulk cohesion dominates and assembly is favored\footnote{Specifically, critical nucleus denotes a structure for which either complete disassembly or growth to a large aggregate are equally probable. In general, for a particular system there will be an ensemble of critical nuclei that have different structures and (if size is not a complete reaction coordinate) different sizes,see e.g. \cite{Pan2004}}.

Because forward assembly of pre-nuclei aggregates is unfavorable, growth of an aggregate to the critical nucleus size is improbable and nucleation is a rare event. In particular, we will see that productive assembly requires that nucleation is a rare event on the timescale of typical subunit-subunit association. This gives rise to a separation of timescales --- pre-nucleated aggregates rapidly reach a quasi-equilibrium on timescales much shorter than the overall timescale required for the assembly process to approach equilibrium. Thus, based on this assumption, the aggregation distribution for sizes below the critical nucleus size $\nCrit$ can be modeled by a variant of the law of mass action
\begin{align}
\phi_n = n\Big( \phi_1 e^{- \beta \epsilon(n) } \Big)^n = n e^{-\beta  \Omega(n)} \qquad \mbox{for } n<\nCrit
\label{eq:CNT}
\end{align}
where $\Omega(n)=n\big[\epsilon(n)-\mu\big]$ with $\mu=-\kt \ln \phi_1$ is the size-dependent {\it grand free energy} of $n$-mers (sometimes referred to as the `excess' free energy) that accounts for the inter-aggregate interactions as well as the entropy cost incurred by subunits joining an aggregate.  It is important to point out that this nonequilibrium description is a departure from from the thermodynamic one introduced in Sec.~\ref{sec: thermo}, in which $\mu$ and $\phi_1$ are thermodynamically defined by the total concentration and temperature.  Here, the chemical potential is used to define only the partial equilibrium of pre-nuclei with free monomer.  Hence, in this usage, $\phi_1$ and hence $\mu$ and $\Omega$ should be understood as time-dependent quantities according to the depletion of free monomers as assembly proceeds.  However, the quasi-equilibrium approximation assumes that these quantities vary slowly in comparison to the timescale required for the pre-nuclei aggregate distribution to reach this form.

By substituting  Eq.~\eqref{eq:epsilonUnlimited} for the aggregate energy into Eq.~\eqref{eq:CNT}, we see that (for $d>1$) there will be a maximum in the grand free energy $\Omega(n)$ at a size $\nCrit=\left(\frac{d-1}{d}\frac{\Delta_0}{-\eMin-\mu}\right)^d$, owing to the competing drives of (negative) bulk assembly and (positive) surface growth.
This corresponds to the critical nucleus size, since aggregate growth beyond $\nCrit$ will decrease the free energy. Here, the factor $\frac{\Delta_0}{-\eMin-\mu}$ gives a ratio of the unfavorable surface energy $\Delta_0$ that impedes assembly to the net thermodynamic driving force for assembly $-\eMin-\mu$.   This provides a natural measure for how far out of equilibrium are initial conditions 
\begin{align}
 \DeltaMu \equiv -\eMin-\mu = \kt \ln \Big( \frac{\Phi}{\phiSat} \Big) \geq 0
\label{eq:supersaturation}
\end{align}
where we have taken $\phi_1 \simeq \Phi$ since we are considering the initial conditions where nearly all subunits are free. We have defined $\phiSat\equiv e^{\beta\eMin}$ so the that ratio $\Phi/\phiSat$ approximately measures how far the total subunit concentration exceeds the CAC and is often referred to as the {\it supersaturation}.

The key argument of classical nucleation theory is that because aggregates of the critical nucleus size are rarefied, the nucleation timescale grows exponentially with the barrier height
\begin{align}
\tNuc \thicksim e^{-\beta\Omega(\nCrit)}.
\label{eq:CNTBarrier}
\end{align}
There are several important points to make here. First, this is the \textit{initial} nucleation timescale at the inception of the assembly; the nucleation timescale increases as assembly proceeds because free subunits are depleted and thus $\mu$ decreases, in turn, increasing $\Omega(\nCrit)$ . Second, Eq.~\eqref{eq:CNT} only applies up to $\nCrit$; above the critical nucleus size the decreasing free energy implies that growth is relatively rapid and thus post-nuclei aggregates do not reach a quasi-equilibrium. Instead, there is a predominant flux of monomers (via association to intermediates) from the population of pre-nuclei toward larger aggregates, either toward system-sized aggregates in the case of unlimited assembly or toward a population of target-sized structures in SLA. Third, if the barrier height becomes too small (e.g. $\Omega(\nCrit) \lesssim10\kt$ depending on the relevant timescales), the separation of timescales that enabled the quasi-equilibrium approximation breaks down, and Eq.~\eqref{eq:CNTBarrier} will under-predict nucleation timescales since free subunits are rapidly depleted. Fourth, for SLA, additional corrections to this classical nucleation picture will arise (such as an apparent size-dependent surface tension) if the critical nucleus size approaches the finite system size due to the higher-order effects captured by the excess energy~ \cite{Mayer1965,Thompson1984,Alder1962,Reguera2003}. Fifth, this analysis has assumed that aggregate size $n$ is a good `reaction coordinate', meaning that it accounts for all relevant slow degrees of freedom and thus the dynamics and probability of an aggregate successfully nucleating can be determined as a function of $n$. In practice, a complete reaction coordinate must include other aggregate characteristics such as its surface area \cite{Pan2004}. Sixth, a variety of other extensions to classical nucleation theory have been investigated, but it remains highly challenging to quantitatively predict nucleation rates~ \cite{Prestipino2012,Auer2001,Loeffler2013,Statt2015,Mcgraw1997,Peters2009,Knott2012,Joswiak2013,Zimmermann2015,Jacobson2010}.

While the discussion so far has considered an aggregate energy form that drives unlimited assembly, in general the results are qualitatively similar for SLA with large target structures, since the interaction terms that eventually limit assembly grow superextensively and thus are small for small aggregates. As an example, here and throughout the remainder of this section, we consider the fluid capsid model of section~\ref{sec: capsules} for which self-closing leads to a finite assembly size $\nT$. To simplify the presentation, we assume the limit of high bending modulus $B \rightarrow \infty$, so the curvature radius of the assembling shell is fixed to $\RT = \sqrt{a_0\nT/4\pi}$, and the free energy becomes
\begin{align}
    \epsilon(n) =-\eMin+ \lamTilde \sqrt{\frac{\nT-n}{n}}
    \label{eq: epsilonCapsid}
\end{align}
with the effective line tension associated with the boundaries of incomplete shells (which gives rise to the nucleation barrier) given by $\lamTilde = \sqrt{4\pi a_0 /\nT} \lambda$, with $\lambda$ the bare line energy. Notice that in the limit of small $n$, Eq.~\eqref{eq: epsilonCapsid} reduces to a bulk cohesive energy $\eMin$ and a surface term $\propto \sqrt{n}$ as anticipated above.

This specific ($B \rightarrow \infty$) model has been considered in the context of viral capsid assembly in Refs~\cite{Zandi2006,Hagan2010}. The grand free energy that corresponds to Eq.~\eqref{eq: epsilonCapsid} is given by 
\begin{align}
    \Omega(n) =\DeltaMu n + \lamTilde \sqrt{n\left(\nT-n\right)}
    \label{eq: omegaCapsid}.
\end{align}
which corresponds to a barrier height~\cite{Zandi2006} of

\begin{align}
\Omega(\nCrit) = \frac{\nT \lamTilde}{2}\left(\sqrt{\Gamma^2+1} - \Gamma\right)
\label{eq:ZandiCNTBarrier}
\end{align}
where $\Gamma = \DeltaMu / \lamTilde$ defines a measure of the dimensionless `quench depth'; i.e., the driving force for assembly in the initial state compared to line energy that impedes assembly~\cite{Zandi2006}.

\begin{figure}
\begin{center}
\epsfxsize=0.47\textwidth\epsfbox{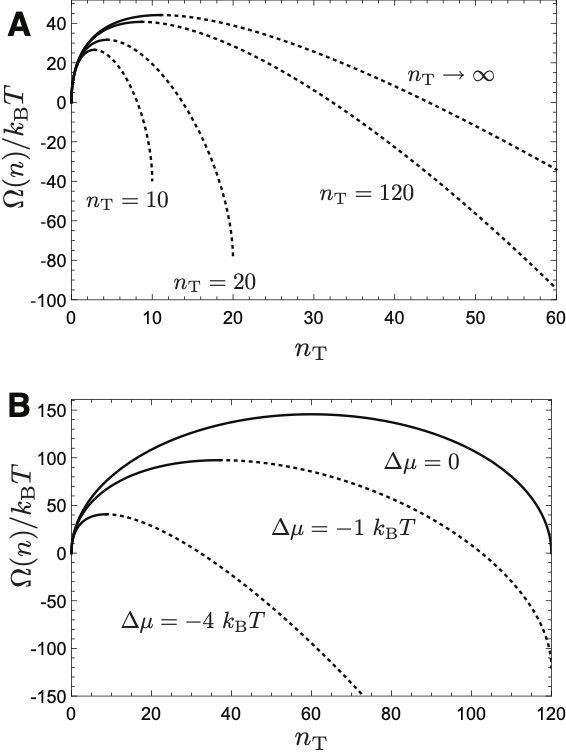}
\caption{  The grand free energy $\Omega$ for the capsid model with fixed curvature $\RT$ (Eq.~\eqref{eq:ZandiCNTBarrier}), as a function of partial capsid size $n$.  {\bf (A)} shows $\Omega(n)$ for indicated complete capsid sizes $\nT$, with the limit $\nT\rightarrow\infty$ corresponding to unlimited assembly (a flat disk). The calculation is performed for a chemical potential difference $\DeltaMu=-\eT-\mu = -4\kt$, and a per-subunit binding free energy $\eT=-15\kt$, corresponding to a subunit-subunit contact energy of $7.5 \kt$ and tetravalent subunits (e.g. \cite{Zlotnick2000,Ceres2002}). We set the line energy to $\lambda = \eT / 2 a_0^{1/2}$, corresponding to one unsatisfied contact per subunit on the partial shell rim.
{\bf (B)} shows $\Omega(n)$  as a function of the chemical potential difference $\DeltaMu$ for $\eT=-15\kt$ and complete capsid size $\nT=120$. In (A) and (B), the solid lines correspond to sizes for which the quasi-equilibrium approximation described in the text applies (i.e., $n\le\nCrit$) and thus the concentration of intermediates is approximately given by $\phi(n)\propto e^{-\beta \Omega(n)}$. The dashed lines correspond to sizes $n>\nCrit$ for  which this assumption is not valid and the intermediate concentrations cannot be described by a quasi-equilibrium (except approximately for the case $\DeltaMu=0$). }
\label{fig:grandFreeEnergy}
\end{center}
\end{figure}

Fig.~\ref{fig:grandFreeEnergy}a shows the grand free energy as a function of aggregate size for several values of the target shell size $\nT$, as well as for unlimited assembly into a flat disk, $\nT\rightarrow \infty$. Here we have plotted the portions of the free energy profiles below(above) the critical nucleus size as solid(dashed) lines to emphasize that the quasi-equilibrium assumption only applies below the critical nucleus. That is, $\phi(n)\propto e^{-\beta{\Omega(n)}}$ for $n\le\nCrit$, but for $n>\nCrit$ there is a constant flux of subunits toward complete shells and the nonequilibrium distribution of intermediate concentrations does not follow the law of mass action. 

Notice that the free energy profiles for the self-limited cases are qualitatively similar to that of the unlimited case at small aggregate sizes, consistent with the analysis above. However, the shell geometry that gives rise to SLA does have quantitative effects --- the shell curvature causes the length of the unfavorable free boundary to decrease relative to a flat disk (i.e. of equal area). Thus, the critical nucleus size and barrier height increase with target size, asymptotically approaching the flat disk limit. Note that, while in this case the self-closing physics giving rise to SLA decreases the nucleation barrier relative to the unlimited case, SLA effects can in general shift it in either direction. For example, the strain energy that gives rise to GFA in section~\ref{sec:OBA} will increase the barrier beyond the surface costs.  We reiterate though that the effects of SLA on nucleation are typically quantitative rather than qualitative.

Fig.~\ref{fig:grandFreeEnergy}b shows the grand free energy as a function of aggregate size for several values of the supersaturation, at fixed target size and line tension. We see that as the supersaturation decreases, the critical nucleus size and corresponding barrier height increase, since the translational entropy of free monomers increases and thus the net assembly driving force decreases. Note that the different curves corresponding to different supersaturation levels in Fig.~\ref{fig:grandFreeEnergy}a can be viewed from two perspectives. On one hand, each supersaturation level can correspond to the initial condition of a separate experiment with a different total subunit concentration $\Phi$, with each curve corresponding to the aggregate free energy at the beginning of the experiment.  In this interpretation, Eq.~\ref{eq:CNTBarrier} describes the \textit{initial} nucleation rate and its dependence on quench conditions. On the other hand, as a single experiment proceeds, the supersaturation is continually decreasing as free monomers are depleted by assembly, and curves at decreasing supersaturation levels reveal the instantaneous nucleation rate as the experiment proceeds (and subunits are effectively removed from the pre-nuclei pool). Notice, then, that the barrier height increases as the reaction proceeds, and thus the nucleation rate decreases over time. For large target sizes $\nT\gg1$, as the reaction proceeds the nucleation barrier eventually becomes so large in comparison to the thermal energy that assembly ceases on relevant timescales; thus, the reaction only asymptotically approaches equilibrium. 

Finally, the curve corresponding to no supersaturation, $\Delta \mu=0$, corresponds to the equilibrium state with coexistence of shells and free monomers (hence the entire curve is plotted with a solid line). In this case the critical nucleus size is given by $\nCrit = \nT/2$, corresponding to a half shell. Notice that at equilibrium, intermediates are higher in free energy than free monomers or complete shells due to the unfavorable line energy at the boundaries and are thus present only at low concentrations (see also section~\ref{sec:maxClusterSize}).  

\subsubsection{Growth} 
We define growth as the process by which a critical nucleus assembles to its final state, commensurate with the optimal aggregation size. In contrast to unlimited assembly (e.g. bulk crystal growth), for SLA there is a well-defined mean timescale for growth of the aggregate, since about $\nT-\nCrit\approx\nT$ subunits must associate to reach the optimal size\footnote{However, even for a system with no minimum in its aggregation free energy, one can define a growth timescale to reach a given finite-size, which can be analyzed as described here.}. 

The physical process of growth differs fundamentally from nucleation. Nucleation is a highly cooperative process --- from Eq.~\eqref{eq:ZandiCNTBarrier}, we see that the nucleation timescale decreases exponentially with subunit interaction strength $\eMin$ and decreases with subunit concentration according to $\phi_1^{\nCrit}$. The latter condition reflects the fact that $\nCrit$ subunits must come together within a short timescale in order to create a stable nucleus. In contrast, because post-nucleus intermediates are relatively stable, the growth phase can proceed through independent additions of individual subunits or small oligomers. Thus, it can generally be expected that growth timescale $\tElong$ depends only weakly on $\eMin$.  Moreover, it can be expected that $\tElong$ varies inversely with subunit concentration $\tElong \propto \phi_1^{-1}$, since the rate of addition should increase in proportion to concentration unless growth requires overcoming any secondary nucleation barriers.  This latter prediction has been tested and confirmed experimentally for Hepatitis B capsid assembly by Selzer et al \cite{Selzer2014}.

In general, the mean growth timescale can be estimated as
\begin{align}
\tElong = \frac{\nT^\alpha}{ \phi_1 \fAssem}
\label{eq:tElong}
\end{align}
where we have assumed $\nT\gg\nCrit$ so that $\nT-\nCrit \approx \nT$. The quantity $\fAssem$ is the association rate constant for subunit addition, averaged over the growth phase since it may vary with aggregate size. The factor in the numerator indicates that the growth timescale generically {\it increases with optimal aggregate size} (i.e. $\alpha>0$) since $\mathcal{O}(\nT)$ independent subunit additions must occur. The value of the exponent $\alpha$ will depend on factors such as the dimensionality, the aggregate geometry, and the relative stability of intermediates. For example, for growth of a globular aggregate in 3D (i.e. a crystal or the SALR system of section~\ref{sec:SALR}), we expect $\alpha\approx1/3$ if growth is proportional to the diffusion limited rate.  For the capsid example that we consider in this section, we expect $1/2 \le \alpha \le 2$. In particular, if assembly is strongly biased over disassembly during growth and the assembly rate is proportional to the perimeter of the free boundary, we obtain $\alpha=1/2$ \cite{Hagan2010}. For moderately biased assembly, growth tends to occur along a single point on the perimeter, giving $\alpha=1$, while for weakly biased assembly (near the reversible limit), the growth timescale approaches that of a random walk, giving $\alpha=2$.

\subsubsection{Beyond nucleation and growth} 
It is important to note that not all assembly processes can be adequately described by the nucleation and growth mechanism. In some systems there are additional timescales which may become rate-limiting. These include subunit conformational changes, or cooperative global rearrangements required to achieve the optimal self-limited structure. For example, strains must propagate across scales on the order of the size of the structure in the frustrated open-boundary assemblies described in section~\ref{sec:OBA}. Similar behaviors may occur in self-closing aggregates; for example recent evidence suggests that assembly of empty HBV capsids proceeds by nucleation, growth into large but defective or disordered intermediates, followed by a `completion phase' in which the intermediate rearranges into the icosahedral capsid structure \cite{Chevreuil2020}. Additional multi-step mechanisms are possible if we move beyond the scope of this review (SLA from a single species). For example, computation and experiments have shown that assembly around a substrate or template can proceed by an alternative `en masse' pathway \cite{Garmann2014,Garmann2016,Panahandeh2020,Tsvetkova2012,Hagan2008,Elrad2010,Perlmutter2014}. In this process, subunits rapidly adsorb on a substrate in a disordered manner, and then cooperatively rearrange to form an ordered aggregate. While that process involves condensation and assembly of individual substrates, it is also possible for the assembling components to undergo bulk phase separation into a metastable liquid phase prior to assembly. For example in many virus families the host cell undergoes liquid-liquid phase separation to form a domain that is concentrated in viral proteins and nucleic acids, within which the nucleocapsid (capsid assembled around the viral RNA) assembles, e.g. \cite{Brocca2020,Carlson2020,FernandezdeCastro2020,Guseva2020,Kieser2020,Nikolic2017,Savastano2020,Schoelz2017}. A related mechanism occurs for unlimited assembly ---  it has been shown that crystallization can proceed by a two-step mechanism in which subunits first condense into a metastable liquid phase, followed by formation of an ordered crystal, see e.g. \cite{Nicolis2003,Sear2009,Whitelam2010,Wolde1997,Fortini2008,Basios2008,Gliko2005,Schubert2017,Veesler2006}.

\subsection{Interplay between thermodynamic stability, assembly rates and kinetic traps}
Achieving productive assembly requires thermodynamic stability of the target structure, which implies that subunit interactions must be strong enough to overcome the translational and rotational entropy losses incurred by the subunits forming an aggregate. Achieving productive assembly in finite time places even more restrictive conditions on interactions, based on the kinetics of assembly.  Interactions must be strong enough to ensure that the nucleation timescale discussed above falls within relevant timescales. However, overly strong subunit interactions lead to \emph{kinetic traps}, or metastable states which evolve toward equilibrium very slowly.  These kinetic traps can be broadly classified into two categories.  We discuss each category and its effects in turn, followed by a brief survey of open questions for optimizing assembly kinetics.

\subsubsection{Overnucleation (i.e. monomer starvation)} 
First, the {\it monomer starvation} trap arises when nucleation timescales are short in comparison to \emph{growth} timescales, or the time required for a nucleated aggregate to grow to its equilibrium size. In this situation, so many nuclei form that the system becomes depleted of monomers before most nuclei grow to completion. Subsequent evolution to equilibrium requires either redistribution of subunits from smaller to larger aggregates (Ostwald ripening), which incurs significant free energy barriers, or coalescence of large intermediates, which is rare and frequently leads to mis-assembled structures. 

In the context of SLA, this condition becomes more stringent as the target assembly size increases, since about $\nT$ monomers will eventually be depleted during the growth of each nucleus, and the growth timescale typically increases with target size since the critical nucleus size depends at most weakly on $\nT$. The parameter regimes that give rise to the monomer starvation trap can be understood from the different dependence of nucleation and growth timescales on subunit-subunit interaction strengths and subunit concentrations \cite{Hagan2010,Zlotnick1999,Endres2002}. In particular, from Eq.~\eqref{eq:CNTBarrier}  the \textit{initial} nucleation timescale for the capsid model is $\tNuc(\Phi) \thicksim  \exp\left[-\beta\Omega(\nCrit(\Phi))\right]$
%
%
 However, as noted above the nucleation timescale increases as the reaction proceeds due to monomer depletion. By evolving aggregates according to the kinetics described by Eq.~\eqref{eq:tNucZandiApp}, we can integrate the cumulative depletion of monomers as a function of time, from which one can obtain median assembly time $\tHalf$ (defined as the time required for half of the subunits to be assembled, $\phi_1(\tHalf)=\Phi/2$) \cite{Hagan2010} (see Appendix~\ref{sec:KineticsAppendix} for details):
 \begin{align}
 \tHalf \thicksim  \frac{\tNuc(\Phi)}{\nT} .
 \label{eq:tHalfZandi}
 \end{align}
\begin{figure}
\begin{center}
\epsfxsize=0.47\textwidth\epsfbox{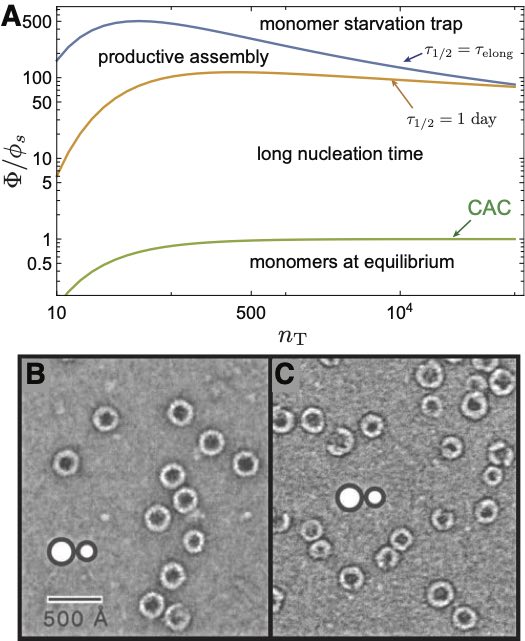}
\caption{  {\bf (A)} Assembly `phase diagram' for the capsid model ($B \to \infty$ model of the fluid capsule model). The boundaries between the different kinetics regimes discussed in the text are shown as a function of capsid size and supersaturation $\Phi/\phiSat = e^{-\beta \DeltaMu}$. The calculation was performed using Eqs.~\ref{eq: epsilonCapsid}, \ref{eq: lma}, \ref{eq:tElong} with $\alpha=1$, and \ref{eq:tHalfZandiApp}. We set $\eMin=-15\kt$ and $\lambda = \eT / 2 a_0^{1/2}$ as in  Fig.~\ref{fig:grandFreeEnergy}.
 {\bf (B), (C)} TEM images of in vitro assembly of empty capsids from CCMV capsid proteins. (B) corresponds to productive assembly, while (C) corresponds to assembly of long-lived partial shells (the monomer starvation trap) that occurs under stronger subunit-subunit interactions. Images in (B) and (C) reproduced from \cite{Zlotnick2000}.
}
\label{fig:KineticTrap}
\end{center}
\end{figure}

The boundary between productive assembly (which we define as having initial nucleation times shorter than, say, 1 day) and monomer starvation can be estimated by the locus in parameter space at which the median assembly time and growth time are equal. Fig.~\ref{fig:KineticTrap} shows this and boundaries between other kinetic regimes as a function of supersaturation and $\nT$ for the capsid model described above. For any optimal size, we see that as the total subunit concentration increases, the system transitions from a monomer-dominated equilibrium phase, to a state in which the monomer rich phase is only metastable with respect to aggregates, but assembly does not occur on relevant timescales due to a large nucleation barrier, to a window of productive assembly, and finally to the monomer starvation regime. 
Notice that the concentration at which assembly becomes kinetically accessible can significantly exceed the CAC, showing that accurately inferring the CAC from experimental measurements can be challenging. Moreover, the region of productive assembly between the nucleation threshold and the monomer starvation trap narrows with increasing target size, due to the different dependencies of the nucleation and elongation timescales on $\nT$. In particular, the median assembly time increases with $\nT$ a small sizes due to rim curvature as discussed above, but decreases as $1/\nT$ at large sizes because the nucleation time saturates while each nucleus consumes $\nT$ subunits (see  Eq.~\eqref{eq:tHalfZandi}). Further, the growth time increases with $\nT$ (Eq.~\eqref{eq:tElong} with $\alpha=1$)\footnote{Note that a similar analysis can be performed for crystallization in a finite-sized system, to tune nucleation and growth rates so that a single nucleus forms on accessible timescales but grows to system-size before additional nuclei arise.}.

\subsubsection{Malformed assemblies} 

In the second class of kinetic traps, mis-assembled structures arise when incorrectly bound subunits do not have time to anneal before becoming trapped within the aggregate by binding of additional subunits. That is, subunit interactions must be reversible on the timescale of subunit association, even in the growth phase of aggregate assembly. This condition depends on both the cohesive interaction strength and initial monomer concentration, since the annealing time increases exponentially with interaction strength while the time interval between subunit association events decreases inversely with monomer concentration. Moreover, in the case of open-boundary assembly (and possibly some instances of curvature-controlled assembly), subunit association must occur sufficiently slowly and close to reversibility that strains due to frustration have time to propagate across the structure. Otherwise, defects or cracks that allow the system to escape frustration may occur. The conditions leading to such escape scenarios remain an open question in open-boundary assembly.

A similar trade-off between thermodynamic and kinetic considerations applies to the orientational specificity of subunit interactions. While some degree of orientational specificity is required to stabilize the target structure over other competing morphologies and to avoid mis-assembly, overly high specificity (e.g. extremely precise lock-and-key interactions) leads to small kinetic cross-sections for subunit association and thus low assembly rates \cite{Whitelam2009}.

\subsubsection{Non-equilibrium protocols for optimal kinetics} 

Finally, we briefly note that it may be possible to achieve faster assembly and/or higher assembly yields of SLA over a larger region of parameter space by using non-equilibrium protocols in which assembly driving forces are varied over time, for example by varying the temperature, solution conditions, or concentration. An intuitive approach is to use strong subunit interactions for sufficient duration to rapidly form a desired number of nuclei, and then to reduce subunit interactions to a level where post-nuclei intermediates can undergo growth but further nucleation is suppressed by a large nucleation barrier. Several approaches have been developed to use feedforward for feedback control to optimize such time varying protocols, e.g. \cite{Grover2019,Tang2016a,Tang2017,Green2019,Klotsa2013,Pineros2018}. A simpler but highly effective strategy is to seed nucleation of particular structures, e.g. \cite{Mohammed2013}. This is related to the more general context of harnessing non-equilibrium assembly pathways  to achieve size-controlled aggregates \textit{out of equilibrium}, which we return to in detail in the next section.

\section{Finite sizes by other means}

\label{sec: othermeans}

In Secs.~\ref{sec:self-limiting} and \ref{sec: models}, we overviewed the statistical thermodynamics of and known physical mechanisms for identical subunits to achieve equilibrium states with well-defined and finite dimensions.  Here, we briefly survey two broad classes of mechanisms for achieving size-controlled assemblies that fall outside of this basic paradigm:  non-equilibrium size control and programmable assemblies multiple species.

\subsection{Non-equilibrium mechanisms of size-controlled assembly}

\label{sec: nonequilibrium}

While Sec.~\ref{sec: kinetics} showed that kinetic effects introduce limitations to the practical ability to achieve equilibrium SLA, in this section we consider how non-equilibrium effects can be exploited to achieve size-control of finite assemblies even when the equilibrium states are non self-limiting.

Such nonequilibrium mechanisms leading to size-controlled assembly distributions can be further classified into two categories. (i) In `kinetically-controlled' assembly reactions, kinetic effects drive a system to a well-defined metastable state which is either sufficiently long-lived for practical applications or can be subsequently stabilized by additional reactions. Examples of such kinetically-controlled reactions include polymer/particle synthesis processes \cite{dePablo2019}, nonequilibrium formation of finite size droplets in microemulsions \cite{Woltornist2017,Woltornist2015}, flow-driven aggregate breakup \cite{Conchuir2013} and kinetically arrested coarsening \cite{Siggia1979}.
(ii) The second class, typically referred to as `non-equilibrium assembly' mechanisms, requires continual energy input into the system to stabilize the self-limited size distribution. These mechanisms use energy consumption to modify the aggregate-size-dependence of subunit association and/or dissociation rates. In both equilibrium and nonequilibrium mechanisms, a stable assemblage requires that association and dissociation rates are equal at a finite aggregate size. The resulting fixed point must also be stable, requiring that dissociation rates exceed association rates at larger aggregate sizes, and association rates are larger for smaller aggregates. These conditions are guaranteed in an equilibrium system by the criteria for self-limited assembly sizes discussed in section~\ref{sec:maxClusterSize}. However, additional nonequilibrium mechanisms to modify association or dissociation rates can also lead to stable finite sizes. For example, a number of biological mechanisms have been proposed in which assembly sizes are regulated by energy-consuming processes, such as active assembly and disassembly by molecular machines, or coupling between protein conformational states and phosphotransfer reactions (e.g. changes in a protein’s phosphorylation state or hydrolysis of a bound, nucleotide triphosphate). Similarly, assembly/disassembly rates can be modulated by externally controlled gradients in monomer concentrations or nucleation factors, as occurs during embryogenesis \cite{Briscoe2015}.

In a well-studied example, the \emph{lengths} of microtubules or actin filaments are regulated by a suite of accessory proteins that modulate assembly and disassembly rates at filament ends, as well as molecular motors that actively remove subunits from filament ends or sever filaments in their interiors \cite{Desai1997,Pollard2016,Mohapatra2016}. Moreover, the subunits themselves undergo conformational changes powered by hydrolysis of nucleotide triphosphates (ATP or GTP). The hydrolysis event shifts subunits into geometries that are incompatible with the global filament structure, thus inhibiting further assembly and/or weakening the existing structure. These nonequilibrium processes not only allow stable finite-length distributions, but also structural dynamics, such as actin treadmilling and microtubule dynamical instability, that allow the cellular cytoskeleton to rapidly respond and reconfigure to environmental cues.
Using models ranging from idealized one-dimensional filaments to geometrically realistic particle-based dynamical simulations~ \cite{Mohapatra2016,Bollinger2018,Hemmat2020,Fai2019,Tong2020}, researchers have identified multiple mechanisms by which active filament assembly/disassembly processes and energy-driven subunit conformational changes can lead to 1D filaments which exhibit dynamical instabilities and/or with well-defined stable sizes. In contrast, recall from section~\ref{sec: unlimited} that equilibrium 1D filaments generically exhibit exponential length distributions.

Other biological structures thought to be subject to non-equilibrium size regulation include  COP protein bound vesicles in the eukaryotic secretory system \cite{Foret2008}, neuronal synapses~\cite{Liu2017,Lisman2006,Lisman2015,Broadhead2016,Tang2016,Burlakov2012,Miermans2017,Shomar2017},  transcriptional regulatory complexes (enhancers)~\cite{Cho2018,Chong2018,Hnisz2017,Sabari2018}, and other phase-separated liquid domains~\cite{Weber2019,Zwicker2014,Zwicker2015}. The key characteristic of all of these systems is that the finite size of the assembled structure depends on continual energy consumption; e.g. through the replacement of `inactive' subunits with `active' ones by  the disassembly of GDP bound tubulin subunits and reassembly of GTP bound tubulin at a microtubule end, or the continual dephosphorylation and re-phosphorylation of subunits by phosphatases and kinases within a neuronal synapse \cite{Lisman2006,Lisman2015}. 

More broadly, it has been known since Turing's seminal paper \cite{Turing1952} that combining imbalances in diffusion rates with interconverting molecular species (e.g. through chemical reactions) can lead to compositional inhomogeneities with well-defined, steady state sizes, e.g.  \cite{Halatek2018,Haselwandter2011,Haselwandter2015}. More recently, advances in stochastic thermodynamics~\cite{Seifert2008,Seifert2012} have demonstrated that active processes   can play important roles in regulating the structures and functions of assembly and self-organization~\cite{Nguyen2016, Marsland2018}. Moreover, theoretical and experimental studies suggest that spatiotemporal patterns with well-defined domain sizes can occur in some active matter systems, whose constituent components consume energy at the particle scale to drive motion, e.g. \cite{Baer2020,Needleman2017,Marchetti2013,Doostmohammadi2018,Bechinger2016,Shaebani2020}.

\subsection{Addressable assembly of programmable subunit mixtures}
\label{sec: AAs}
In Sec. \ref{sec: models} above, we reviewed two broad categories of self-limiting assemblies, both of which achieve equilibrium finite-size assemblies from a single species of subunit.  Here, we briefly describe an emerging class of self-assembling systems that also realizes equilibrium finite-size assembly, but which falls (at least partly) outside of these two categories.

We refer to this class of systems as {\it  addressable assemblies} (AAs) following the terminology introduced in \cite{Jacobs2015, Jacobs2016}.  AAs are formed by mixtures of multiple assembling subunit species (say species $A, B, C, D$ {\it etc.}), each with specific interactions that selectively bind to a subset of all subunit species (e.g. $A$ binds selectively only to $D$, while $B$ binds to itself as well as $C$).  The core design principle of AAs is: one can ``program'' the matrix of species interactions to match the 3D adjacency matrix of a desired structure, perhaps uniquely, such that this target structure becomes the equilibrium assembly state in mixtures of controlled subunit stoichiometry~\cite{Hormoz2011}. That is, each particle has an ``address'' (or set of addresses) where it sits in the 3D target assembly.  

Examples of AAs include colloids and nanoparticles functionallized by single-stranded DNA tethers that mediate interactions via complementary base pairing~\cite{Jones2015}, as well as ``DNA bricks'' assembled by  mixtures of oligmeric DNA strands whose sequences are designed to interleave ends into 3D patterns via complementary base pairing~\cite{Ke2012}.  Each of these systems has been considered and studied for its potential to selectively design and assemble target superstructrures that terminate at specifically predetermined dimensions~\cite{Zeravcic2014}.  In the sense that the self-limiting target structure can be designed to be stress-free with each cohesive bond ideally satisfied, finite AAs can be thought of as multi-species analogs to self-closing structures, albeit with more complex bond networks.  As an example, rectangular beams with precisely defined and finite cross-sectional dimensions were assembled using the programmable DNA bricks \cite{Ke2014}.  The different values of target finite dimension each required  mixtures composed of different numbers of distinct oligomeric species.  For example, finite width beams of $6 \times 6$, $8 \times 8$ and $10 \times 10$ dsDNA helices across were assembled, respectively, from mixtures containing 60, 112 and 180 distinct oligomers.  Generically, equilibrium termination at specific size via AA requires a number of distinct subunit species $n_{\rm S}$ that grows with the finite target size $W_*$~\cite{Ong2017}, presumably with some power law $n_{\rm S} \sim W_*^{\beta}$ (e.g. $\beta =2$ if distinct subunits are required in every 2D cross-section of a beam).  

The unbounded growth of the number of subunit species with target size for such an implementation of AA would seem to limit its practical applicability, due to the cost (in terms of design, synthesis, and processing) and limited scalability of the programmable mixture. This raises a basic question about AA: Are there optimal strategies that minimize the complexity (e.g. number of distinct species) of a subunit mixture needed to achieve self-limiting AAs of a given topology, and how do these scale in the limit of large target size?  The self-assembly of viral shells, or capsids, may provide clues for how to approach this question.

Capsids are ``crystalline'' shells composed of protein subunits (capsomers) that self-assemble to enclose the viral genome.  Since the enclosed genome has to code for the capsomers themselves, it has been understood since Crick and Watson's seminal paper \cite{Crick1956} that such assemblies should be {\it economical}. I.e., a viral capsid is under selective pressure to enclose the largest possible volume using the minimal number of distinct capsomer types.  Caspar and Klug (CK) proposed that viruses achieve this optimization by exploiting symmetry principles \cite{Caspar1962}. In their well-known construction, quasi-spherical capsid structures are mapped to high-symmetry triangulations of the sphere, in which each triangle is constituted by 3 capsomeric subunits~\cite{Prasad2012}.  Based on this reasoning, CK conjectured that ``optimal'' capsids correspond to sub-triangulations of the icosahedron.  Each CK structure can be classified by the number $T=h^2+k^2+hk$ of sub-triangles per each of the 20 triangular faces of the icosahedron, where $h$ and $k$ are positive integers.  Since  regular triangulations of the sphere are not possible beyond those corresponding to Platonic solids, CK noted that higher triangulation numbers require capsomers to accommodate different local environments, corresponding to variations in neighbor spacing, orientation, and the number of neighbors.  In CK capsids, the triangulation number $T$ is the number of symmetry-inequivalent capsomer positions. 

In this way, we might view CK capsids as a highly-symmetric and economical limit of AA.  While the complete shell could be assembled from $60 T$ distinct and specifically interacting capsomeric subunits, the large number of symmetry elements of the icosahedral net implies that there are many redundancies in such a design, and in fact, the same unique target structure could be realized from specific interactions of only $T$ distinct units.  This approach has been adopted by the protein design community, enabling researchers to engineer proteins that assemble into icosahedral shells of various sizes \cite{Bale2016,Butterfield2017,King2014,Lai2014,Mosayebi2017}.
Recently, the CK design principles have been repurposed for the de novo design of triangular DNA origami particles with precisely defined geometry and edge interactions that selectively assemble into $T$ icoshedral shells~\cite{Sigl2020}. Notably they show such capsids can be uniquely programmed and assembled from even a fewer number of distinct subunit species, $\lceil T/3 \rceil$, because each triangle can have three inequivalent edges.

In its simplest implementation, a CK capsid can be assembled by synthesizing a distinct subunit species for each of the $T$ symmetry-equivalent positions. While some small viruses follow this approach, it becomes increasingly impractical as the target capsid size grows, since the number of distinct species scales linearly with the capsid area. In practice, there are two mechanisms to reduce the number of distinct subunits that need to be synthesized by a virus (or by any other manufacturer). In many viruses, the capsid protein  interconverts between different `quasi-equivalent' conformations with slightly different interaction geometries that accommodate the different local symmetry environments in the capsid, thus enabling assembly of capsids with $T>1$ from a single subunit species \cite{Caspar1962,Johnson1997}. More broadly, it has been shown that such high symmetry constructions correspond to free energy minima of assembled structures with spherical topologies with relatively generic types of short-range cohesive subunit-subunit interactions, although the free energy minimum symmetry depends on the size of the assembled structure \cite{Chen2007,Zandi2004}.  Correspondingly, CK-like capsids emerge naturally from the assembly of elastic structures from subunits resembling the tapered subunits discussed in section~\ref{sec: amphiphiles}~\cite{Chen2007,Fejer2010,Reguera2019,Lazaro2018,Lazaro2018a}, and many of the principles discussed in that section can be extended to systems that form CK capsids.
However, the assembly dynamics of such structures remains an open question --- how does the `right' subunit conformation end up in the appropriate location within an assembling capsid~\cite{Berger1994,Elrad2008,Panahandeh2020,Panahandeh2018,Li2018,Zandi2020,Perlmutter2015,Twarock2018,Stockley2013,Perkett2016,Morton2010}? More recently, the CK construction has been extended to account for other viruses in which capsomers accommodate more extreme differences in local environments and other symmetry classes~\cite{Twarock2004,Luque2010a,Twarock2019}. In a second strategy, finite shells are assembled from a single (or few) subunit types which do not adopt explicitly distinct conformations, but are sufficiently deformable to accommodate different local symmetry environments by the formation of inhomogeneous strains within the capsid structure. It is easy to imagine that these two strategies (different conformations with specific interactions or subunits with deformable interaction geometries) could be combined to extend the size and complexity of a shell that can be assembled by the CK mechanism.

Returning to the context of AA, CK constructions and their extensions can be viewed as limiting cases of AA in which one uses symmetry to putatively minimize the number of distinct subunits types need to enclose a given volume. The CK framework suggests that there is a trade-off between the complexity of the mixture of distinct subunits and the complexity, or asymmetry,  of the target assembly (i.e. fewer symmetry elements in the target structure implies more subunit species). This idea can be extended more broadly to designing self-limiting AAs that target other, non-spherical topologies. In this context, a more precise, and potentially useful notion of ``economy'' may be to consider the ratio of target finite size to number of subunit types $W_*/n_{\rm S}$, and ask what are the analogs to CK designs for arbitrary topologies that maximize this ratio in the limit of increasing $W_*$.

\section{Concluding Remarks: Self-Limiting Assembly by Discovery and Design}

In this review we have attempted to provide a unified theoretical perspective on assembly processes that, while occurring in chemically and physically diverse systems, share the common thread of autonomously terminating at a well-defined equilibrium finite size. Starting from the framework of ideal aggregation theory, we showed that the necessary conditions for such self-limited assembly are linked to the existence of a minimum in the size-dependent aggregation energetics $\epsilon(n)$. In contrast, systems which do not meet these conditions generically assemble bulk structures such as crystals. We saw that properties of self-limited assembly reactions can be identified from the functional form of $\epsilon(n)$, including the onset of aggregation with increasing subunit concentration and the size of fluctuations around the optimal aggregate size. Systems with multiple local minima in $\epsilon(n)$ can exhibit polymorphic self-limited assembly, with concentration-dependent transitions (secondary CACs) between different aggregation states. We saw that the small, but non-zero, translational entropy of the aggregates plays a key role in driving secondary CACs. 

The existence of local minima at non-trivial sizes $n_{\rm T} \gg 1$ implies that the energetics of assembly, as mediated by the shape and interactions between subunits themselves, requires the ability to ``sense'' the aggregate size  on  scales comparable to the optimal size.  Surveying known examples of equilibrium SLA from identical subunits, we argued that there are two broad classes of physical mechanisms that achieve this.  Either the subunit interaction geometries encode a target assembly geometry that ``returns to itself'' after a characteristic number of subunits, or instead, there is a source of intra-assembly stress gradients that can propagate up to the finite-size scale of the assembly. An important feature of the latter mechanism is that it enables finite-sized aggregates that have open boundaries. We presented generic descriptions for understanding each of these mechanisms, including how to assess the limits to the finite size that can be achieved. Toward that end, we considered mechanisms by which the system can ``escape'' finite size, resulting in bulk assembly products. For example, a system can curtail the accumulation of stress gradients with assembly size by expelling strain to its periphery or by locally relaxing strain through defects. We presented distinct physical mechanisms that can give rise to accumulating stress gradients: the interplay between short-range attractions and long-range repulsions, or an incompatibility between the preferred local subunit packing and the large-scale assembly geometry. We also presented example experimental systems that may correspond to each of these SLA mechanisms.

In part, our purpose in spotlighting the relatively rarefied conditions required for equilibrium self-limitation is to reframe broadly open challenges in understanding and engineering SLA.  One such challenge could be described as the experimental inference of SLA.  That is, for a given set of experimental observations of assembly, is it possible to determine if finite aggregates are the result of equilibrium self-limitation?  This is a particularly vexing issue for experimental observations of both synthetic and biological assemblies that appear to be finite and well-defined.  While it is often desirable to link the observations to specific microscopic models that recapitulate aspects of finite assembly {\it a posteriori}, such models often require assumptions about the interactions and energetics of complex subunits that are poorly understood.  Thus, it is challenging to rule out alternative mechanisms of kinetic trapping of assembly in such systems.  As one example, in living tissues collagen forms fibrillar assemblies that appear to have well-defined diameters reaching up to microns, well beyond the nm-scale width of a single pro-collagen molecule~\cite{Ottani2002}. Moreover, the mean diameter varies considerably between tissue types.  Fibers in tendons have mean diameters in excess of $\gtrsim 1 ~{\rm \mu m}$, while those found in corneal tissue have a tighter distribution around $\lesssim 50 {\rm nm}$~\cite{Wess2008}.  Such observations combined with the functional needs of these different tissues~\cite{Meek2009}, high stiffness vs. optical transparency respectively, suggest a need to regulate the assembly of the same subunits to form architectures of \emph{tunable} finite size.  Indeed, physical models have proposed mechanisms of geometric frustration deriving from the chiral organization within fibers as a means of imposing a self-limited diameter~\cite{Turner2003, Grason2007, Yang2010, Brown2014}.  While these explanations are plausible and broadly consistent with observations of finite-diameter collagen fibers and other fibrous biofilament structures, these physical models require knowledge of parameters describing chiral inter-molecular forces and inter-molecular mechanics, which are difficult to predict under conditions relevant to assembly~\cite{Grason2020}.  Without direct knowledge of these inter- and intra-molecular parameters, not to mention the non-equilibrium conditions of assembly, at best, such models can plausibly explain the finite size of fiber diameters. Developing generalizable experimental methodologies that strictly prove (or disprove) mechanisms of equilibrium self-limitation, particularly from complex biomolecular subunits for which the intermolecular aggregation energetics is poorly understood, remains a more distant and unmet goal.

A related challenge is to use advances in synthetic techniques to design and engineer self-limiting assemblies that target {\it a priori} finite dimensions.  Great advances have been made in designing shape-controlled particles~\cite{Glotzer2007, Sacanna2011} whose symmetries and interactions direct assembly to targeted structures. However, assembly targets have thus far been largely restricted to various bulk structures (albeit with complex unit cells), or  1D or 2D aggregates of uncontrolled ultimate size.  Alternately, the field of ``supramolecular chemistry'' has leveraged chemical synthesis of an outstanding variety of architecturally- and compositionally-defined macromolecules to direct their assembly. The chemical control over these ``precision amphiphiles'' has significantly increased the ability to rationally design and form periodic mesophases, thermotropic supramolecular crystals, or liquid crystals~\cite{Su2020}.  However, while the symmetries of these phases have become increasingly complex, they remain bulk structures.  Synthetic advances in amphiphile assembly have largely focused on imbuing micellar assemblies with functional properties, such as controlled uptake and release of drugs~\cite{Geng2007, Oltra2014}.  While micellar assemblies are finite-size in terms of diameter, as in the case of traditional surfactants, the finite size remains limited to the size of the molecules that span the aggregate core.  Thus, notwithstanding tremendous advances in synthesizing shape- and interaction-controlled subunits, controlling the finite-size of target superstructures, particularly on size scales much larger than the subunits themselves, remains a relatively unexplored aspect of engineered assemblies.  While recent advances in methods such as DNA nanotechnology seem to pave the way to geometric control of subunits needed to realize bioinspired capsules and tubules~\cite{Rothemund2004,Tian2014,Benson2015,Sigl2020}, it remains to be explored what the experimentally realizable upper limits to finite sizes are, and what mechanisms of self-limitation are needed to reach this limit.

A further challenge is to design supramolecular structures that do not have a single finite size but, rather, can exist in multiple different sizes, all of which are stable. Such classes of structures enable essential functions in biology. For example, recent evidence suggests that a neuronal synapse changes size during long-term memory storage, but then must remain stable at that size over the lifetime of a memory \cite{Tang2016,Lisman2006}. Designing such variable-size stable structures is also becoming of interest to nanomaterials science, since materials capable of ``learning'' or ``remembering'' multiple stable configurations~\cite{Murugan2015, Zhong2017} could adapt their structures to store information, self-heal, or respond to environmental cues. Despite this interest and insights from biology, the principles underlying such variable-size stable structures remain far from clear. In this context, it would be of interest to extend the considerations in section~\ref{sec:SecondCMC} of secondary CACs, to understand more broadly how the interplay between aggregate translational entropy and interaction energies can lead to controllable transitions between structures with different finite number sizes and/or dimensionalities. Similarly, can these principles be combined with the concepts of nonequilibrium assembly to design subunits that are preprogrammed to organize into nanoscale machines capable of autonomously manipulating matter or performing other functions currently found only in living organisms?

\begin{acknowledgments}
The authors are grateful to D. Hall, N. Hackney, B. Tyukodi, L. Tsidilkovski, B. Rogers, M. Lenz, A. Zlotnick, R. Jack, E. Efrati, P. Charbonneau, R. Zandi, and B. Gelbart for valuable discussions and useful feedback on this manuscript.  Support for this work was provided through the NSF through the the Brandeis Center for Bioinspired Soft Materials, an NSF MRSEC, DMR-1420382 and DMR-2011846  (to M.F.H. and G.M.G.), and grant no. DMR-2028885 (to G.M.G.); and the NIH through Award Number R01GM108021 from the National Institute Of General Medical Sciences (to M.F.H).
\end{acknowledgments}

\bibliography{refsMerged}

\appendix
\section{Polymorphic amphiphile assembly phase diagram}
\label{sec:SurfAppendix}


Here we summarize the calculation of the polymorphic assembly phase diagram for the amphiphile model in  Fig.~\ref{fig:phasecp}.
The aggregate energy for each of the dimensionalities is given by Eq.~\eqref{eq: amphi}. To determine the phase diagram, we calculate the following law of mass action for subunit populations:
\begin{align}
\Phi = \phi_1 + \Phi_3 + \Phi_2 + \Phi_1
\label{eq:}
\end{align}
with $\Phi_{d_{\rm L}}$ the mass fraction of subunits in spherical, cylindrical, or planar aggregates for $d_{\rm L}=3,2,1$ respectively, and $\phi_1$ the free monomer population

Adopting the continuum limit, the mass fraction of subunits in spheres is given by
\begin{align}
\Phi_{3, {\rm cont.}} & = \int_0^\infty dr 4 \pi r^2 n_0 \phi_3(r) \nonumber \\
\phi_3(r) &=\nSPH(r) \exp\left[-\left(\eSPH(r)-\mu\right) \nSPH(r)\right]
\label{eq:PhiSph}
\end{align}
with $\eSPH(r)=\epsilon(r,3)$ from Eq.~\eqref{eq: amphi}, $\nSPH(r)= \frac{4}{3}\pi r^3 n_0$, and $n_0=v_0^2/a_0^3$ a dimensionless number density. 

Likewise, the mass fraction in spherocylinders is
\begin{align}
\Phi_2 &=  \int_\lBar^\infty dL \int_0^\infty dr 2 \pi r n_0 \phi_2(r,L) \nonumber \\
\phi_2(r,L)&=\left(\nSPH(r)+\nCYL(r,l)\right) \times \nonumber \\
& \exp\left[-\left(\eSC(r,L)-\mu\right) \left(\nSPH(r)+\nCYL(r,L)\right)\right]
\label{eq:PhiSphCyl}
\end{align}
with $\lBar$ set by the minimal length of the stable spherocylinder branch (calculated below), $\nCYL(r,L)=n_0 \pi r^2 L$, and 
\begin{align}
\eSC(r,L)=\frac{\nSPH(r)\eSPH(r) + \nCYL(r,l)\eCYL(r)}{\nSPH(r)+\nCYL(r,L)}
\label{eq:eSC}
\end{align}
with $\eCYL(r)=\epsilon(r,2)$ from Eq.~\eqref{eq: amphi}. Note that the first term in the numerator of Eq.~\eqref{eq:PhiSphCyl} corresponds to the ``endcap energy'' arising from the hemispherical cap at either end of the spherocylinder. We make the simplest assumption, that the radius of the hemispherical cap is equal to that of the cylindrical portion of the micelle, so that solvophobic tails remain shielded from solvent contact at the cylinder-endcap connection.  More realistic models consider lower energy shapes that smoothly connect ``bulbous" ends to cylindrical cores~\cite{May2001}.

Finally, $\Phi_1$ is the mass fraction in layers, and is calculated below.

To proceed, recall from section~\ref{sec:maxClusterSize} that for the case of a minimum in the aggregate size around an optimal size $n_{\rm T}$, fluctuations vanish in the limit of large $n_{\rm T}$ or $\epsilon''|_{n_{\rm T}}$. Specifically, consider a generic aggregate energy function $\epsilon(n)$ with a minimum $\epsilon_*$ at the optimal size $n_{\rm T}$. The mass fraction of subunits in aggregates is then given by 
\begin{align}
\Phi_{\rm T} = \int_0^\infty dn\, n \exp[\mu-\epsilon(n)].
\label{eq:PhiT}
\end{align}
Performing a saddle point as in section~\ref{sec:maxClusterSize} then results in
\begin{align}
R_\text{fluc}=\frac{\Phi_{\rm T}}{\Phi_{\rm T}(n_{\rm T})} = \sqrt{\frac{2 \pi}{\epsilon''|_{n_{\rm T}} n_{\rm T}}}
\label{eq:Rfluc}
\end{align}
with $\Phi_{\rm T}(n_{\rm T})=n_{\rm T} e^{-(\epsilon_*-\mu)n_{\rm T}}$ the mass of subunits in aggregates if fluctuations are neglected. Thus, we see that when $\epsilon''|_{n_{\rm T}} n_{\rm T} \gtrsim 1$ the width of the distribution smaller than a subunit, and fluctuations are negligible. For spherocylinders, we see that the contribution due to fluctuations in the radial direction $r$ diminishes with length as $R_\text{fluc}\thicksim L^{-1/2}$. Thus, even for narrow spherocylinders, fluctuations become negligible in the large-length limit.

With this in mind, we account for polydispersity in micelles as follows. We first calculate the optimal radius for spherical aggregates, $\rSphOpt$ by minimizing $\eSPH(r)$ for given values of $k$ and $P$. We then numerically calculate the location of the barrier between the spherical and spherocylindrical branches of Eq.~\eqref{eq:eSC}, as $\lBar=\argmax{L}\left(\underset{r}{\min}\, \eSC(r,L)\right) $; i.e., the length at which a spherocylinder of optimal radius has a maximum energy per particle. The number of particles at the barrier is then given by $\nBar=\nSPH(\rBar)+\nCYL(\rBar)$ with $\rBar=\argmin{r}\eSC(r,\lBar)$ the optimal radius at the barrier.

To calculate the mass of spherical micelles, we numerically integrate the expression in Eq.~\eqref{eq:PhiSph}. To maintain the assumptions of section~\ref{sec: amphiphiles}, we perform the integral over the range $n\in [\nSphOpt,\nBar]$ with $\nSphOpt=\nSPH(\rSphOpt)$, although the result is largely insensitive to increasing these integration bounds. We then include fluctuations only when they exceed the size of a single subunit by setting 
\begin{align}
\Phi_3=\max\left[\Phi_{3,{\rm cont.}}, \nSphOpt e^{-\left(\eSPH(\nSphOpt)-\mu\right) \nSphOpt}\right].
\label{eq:Phi3Max}
\end{align}
where the second argument is simply the concentration of micelles at the optimal size.

For spherocylinders we make the simplifying assumption that radial fluctuations can be neglected at all lengths, and take the optimal radius $\rSCOpt(L)=\argmin{r} \eSC(r,L)$ as a function of spherocylinder length $L$. In practice, we found that the numerics are more tractable if the integral is performed over particle number $n$ rather than spherocylinder length $L$, so we calculate the optimal spherocylinder length $\lSCOpt(n)= \argmin{L} \eSC(\rSCOpt(n,L),L)$ with $\rSCOpt(\hat{n},L)$ determined from the volume of the spherocylinder: $\nSPH(\rSCOpt)+\nCYL(\rSCOpt,L)=\hat{n}$. To make the integral numerically tractable, we perform the integral to a predefined (large) size $\nMax$, beyond which we assume that the effect of changing radius from the hemispherical caps is negligible, so that the optimal radius is given by the minimum of the cylinder energy, $\rSCOpt \approx \rCylOpt=\argmin{r} \eCYL(r)$, and the energetics becomes simply the 1D energetics of the form in sec. \ref{sec: unlimited}:
\begin{align}
\Phi_2 \approx & \int_{\nBar}^{\nMax} dn\, \Phi_2(\rSCOpt(n, \lSCOpt(n)), \lSCOpt(n)) + \Phi_\infty(\nMax)
\label{eq:Phi2Approx}
\end{align}
with the contribution from the spherocylinders with sizes larger than $\nMax$ given by
\begin{align}
\Phi_\infty(\nMax) = e^{-\eCap} e^{\left(\mu-\eCylOpt\right) \nMax } \frac{1 + (\eCylOpt-\mu) \nMax} {\left(\eCylOpt-\mu\right)^2}
\label{eq:PhiInfty}
\end{align}
with $\eCylOpt=\eCYL(\rCylOpt)$ the energy per particle within the cylindrical region and $\eCap = \nSPH(\rCylOpt) \left( \eSPH(\rCylOpt) - \eCylOpt\right)$ the total \emph{extra} energy that arises due to the unfavorable hemispherical caps. 

Finally, the fraction of subunits in layers $\Phi_1$ is calculated by noting that the free subunit chemical potential can never exceed the chemical potential of a subunit in a sheet, $\phi_1 \le e^{\eLamOpt}$ with $\eLamOpt=\underset{r}{\min}\, \epsilon_{\rm layer} (r)$, where $\epsilon_{\rm layer} (r)= \epsilon(r,1)$ from eq. (\ref{eq: amphi}). Thus, the amount of subunits in layers and the corresponding free subunit concentration are given by mass conservation as
\begin{align}
\Phi_1 = & \max\left[ \Phi - \Phi_3 - \Phi_2 - e^{\eLamOpt}, 0 \right] \nonumber \\
\phi_1 = & \Phi - \Phi_3 - \Phi_2 - \Phi_1,
\label{eq:Phi1}
\end{align}
effectively treating layers as an unlimited bulk phase, with negligible edge energy.

Phase boundaries in Fig.~\ref{fig:phasecp}A are calculated from Eq.~\eqref{eq:PhiSph} - Eq.~\eqref{eq:Phi1} by determining the total subunit concentration at which the fraction of subunits in an aggregate of a given dimensionality exceeds 50\%. That is, the concentrations corresponding to transitions between monomers/layers, layers/spherocylinders, and spherocylinders/spheres are calculated as the lowest total concentration at which $\Phi\left(\Phi_1=0.5\right)$, $\Phi\left(\Phi_2=0.5\right)$, and $\Phi\left(\Phi_3=0.5\right)$ respectively. The concentrations in Fig.~\ref{fig:phasecp} are normalized by the CAC for cylinders, $\phi^*=e^{\eCylOpt}$. Note that $\phi^*$ is the only result within this section that depends on the cohesive energy strength $\epsilon_0$; the relative concentrations corresponding to the transitions depend only on the elastic and boundary energy terms.

The CAC ratio shown within the spheres/spherocylinders coexistence region in Fig.~\ref{fig:phasecp}B is computed as $\Phi\left(\Phi_3=0.5\right) / \Phi\left(\Phi_2=0.5\right)$. 
The infinite-concentration transitions, shown as solid red and blue lines in Fig.~\ref{fig:phasecp}B, are calculated respectively from $\eLamOpt(\bar{k},P)=\eCylOpt(\bar{k},P)$ and $\eCylOpt(\bar{k},P)=\eSphOpt(\bar{k},P)$. 

Finally, the boundary of the  spheres/spherocylinders coexistence region (solid green line in Fig.~\ref{fig:phasecp}B) is estimated as the minimum concentration at which either the height of the barrier between the sphere and spherocylinder branches goes to zero (i.e. corresponding to $\delta=0$ in Eq.~\eqref{eq: gap} of section~\ref{sec:SecondCMC}) or the point at which the transition concentration from spheres to spherocylinders becomes equal to the threshold concentration for assembling cylinders in the absence of other aggregates Eq.~\eqref{eq:deltaStar}.   

\section{Continuum elastic theory of frustrated chiral ribbons}
\label{sec: ribbon}

Here we present details of the ``narrow ribbon'' theory of frustrated chiral ribbons.  As the model has been described elsewhere~\cite{Ghafouri2005, Armon2014, Grossman2016}, our primary aim is to present provide more details on the physical ingredients of the model, and further to describe how the elastic instability of wide helicoids quantitatively alters the picture of ``frustration escape'' presented in the Sec.~\ref{sec: GFA}.

Following the approach of \cite{Ghafouri2005}, we consider a simplified theory that describes the shape of ribbons in terms of the surface curvature tensor $C_{ij}$ along the mid-line of ribbons, written in terms of coordinate directions $\hat{x}_e$ and $\hat{x}_p$ that point, respectively, along and perpendicular to the wide direction of the ribbon (see e.g. Fig.~\ref{fig: ribbons}A).  Specifically, this assumes that the in-plane curvatures vary little away from the mid-line of the ribbon, which is strictly valid when the ribbon widths are narrow with respect to their curvature radii.  Since the ribbons effective ``flatten'' in shape as they grow wider, this narrow-ribbon approximation provides at least a qualitative picture of the ribbons' thermodynamics over the entire range of the widths.

The excess energy derives from two elastic contributions,
\begin{equation}
    E_{\rm elast} =  E_{\rm intrinsic} + E_{\rm extrinsic} ,
\end{equation}
where the first term depends on the intrinsic geometry, or metric distortions away from a planar 2D lattice, while the second term describes the elasticity of the extrinsic geometry of the ribbon, i.e. , a generalized form of its bending energy expressed as quadratically in terms of curvature elements $C_{ij}$.  The former term is captured by a 2D elastic energy,
\begin{equation}
    E_{\rm intrinsic} = \int dA~ \sigma_{ij} u_{ij} ,
\end{equation}
with $u_{ij}$ and $\sigma_{ij} \approx Y u_{ij}$ as the respective in-plane stress and strain of the 2D crystal ribbon order, and $Y$ as the 2D Youngs modulus of the crystal~\cite{Seung1988}.  Assuming that the crystalline packing favors uniform inter-subunit spacing, there is a geometrical and mechanical coupling between in-plane stress and out-of-plane deflections described by the so-called compatibility equation,
\begin{equation}
    \nabla_\perp^2 \sigma_{ii} = -Y \KG ,
\end{equation}
where $\sigma_{ii}=\sigma_{pp}+\sigma_{ee}$ and $\KG \simeq C_{ee} C_{pp} - C_{ep}^2$ is the Gaussian curvature (neglecting variations of $\KG$ across the ribbon width). Assuming approximately uniform negative curvature, it is straightforward to show that $E_{\rm intrinsic}/(W L) =Y \KG^2 W^4/1440$~\cite{Ghafouri2005, Grason2016}.  

The extrinsic energy takes the form of the generalized bending energy
\begin{equation}
    E_{\rm extrinsic} = \frac{1}{2} \int dA~ \Big[B_{pp} C_{pp}^2+B_{ee} C_{ee}^2+2 B_{ep} (C_{ep}-\Omega_0)^2 \Big]   
\end{equation}
where $B_{ij}$ are bending coefficients for different curvature elements.  Here we consider the case of $B=B_{pp}=B_{ee}=B_{ep}$; relaxing this restriction does not alter the qualitative behavior.   Symmetry considerations (i.e. lack of inversion symmetry) argue that chirality at the subunit scale generates a linear coupling to the off-diagonal curvature component, which we define as the ribbon twist,
\begin{equation}
    \Omega \equiv C_{ep}
\end{equation}
and hence $\Omega_0$ can be associated with the preferred rotation, or {\it twist}, of the tangent plane along the edge or pitch axis~\cite{Helfrich1988}. At a subunit scale, this preferrence for mesoscopic twist derives form a energetic preferrence for locally skewed packing of molecules in the membrane~\cite{Zhang2019}, although a predictive understanding of the relationship between preferred pitch and structure of constituent chiral molecules is a notoriously complex and long-standing issue in and of itself, see e.g.~\cite{Harris1999}.
  
Combining these together and dividing by the number of subunits per ribbon $n = W L/a_0$ gives the excess energy
\begin{multline}
    \epsilon_\text{ex}  (W)= \frac{Y a_0 }{1440} \big(C_{pp}C_{ee}-C_{ep}^2\big)^2 W^4  \\ + \frac{B a_0}{2} \big[C_{pp}^2+C_{ee}^2+2 (C_{ep}-\Omega_0)^2 \big] ,
    \label{eq: ribbon}
\end{multline}
where again, the curvature components in this expression are taken to correspond in to the ribbon shape at the mid-line, and further assumed to be constant along the length of the ribbon.  Reconsidering the comparison to the heuristic description in Eq. (\ref{eq: epex2}), inspection of the chirality-frustrated ribbon energy in Eq. (\ref{eq: ribbon}) shows that Gaussian curvature plays the role of frustration strength (i.e. $f \to \KG\simeq {\rm det} (C_{ij})$), while the shape parameter can be captured by the curvature tensor (i.e. $s \to C_{ij}$) with a preferred (tensorial) shape component that is non-zero only along the off-diagonal elements.  

The shape equilibrium of the ribbon varies with width from the roots of the equations
\begin{equation}
    \frac{ \partial \epsilon_\text{ex} }{ \partial C_{ee} } =   \frac{ \partial \epsilon_\text{ex} }{ \partial C_{pp} } =  \frac{ \partial \epsilon_\text{ex} }{ \partial C_{ep} } =0
\end{equation}
for  fixed $W$.  The shape equilibrium is characterized by two branches.  The first is the {\it helicoidal} branch,
\begin{equation}
\left.
\begin{array}{c}C_{ee}=C_{pp} = 0 \\ \\
 C_{ep} +\frac{Y W^4}{720 B} C_{ep}^3 =\Omega_0  \end{array} \right\} \ {\rm ( helicoid )}.
\end{equation} 
This is the branch of equilibria discussed in the main text.  Helicoid twist tends to its preferred value for narrow ribbons $C_{ep}(W \ll W_{\rm un}) \simeq \Omega_0$; and the helicoid unwinds in the wide limit as $C_{ep}(W \gg W_{\rm un}) \simeq \Omega_0 (W_{\rm un}/W)^{4/3}$, where the {\it unwinding size} $W_{\rm un} \equiv \big( 720 B/Y \Omega_0^2\big)^{1/4}$, defined in eq. (\ref{eq: unwinding}, characterizes the crossover width between the these two regimes.

The second branch corresponds to a symmetry-breaking transition to a {\it spiral ribbon} shape, 
\begin{equation}
\left.
\begin{array}{c}C_{ee}=C_{pp}= \pm \sqrt{ C_{ep}^2 - \Omega_0^2 ( W_{\rm un}/W )^4 }\\ \\
C_{ep} =\Omega_0/2 \end{array} \right\} \ {\rm ( spiral \ ribbon)} .
\end{equation} 
Note that this branch only exists above a critical width $\Wc = \sqrt{2} W_{\rm un}$, for which $C_{ee}$ and $C_{pp}$ are real.  

The critical value corresponds to an elastic instability.  For $W<\Wc$ the helicoid branch is stable, and no spiral equilibrium exists.  For $W\geq \Wc$, the helicoid branch becomes unstable, and the stable branches become the two degenerate spiral states, which differ by signs of $C_{ee}$ and $C_{pp}$.  The shape equilibria and excess energy of both branches are plotted in Fig. ~\ref{fig: ribbons}B and C respectively.

Notice that for wide ribbons, the Gaussian curvature of both branches vanishes:  for (unstable) helicoids, $\KG=-C_{ep}^2(W \gg W_{\rm un}) \approx - \Omega_0^2 (W_{\rm un}/W)^{2/3}$; and for (stable) spiral ribbons, $\KG(W \geq W_c)=- \Omega_0^2 (W_{\rm un}/W)^{4}$.  The difference in power law suggests a much more rapid expulsion with Gaussian curvature with increasing width of spirals.  Notwithstanding the faster ``frustration escape'' of spiral ribbons, the accumulant analysis shown in the inset of Fig. ~\ref{fig: ribbons}C predicts that the maximum self-limiting size preempts the mechanical stability with $W_{\rm max} = 0.85 W_{\rm un}< \Wc = \sqrt{2} W_{\rm un}$.  Hence, a generic prediction of this narrow ribbon model is that self-limiting ribbons can only be helicoidal in shape \cite{Ghafouri2005, Armon2014}. 
\section{Assembly timescales and kinetic traps}
\label{sec:KineticsAppendix}

From Eq.~\eqref{eq:CNTBarrier}  the initial nucleation timescale for the capsid model is
\begin{align}
\tNuc(\phi) \approx \frac{\exp\left[-\beta\Omega(\nCrit(\phi))\right]}{Z \phi \fAssem(\nCrit(\phi))}.
\label{eq:tNucZandiApp}
\end{align}
with $Z=\sqrt{\frac{\beta \lamTilde}{\pi \nT}}\left(1+\Gamma^2\right)^{3/4}$ the Zeldovich factor that accounts for the time the system spends in the vicinity of the critical nucleus and $\fAssem(\nCrit)\approx \fAssem$.

The median assembly time is then given by using Eq.~\eqref{eq:tNucZandiApp}  to integrate  the cumulative depletion of monomers as a function of time, approximately accounting for reversibility of the reaction as in \cite{Hagan2010}, resulting in 
\begin{align}
 \tHalf \approx 2^{\nCritZero-1} / \left( \nCritZero-1\right) \fc  \frac{\tNuc(\Phi)}{\nT}
 \label{eq:tHalfZandiApp}
 \end{align}
in which we made the approximation that the critical nucleus size remains constant over time, $\nCrit(\phi_1(t))\approx \nCritZero\equiv\nCrit(\Phi)$.

\end{document}